\documentclass[preprint,12pt]{elsarticle}

\usepackage{amssymb}
\usepackage{graphicx}
\usepackage{epstopdf, epsfig}
\usepackage{lineno,hyperref}
\usepackage{booktabs} 
\usepackage{array} 
\usepackage{amsmath}
\usepackage{mathtools}
\usepackage{graphicx,caption,subcaption,color} 

\journal{Computers and Fluids}

\begin{document}

\begin{frontmatter}

\title{A comparative study of immersed boundary method and interpolated bounce-back scheme for no-slip boundary treatment in the lattice Boltzmann method: Part I, laminar flows}

\author[label1,label2]{Cheng Peng\corref{cor1}}
\ead{cpengxpp@udel.edu}
\author[label3]{Orlando M. Ayala}
\ead{oayala@odu.edu}
\author[label1,label2]{Lian-Ping Wang}

\address[label1]{Department of Mechanics and Aerospace Engineering, Southern University of Science and Technology,
Shenzhen 518055, Guangdong, China}
\ead{lwang@udel.edu}

\address[label2]{126 Spencer Lab, Department of Mechanical Engineering, University of Delaware, Newark, DE, USA, 19716}

\cortext[cor1]{Corresponding author}

\address[label3]{111A Kaufman Hall, Department of Engineering Technology, Old Dominion University, Norfolk, VA, 23529, USA}

\begin{abstract}
The interpolated bounce-back scheme and the immersed boundary method are the two most popular algorithms in treating a no-slip boundary on curved surfaces in the lattice Boltzmann method. While those algorithms are frequently implemented in the numerical simulations involving complex geometries, such as particle-laden flows, their performances are seldom compared systematically over the same local quantities within the same context. In this paper, we present a systematic comparative investigation on some frequently used and most 
state-of-the-art interpolated bounce-back schemes and immersed boundary methods, based on both theoretical analyses and numerical simulations 
of four selected 2D and 3D laminar flow problems. Our analyses show that immersed boundary methods (IBM) typically yield a first-order accuracy when the regularized delta-function is employed to interpolate velocity from the Eulerian to Lagrangian mesh, and the resulting boundary force back to the Eulerian mesh. 
This first order in accuracy for IBM is observed for both the local velocity and hydrodynamic force/torque, apparently different from the second-order accuracy
sometime claimed in the literature.
Another serious problem of immersed boundary methods is that the local stress within the diffused fluid-solid interface tends to be significantly underestimated. On the other hand, the interpolated bounce-back generally possesses a second-order accuracy for  velocity, hydrodynamic force/torque, and local stress field. The main disadvantage of the interpolated bounce-back schemes is its higher level of fluctuations in the calculated hydrodynamic force/torque when a solid object moves across the grid lines. General guidelines are also provided for the necessary grid resolutions in the two approaches in order to accurately simulate  flows over a solid particle.
\end{abstract}


\end{frontmatter}


\section{Introduction}
\label{sec:introduction}
Over the last thirty years, the lattice Boltzmann method (LBM) has been actively developed and has become a reliable tool for simulating flow problems with complex geometries, such as flow in porous media~\cite{pan2006evaluation}, fluid structure interaction~\cite{tian2011efficient} and particle-laden turbulent flows~\cite{wang2016latticeb,eshghinejadfard2017immersed}. In these applications, the treatment of the no-slip boundary condition at the fluid-solid interfaces is often an
important issue that affects the overall accuracy,  numerical stability, and computational efficiency of the lattice Boltzmann method.

As a mesoscopic method based on the Boltzmann equation but with the goal to solve the macroscopic Navier-Stokes equations, the treatment of a no-slip boundary within the
LBM can be flexible as either the no-slip schemes used in conventional computational fluid dynamics (CFD) or the microscopic properties in the Boltzmann equation may be applied and implemented. There are mainly two categories of no-slip boundary treatment in LBM simulations. The first  is the immersed boundary method (IBM). IBM is a popular no-slip boundary treatment developed in conventional CFD~\cite{peskin2002immersed,uhlmann2005immersed,breugem2012second}, but it can be easily incorporated within the LBM algorithm~\cite{feng2005proteus,wu2009implicit}. The idea of IBM is to represent the effect of the no-slip condition as a boundary force applying to the neighboring region of the fluid-solid interface. In order to ensure that the no-slip condition is enforced at precisely the location of the boundary, a body-fitted Lagrangian grid is usually attached to the surface of each solid object besides the Eulerian grid covering the whole computational domain. A regularized delta function is employed to interpolate information between the Eulerian and Lagrangian grids~\cite{peskin2002immersed,uhlmann2005immersed}. Depending on how the boundary force that enforces the no-slip condition is calculated, IBM can be grouped as penalty IBM~\cite{goldstein1993modeling} or direct-forcing IBM~\cite{fadlun2000combined}. For problems involving only non-deformable rigid surfaces, direct-forcing IBM is preferred due to its clearer physical picture and better numerical stability.  

The second category of no-slip boundary treatment in LBM is to directly construct the unknown distribution functions at the boundary nodes using the known ones while observing the hydrodynamic constraints. This type of algorithm is known as bounce-back schemes. The early bounce-back scheme such as that proposed by Ladd~\cite{ladd1994numerical1} approximates a curved surface as a staircase shaped polylines when applied to a complex geometry. The improved bounce-back schemes were developed later to address this deficiency~\cite{bouzidi2001momentum,mei2000lattice,guo2002extrapolation,yu2003viscous,ginzburg2003multireflection}. While the detailed algorithms are not unique, the idea of these improved schemes are similar, which is to construct the unknown distribution functions to have at least a second-order accuracy. These schemes are typically referred  to as the interpolated bounce-back (IBB) schemes.
It is known that the hydrodynamic equations can be obtained
from the Chapman-Enskog expansion of the Boltzmann equation,
however, it is not completely clear whether the IBB schemes
are consistent with the Chapman-Enskog expansion at the boundary nodes. The accuracy and numerical stability
of the IBB schemes are typically examined only by numerical tests. 

In the past, both IBM and IBB were extensively used by the LBM community in a wide range of applications. Although each method is validated in a few numerical tests on its own, systematic comparative studies between the two sets of methods are rare.  Peng \& Luo~\cite{peng2008comparative} compared performances of Bouzidi {\it et al.}'s quadratic IBB scheme~\cite{bouzidi2001momentum} and Feng \& Michaelides's direct-forcing IBM-LBM~\cite{feng2005proteus},  focusing on evaluating the drag and lift coefficients of a cylinder placed at different location facing a uniform stream. They observed that while the numerical error in the integrated force evaluation generally followed a second-order convergence rate, the results from the IBB scheme are much more accurate than those from IBM-LBM. As will become clearer later with the present work, although in certain cases the hydrodynamic force/torque evaluation does possess a second-order accuracy, such observation may not be generalized for arbitrary flows.
Chen {\it et al.}~\cite{chen14comparative} compared a few IBB schemes and IBM-LBM algorithms in simulating the acoustic waves scattering on static and moving cylinder surfaces. They reported that while IBB schemes outperformed in accuracy in static cylinder cases, IBM-LBM could be a better choice in cases with moving objects in terms of suppressing the high-frequent fluctuations ({\it i.e.}, the grid jitter problem) associated with objects crossing the grid mesh lines. 

While these previous comparative studies are useful, a re-examination of the inter-comparison of the two treatments is still necessary, for several reasons. First, in the aforementioned studies, the benchmark results used as standards are usually from other simulations, rather than from the theory. This brings difficulty to rigorously gauge the accuracy of a method. For example, in the study of Peng \& Luo~\cite{peng2008comparative}, as will be shown, the IBM-LBM method is of only first-order accuracy; 
it remains a puzzle that the first-order accurate IBM-LBM could lead to second-order converged drag and lift force evaluations. In many validation studies of IBM, the Taylor-Green flow without a solid-fluid interface was employed~\cite{uhlmann2005immersed,kang2011comparative}. This validation is not so meaningful since the accurate flow field can be obtained with or without the IBM. Second, it is important to follow the recent developments in both categories of methods in order to make unbiased conclusions. For example, Breugem~\cite{breugem2012second} proposed an improved IBM by retracting the locations of the Lagrangian grid points from the surface of a solid object towards the interior of the solid object. It is claimed this retraction could improve the accuracy of IBM from first-order to second-order. Zhou \& Fan~\cite{zhou2014second} incorporated this improvement to LBM that seemed to reach a similar conclusion. On the other hand, IBB schemes are also under further developments. A good example is the single-node second-order accurate IBB scheme by Zhao \& Yong~\cite{zhao2017single}, which allows the second-order accurate no-slip boundary to be realized using the information only on the boundary node itself. This scheme is particularly useful for cases such as dense particle suspension where the gap between two solid surfaces is too narrow for other IBB schemes to be executed. Whether these new developments would alter the conclusions made in the previous comparative studies is yet to be examined.

In this paper, we examine the performance of several selected IBM algorithms and IBB schemes in flows with reliable benchmark results. Those IBM algorithms and IBB schemes are chosen because they have been implemented in complex simulations such as direct numerical simulations of particle-laden turbulent flows~\cite{uhlmann2008interface,picano2015turbulent,wang2016latticeb,eshghinejadfard2017immersed,peng2018study}. In order to assess the reliability of the reported results, it is important to test the accuracy and robustness of these methods in relatively simpler laminar flows that are easier to analyze. The rest of the paper is arranged as the following. In Sec. 2, we briefly introduce LBM and the selected IBB schemes and IBM algorithms to be examined. Then, the performances of these no-slip boundary treatments are compared in some carefully chosen two-dimensional and three-dimensional laminar flow tests in Sec. 3. Finally, the key observations will be summarized in Sec. 4.  
\section{The lattice Boltzmann method and its no-slip boundary treatments}
\label{sec:LBM}
The evolution equation of LBM can be viewed as a fully discrete form of the Boltzmann BGK equation in space and time, with a selected set of particle velocities
\begin{equation}
f_{i}\left({\bf x},{\bf e}_{i}\delta_{t},t+\delta_{t}\right) - f_{i}\left({\bf x},t\right)  = -\frac{1}{\tau}\left[f_{i}\left({\bf x},t\right)-f_{i}^{(eq)}\left({\bf x},t\right)\right] + F_{i}\left({\bf x},t\right),
\label{eq:LBGK}
\end{equation}
where $f_{i}$ is the particle distribution function for the discrete velocity ${\bf e}_{i}$, ${\bf x}$ and $t$ are the spatial coordinate and time, respectively. $f_{i}^{(eq)}$ is the equilibrium distribution of $f_{i}$, $F_{i}$ is the term representing the body force in the Boltzmann equation.
$\tau$ is the non-dimensional relaxation time, which is related to the kinematic viscosity $\nu$ as 
\begin{equation}
\nu = \left(\tau - 0.5\right)c_s^2\delta_{t},
\label{eq:viscosity}
\end{equation}
with $c_{s}$ being the speed of sound. 

Eq.~(\ref{eq:LBGK}) is known as the lattice BGK equation, whose collision operator (right-hand side of Eq.~(\ref{eq:LBGK})) contains only one relaxation time $\tau$. Alternatively, if the collision operator is constructed in the moment space through linear transformation, different moments can be relaxed at different rates, the evolution equation of LBM can then be expressed as
\begin{equation}
{\bf f}\left({\bf x},{\bf e}_{i}\delta_{t},t+\delta_{t}\right) - {\bf f}\left({\bf x},t\right)
= -{\bf M}^{-1}{\bf S}\left[{\bf m}\left({\bf x},t\right) - {\bf m}^{(eq)}\left({\bf x},t\right)\right] + {\bf M}^{-1} {\bf \Psi}\left({\bf x},t\right).
\label{eq:MRTcollision}
\end{equation}
which possesses larger flexibility in the model design. ${\bf f}$ is the vector expression of $f_{i}$. ${\bf m}$, ${\bf m}^{(eq)}$, and ${\bf \Psi}$ are the moment vector, equilibrium moment vector, and the forcing vector, respectively. ${\bf M}$ is the transform matrix that relates the moment vector and vector of distribution functions as ${\bf m} = {\bf M}{\bf f}$ and ${\bf f} = {\bf M}^{-1}{\bf m}$. 
LBM using Eq.~(\ref{eq:MRTcollision}) as the evolution equation is known as the multi-relaxation time (MRT) LBM. More details regarding Eq.~(\ref{eq:LBGK}) and Eq.~(\ref{eq:MRTcollision}) can be found in the textbooks~\cite{guo2013lattice} and other classic articles of LBM~\cite{lallemand2000theory,guo2002discrete}, thus they are not repeated here.

\subsection{Immersed boundary-lattice Boltzmann method}
The standard LBM can be viewed as a mesoscopic alternative of the incompressible Navier-Stokes solver in the weakly compressible limit. The no-slip boundary treatments in conventional CFD may be incorporated in LBM. The most popular method that has been used widely in CFD for the no-slip boundary treatment on arbitrarily shaped surface is the immersed boundary method (IBM). The first incorporation of IBM into LBM has been achieved by Feng \& Michaelides~\cite{feng2004immersed}. Since then, there have been many variations of the method in terms of the calculation the boundary force and the incorporation of this force into the evolution equation of LBM. The boundary force in IBM-LBM can be calculated by the penalty feedback forcing~\cite{feng2004immersed}, direct forcing~\cite{feng2005proteus}, and momentum exchange forcing~\cite{niu2006momentum}. Among these three force calculation methods, the direct forcing is the most popular one due to its simplicity and the capability to use larger CFL numbers~\cite{uhlmann2005immersed}. The direct-forcing IBM has been made particularly efficient to realize the no-slip condition on rigid particle surfaces in particle-laden flows~\cite{uhlmann2008interface,picano2015turbulent,eshghinejadfard2017immersed}. In this study, we focus our attention on the evaluation of direct-forcing IBM algorithms that has been frequently used in the three-dimensional flow simulations with a large number of particles. 
In these algorithms, two sets of grids, a fixed Eulerian grid is used to store the information of the flow field, and a Lagrangian grid attached to the solid surface is used to ensure the no-slip condition is enforced precisely on the physical location. Uhlmann significantly simplified the algorithm of direct-forcing IBM as five key steps~\cite{uhlmann2005immersed}. 
First, the known velocity field stored at the Eulerian grid ${\bf u}^{n}$ is evolved to a temporary velocity field $\tilde{{\bf u}}$ by solving the N-S equations without considering the boundary force.
\begin{equation}
    \rho\frac{\tilde{{\bf u}} - {\bf u}^{n}}{\delta t} = -\rho \left({\bf u}\cdot \nabla\right) {\bf u} -\nabla p + \mu \nabla^{2}{\bf u}. 
    \label{eq:evolveNS}
\end{equation}
Next, this temporary velocity field at the Eulerian grid ${\bf x}$ is interpolated to the Lagrangian grid ${\bf X}$.

\begin{equation}
     \tilde{{\bf U}}\left({\bf X}\right) = \sum_{{\bf x}}\tilde{{\bf u}}\left({\bf x}\right)\delta_{h}\left({\bf x} - {\bf X}\right)h^{3},
     \end{equation}
where $\delta_{h}$ is the interpolation kernel which typically has a form of the regularized delta function~\cite{peskin2002immersed}. By default, the four-point delta-function~\cite{peskin2002immersed} 
\begin{equation}
\begin{split}
    &\delta_{h} = \frac{1}{h^3}\phi\left(\frac{x_{1}}{h}\right)\phi\left(\frac{x_{2}}{h}\right)\phi\left(\frac{x_{3}}{h}\right),\\
   &\phi\left(r\right) = \left \{
\begin{array}{ll}
0,     ~~~~~~~~~~~~~~~~~~~~~~~~~~~~~~~~~~~~~~~~~~~~~~~~~~      |r| \ge 2,\\ 
\frac{1}{8}\left(5-2|r|-\sqrt{-7+12|r|-4r^2}\right),~    1\le|r|<2 \\ 
\frac{1}{8}\left(3-2|r|+\sqrt{1+4|r|-4r^2}\right), ~~~~~    0\le|r|<1,
\end{array}\right .
\end{split}
\label{eq:deltafunction}
\end{equation}
derived by Peskin is used for all the simulations presented below, unless specified  otherwise. $h^3$ is the volume of a Eulerian grid cell. 
By default, we use the uppercase letters to represent the properties on the Lagrangian grid and the lowercase letters to represent the properties on the Eulerian grid. 
Next, the boundary force ${\bf F}({\bf X})$ used to enforce the no-slip condition on the Lagrangian grid should be calculated as

\begin{equation}
     {\bf F({\bf X})}  = \frac{{\bf U}^{d}({\bf X}) - \tilde{{\bf U}}({\bf X})}{\delta t}. 
    \label{eq:directforcing}
\end{equation}
Then, this boundary force is distributed back to the Eulerian grid.
\begin{equation}
     {\bf f}\left({\bf x}\right) = \sum_{{\bf X}}{\bf F}\left({\bf X}\right)\delta_{h}\left({\bf x} - {\bf X}\right)\Delta V,
     \label{eq:redistributeforce}
     \end{equation}
where $\Delta V$ is the control volume of a Lagrangian grid, which is typically chosen as $\Delta V\approx h^3$~\cite{uhlmann2005immersed}.
At last, the obtained force field is used to update the velocity field from $\tilde{{\bf u}}$ to ${\bf u}^{n+1}$.
\begin{equation}
     {\bf u}^{n+1} = \tilde{{\bf u}} + {\bf f}\delta t,
     \label{eq:updatevel}
     \end{equation}

Eq.~(\ref{eq:evolveNS}) to Eq.~(\ref{eq:updatevel}) well describe a direct-forcing IBM algorithm in CFD.
There are different ways to incorporate the above algorithm into the frame of LBM~\cite{dupuis2008immersed,wu2009implicit,kang2011comparative,zhang2016accuracy}. Since the interpolation via delta function only has a first-order accuracy on a general fluid-solid surface (which will be proven later)~\cite{peskin2002immersed,zhang2016accuracy}, the choice of a specific algorithm may not affect the accuracy of the simulation results that much. 
Of course, it is more reasonable to use the mesoscopic forcing terms in the evolution equations of LBM, {\it i.e.}, $F_{i}$ in Eq.~(\ref{eq:LBGK}) or ${\bf \Psi}$ in Eq.~(\ref{eq:MRTcollision}), which ensures a second-order accuracy when applied to a non-uniform force field, as the boundary force redistributed back to the Eulerian grid is a non-uniform force field. When the lattice BGK equation is employed, both Guo's scheme~\cite{guo2002discrete} and Cheng \& Li's scheme~\cite{cheng2008introducing} possess the second-order accuracy when applied to a non-uniform force field. These two schemes are actually identical (proven in \cite{kang2011comparative}). When the MRT-LBM equation is used, the forcing term can be constructed using the inverse design, as demonstrated in~\cite{min2016inverse}.

In the direct-forcing IBM algorithm described in Eq.~(\ref{eq:evolveNS}) to Eq.~(\ref{eq:updatevel}), the boundary force is defined as a correction force that brings the fluid velocity to target one at the next time step $n+1$. The IB-LBM algorithm that corresponds to this algorithm is the implicit velocity correction based IB-LBM developed by Wu \& Shu~\cite{wu2009implicit}. In this algorithm, Guo's forcing scheme~\cite{guo2002discrete} is used. Kang \& Hassan~\cite{kang2011comparative} developed a similar algorithm using Cheng \& Li's forcing scheme~\cite{cheng2008introducing}. The only concern about these correction-based IB-LBM algorithms is whether they are fully consistent with the Chapman-Enskog expansion. When Guo's forcing scheme is used, half of the force is added when calculating the velocity field from the distribution functions~\cite{guo2002discrete}. However, in the correction based IB-LBM, this half force is absent in order to calculate an ``unforced'' velocity field. The same issue can be identified in Kang \& Hassan's algorithm with Cheng \& Li's forcing scheme. The implicit force field that should be added right after the propagation of the distribution functions is postponed after the update of the hydrodynamic properties (density, velocity, etc.) as a correction~\cite{kang2011comparative}. A more consistent algorithm of the correction-based IB-LBM may be the one developed by Zhang et al.~\cite{zhang2016accuracy} recently. In this algorithm, the implicit force field added after the propagation of the distribution functions is obtained through iterations~\cite{zhang2016accuracy}. However, in the simulations with a large number of particles, the iteration is usually undesired.

The specific IB-LBM algorithm we examine in this paper is a relatively simple one. At each step, prior to the evolution of distribution functions, the boundary force is first calculated as Eq.~(\ref{eq:directforcing}). This boundary force is then distributed to the Eulerian grid, and used for evolving the distribution functions according to Eq.~(\ref{eq:LBGK}) or Eq.~(\ref{eq:MRTcollision}). The boundary force in this algorithm is therefore a force responding to the presence of a solid force at the current time, rather than a force that enforce the no-slip condition at the next time step.

\subsection{Interpolated bounce-back schemes}

\begin{figure}
\centering
\includegraphics[width=90mm]{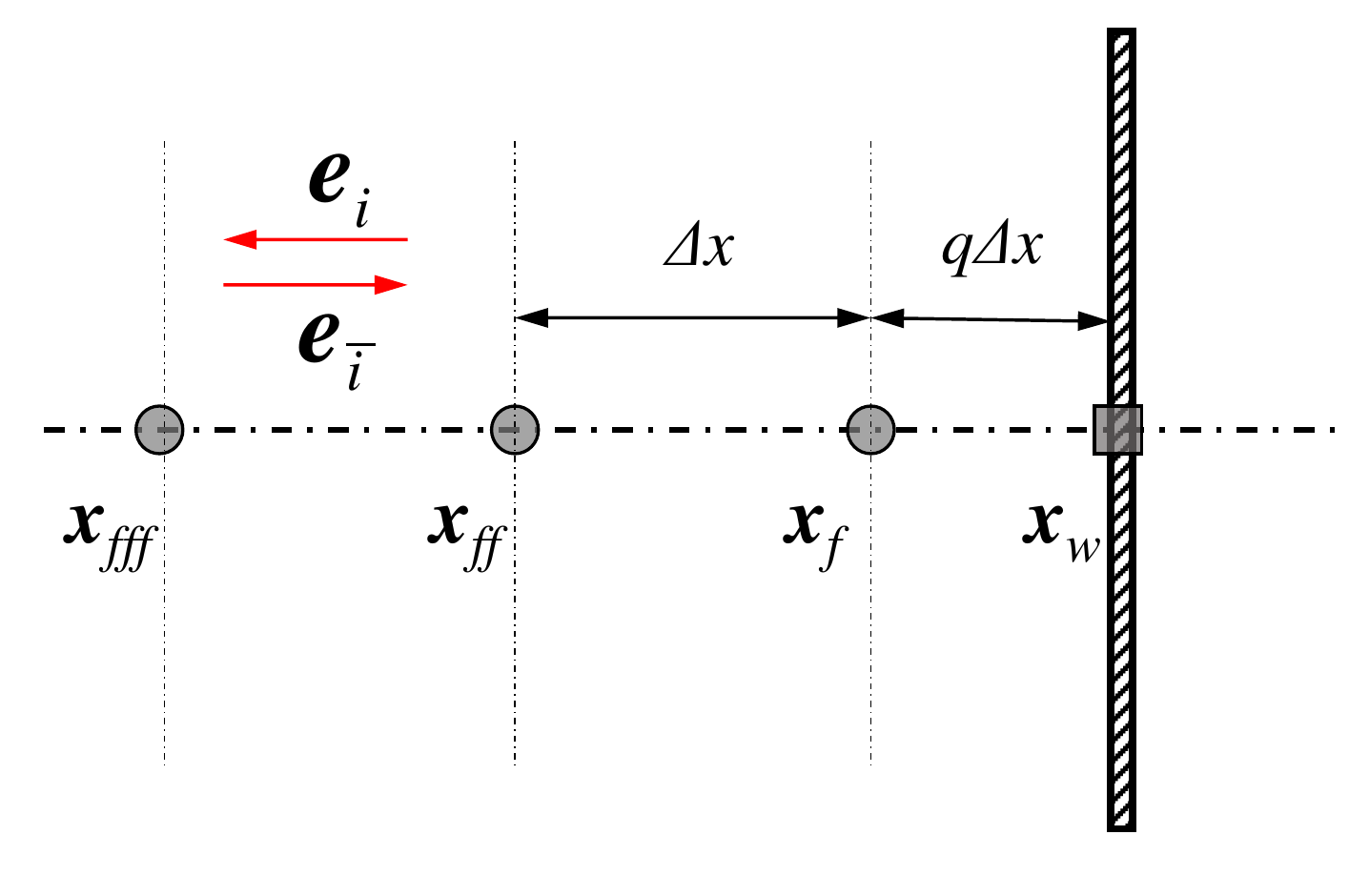}
\caption{A sketch of a fluid-solid interface in a LBM simulation.}
\label{fig:boundary}
\end{figure}

The essence of bounce-back schemes is to directly construct the unknown distribution functions from the known ones and the hydrodynamic constraints at the boundary nodes. With the boundary configuration in Fig.~\ref{fig:boundary}, a simple bounce-back scheme can be written as~\cite{ladd1994numerical1}
\begin{equation}
    f_{i}\left({{\bf x}_{f},t+\delta_{t}}\right) = f_{\bar i}^{\ast}\left({\bf x}_{f},t\right) + 2\rho_{0}w_{i}\frac{{\bf e}_{i}\cdot{\bf u}_{w}}{c_s^2},
\label{eq:bounceback}
\end{equation}
where $f_{i}\left({{\bf x}_{f},t+\delta_{t}}\right)$ and $f_{\bar i}^{\ast}\left({\bf x}_{f},t\right)$ are the bounce-back distribution function and the incident distribution function, both locate at the boundary node ${\bf x}_{f}$ and with ${\bf e}_{i} = -{\bf e}_{\bar i}$. ${\bf u}_{w}$ is the velocity at the wall location ${\bf x}_{w}$. The last term on the right-hand side is used to ensure the no-slip condition when the solid boundary is  moving. Eq.~(\ref{eq:bounceback}) means that the post-collision particles traveling towards a wall return back along the same location after bouncing back from the wall, thus the scheme obtained its name. Since the distribution function travels precisely one grid spacing from $t$ to $t+\delta_{t}$, the particles start from ${\bf x}_{f}$ can end precisely at the same location only when ${\bf x}_{f}$ is half a grid spacing from the wall. In fact, when this condition is not satisfied, the bounce-back scheme of Eq.~(\ref{eq:bounceback}) only has the first-order accuracy, which restricts its application on an arbitrarily shaped surface. 

In order to ensure that the second-order spatial accuracy in a bounce-back process for more general cases, interpolation is usually required. Since the number of unknown distribution functions is usually larger than the number of hydrodynamic constraints, the method to design interpolated bounce-back schemes is not unique. Two representative interpolated bounce-back schemes are the conditional scheme proposed by Bouzidi et al.~\cite{bouzidi2001momentum}, and the unified scheme by Yu et al.~\cite{yu2003viscous}. 
In Bouzidi et al.'s scheme, when the relative distance from the boundary node point to the wall location, {\it i.e.}, $q = |{\bf x}_{f} -{\bf x}_{w}|/|{\bf x}_{f}-{\bf x}_{b}|$, is smaller than 0.5, a virtual distribution function is interpolated first at ${\bf x}_{i}$ so that the molecules represented by this virtual distribution function ends precisely at ${\bf x}_{f}$ after the bounce back from the wall. Apparently, ${\bf x}_{i}$ locates between ${\bf x}_{f}$ and the neighboring fluid node ${\bf x}_{ff}$, thus the virtual distribution function can be  interpolated from the corresponding distribution functions at ${\bf x}_{f}$, ${\bf x}_{ff}$, and ${\bf x}_{fff}$. On the other hand, when $q \ge 0.5$, ${\bf x}_{i}$ locates between ${\bf x}_{f}$ and ${\bf x}_{w}$, the interpolation becomes extrapolation, which could result in numerical instability. To avoid this, the streaming is proceeded first, {\it i.e.}, the distribution function at ${\bf x}_{f}$ first bounce-back from the wall and ends at a temporary location ${\bf x}_{t}$. Then the unknown distribution function at ${\bf x}_{f}$ is interpolated with the corresponding distribution functions at ${\bf x}_{t}$, ${\bf x}_{ff}$, and ${\bf x}_{fff}$. Bouzidi et al.'s interpolated bounce-back scheme can be summarized as
\begin{subequations}
\begin{align}
\begin{split}
&f_{i}\left({\bf x}_{f},t+\delta_{t}\right) = q\left(2q+1\right)f_{\bar i}^{*}\left({\bf x}_{f},t\right)+\left(1+2q\right)\left(1-2q\right)f_{\bar i}^{\ast}\left({\bf x}_{ff},t\right)\\
&-q\left(1-2q\right)f_{\bar i}^{\ast}\left({\bf x}_{fff},t\right)+2\rho_{0}w_{i}\frac{{\bf e}_{i}\cdot {\bf u}_{w}}{c_s^2},~~~~q<0.5,
\end{split}
\label{eq:Bouzidi:A}\\
\begin{split}
&f_{i}\left({\bf x}_{f},t+\delta_{t}\right) = \frac{1}{q\left(2q+1\right)}\left[f_{\bar i}^{\ast}\left({\bf x}_{f},t\right)+2\rho_{0}w_{i}\frac{{\bf e}_{i}\cdot {\bf u}_{w}}{c_s^2}\right]\\
&+\frac{2q-1}{q}f_{i}\left({\bf x}_{ff},t+\delta_{t}\right) - \frac{2q-1}{1+2q}f_{i}\left({\bf x}_{fff},t+\delta_{t}\right),~~~~q\ge 0.5.
\end{split}
\label{eq:Bouzidi:B}
\end{align}
\label{eq:Bouzidi}
\end{subequations}
Alternatively, Yu {\it et al.} designed a unified IBB scheme for all values of $q$ from 0 to 1. Their idea is straightforward. First, a virtual distribution function is interpolated between ${\bf x}_{f}$ and ${\bf x}_{ff}$, which ends exactly at the wall location after streaming a grid spacing towards the wall, {\it i.e.},
\begin{equation}
f_{\bar i}\left({\bf x}_{w},t+\delta_{t}\right) = \frac{q\left(q+1\right)}{2}f_{\bar i}^{*}\left({\bf x}_{f},t\right)
+\left(1+q\right)\left(1-q\right)f_{\bar i}^{\ast}\left({\bf x}_{ff},t\right)-\frac{q\left(1-q\right)}{2}f_{\bar i}^{\ast}\left({\bf x}_{fff},t\right).
\label{eq:Yustep1}
\end{equation}
Next, an instantaneous bounce-back happens right after the virtual distribution function arrives at the wall location
\begin{equation}
f_{i}\left({\bf x}_{w},t+\delta_{t}\right) = f_{\bar i}\left({\bf x}_{w},t+\delta_{t}\right) + 2\rho_{0}w_{i}\frac{{\bf e}_{i}\cdot {\bf u}_{w}}{c_s^2}
\label{eq:Yustep2}
\end{equation}
At last, the unknown distribution function $f_{i}\left({\bf x}_{f},t+\delta_{t}\right)$ is interpolated from $f_{i}\left({\bf x}_{w},t+\delta_{t}\right)$, $f_{i}\left({\bf x}_{ff},t+\delta_{t}\right)$ and $f_{i}\left({\bf x}_{fff},t+\delta_{t}\right)$ as
\begin{equation}
\begin{split}
&f_{i}\left({\bf x}_{f},t+\delta_{t}\right) = \frac{2}{\left(1+q\right)\left(2+q\right)}f_{ i}\left({\bf x}_{w},t+\delta_{t}\right)\\
&+\frac{2q}{1+q}f_{i}\left({\bf x}_{ff},t+\delta_{t}\right)-\frac{q}{2+q}f_{i}\left({\bf x}_{fff},t+\delta_{t}\right).
\end{split}
\label{eq:Yustep3}
\end{equation}
In practice, it is more efficient to combine the above three steps into a single equation involving up to five distribution functions, which reads
\begin{equation}
\begin{split}
&f_{i}\left({\bf x}_{f},t+\delta_{t}\right) = \frac{q}{2+q}f_{\bar i}^{\ast}\left({\bf x}_{f},t\right) + \frac{2\left(1-q\right)}{1+q}f_{\bar i}^{\ast}\left({\bf x}_{ff},t\right)\\ &-\frac{\left(1-q\right)q}{\left(1+q\right)\left(2+q\right)}f_{\bar i}^{\ast}\left({\bf x}_{fff},t\right)
+ \frac{2q}{1+q}f_{i}\left({\bf x}_{ff},t+\delta_{t}\right)\\
&- \frac{q}{2+q}f_{i}\left({\bf x}_{fff},t+\delta_{t}\right)+\frac{4}{\left(1+q\right)\left(2+q\right)}\rho_{0}w_{i}\frac{{\bf e}_{i}\cdot {\bf u}_{w}}{c_s^2}.
\end{split}
\label{eq:Yucombined}
\end{equation}

While these two schemes are constantly used in LBM for no-slip boundary treatment on curved surfaces. A potential issue is that they require not only the information at the boundary node itself, {\it i.e.}, ${\bf x}_{f}$, but also the distribution functions at ${\bf x}_{ff}$ and ${\bf x}_{fff}$ to process the interpolation. When two solid surfaces sit very close, which frequently happens in particle-laden flows with dense particle suspensions, Eq.(\ref{eq:bounceback}) has to be used instead, where the overall accuracy of the boundary treatment might be contaminated. This potential issue is resolved with the recently proposed single-node second-order bounce-back scheme by Zhao \& Yong~\cite{zhao2017single}, which reads
\begin{equation}
f_{i}\left({\bf x}_{f},t+\delta_{t}\right) = \frac{2q}{1+2q}f_{i}^{\ast}\left({\bf x}_{f},t\right) +\frac{1}{1+2q}f_{{\bar i}}\left({\bf x}_{f},t\right)+\frac{2}{1+2q}\rho_{0}w_{i}\frac{{\bf e}_{i}\cdot {\bf u}_{w}}{c_s^2}.
\label{eq:ZhaoYong}
\end{equation}
Unlike the previous two IBB schemes that construct $f_{i}\left({\bf x}_{f},t+\delta_{t}\right)$ purely from the post-collision distribution functions. Zhao \& Yong's scheme utilize both the pre-collision and post-collision distribution functions to fulfill the ``interpolation''. The second-order accuracy of this scheme can be rigorously proven by an asymptotic analysis~\cite{zhao2017single}. It is also worth mentioning that an alternative single-node second-order bounce-back scheme was recently  proposed by Tao {\it et al.}~\cite{tao2018one}. We were made aware of this scheme quite late thus it is not included  in our comparisons shown below.

The three IBB schemes, {\it i.e.}, Bouzidi et al's scheme, Yu et al's scheme, as well as Zhao \& Yong's scheme will be used in the numerical examinations in Sec.~\ref{sec:Numerical}. With the use of bounce-back schemes, the natural way to calculate the hydrodynamic force and torque acting on a solid surface is the momentum exchange method~\cite{ladd1994numerical1,mei2002force,wen2014galilean}. Although the combinations of bounce-back schemes and momentum exchange method do not ensure the instantaneous Galilean invariance~\cite{peng2017issues}, their accuracy has been proven to be sufficient in most simulations~\cite{peng2016implementation,tao2016investigation}. In particular, the Galilean invariant momentum exchange method (GIMEM) proposed by Wen et al.~\cite{wen2014galilean},
\begin{subequations}
\begin{align}
&{\bf F}\delta_{t} = \sum_{{\bf x}_{f},i}\left[f_{\bar i}^{\ast}\left({\bf x}_{f},t\right)\left({\bf e}_{\bar i} - {\bf u}_w\right)-f_{i}\left({\bf x}_{f},t+\delta_t\right)\left({\bf e}_{i}-{\bf u}_{w}\right)\right],
\label{eq:GIMEM:A}\\
&{\bf T}\delta_{t} = \sum_{{\bf x}_{f},i}\left({\bf x}_{w}-{\bf Y}_{c}\right)\times\left[f_{\bar i}^{\ast}\left({\bf x}_{f},t\right)\left({\bf e}_{\bar i} - {\bf u}_w\right)-f_{i}\left({\bf x}_{f},t+\delta_t\right)\left({\bf e}_{i}-{\bf u}_{w}\right)\right], 
\end{align}
\label{eq:GIMEM}
\end{subequations}
 will be used in the subsequent numerical examinations to reduce the ``grid locking" (``grid locking" will be discussed in detail later)~\cite{peng2016implementation}.
This is different from the original MEM~\cite{ladd1994numerical1,mei2002force}.

\section{Numerical examinations}
\label{sec:Numerical}

Appropriately chosen validation cases help us better evaluate the performance of the boundary treatment schemes.
In the earlier investigations, the accuracy of IBM was often validated in the flow of a two-dimensional Taylor-Green vortex flow. These tests, in our view, are not so meaningful since the accurate flow field can be obtained with or without the boundary forcing. 
The only information one may obtain from these tests is perhaps that IBM does not contaminate the second-order accuracy of LBM when it is applied to a smooth flow field\footnote{the smooth flow field is defined as a field where the velocity gradient normal to the interface is continuous, see Peskin~\cite{peskin2002immersed}}. Unfortunately, the velocity across a real solid-fluid interface is usually not smooth~\cite{peskin2002immersed}. Another often used test flow is a uniform flow passing a 2D cylinder or 3D sphere at finite Reynolds number. In this case, since the analytic solution is unavailable, while it is safe to validate whether a boundary treatment method is generating reasonable results, it is difficult to assess rigorously the accuracy and compare the results among different methods. 

In this paper, we choose four test flows to benchmark the performances of IBB schemes and IBM algorithms. The two-dimensional circular Couette flow and the three-dimensional laminar pipe flow are chosen since analytic solutions are available in the two flows that can help benchmarking the accuracy of each boundary treatment when a actual curved wall presents. A case of two-dimensional cylinder settling in a quiescent flow is used to examine the performance of each boundary treatment in predicting the motion of the objects in a viscous fluid. At last, a case of a uniform flow passing a static sphere is employed to assess the grid resolution requirement for each boundary treatment in order to obtain reliable hydrodynamic force acting on a spherical particle at different Reynolds numbers.

\subsection{Transient circular Couette flow}\label{sec:CircularCouette}

\begin{figure}
\centering
\includegraphics[width=90mm]{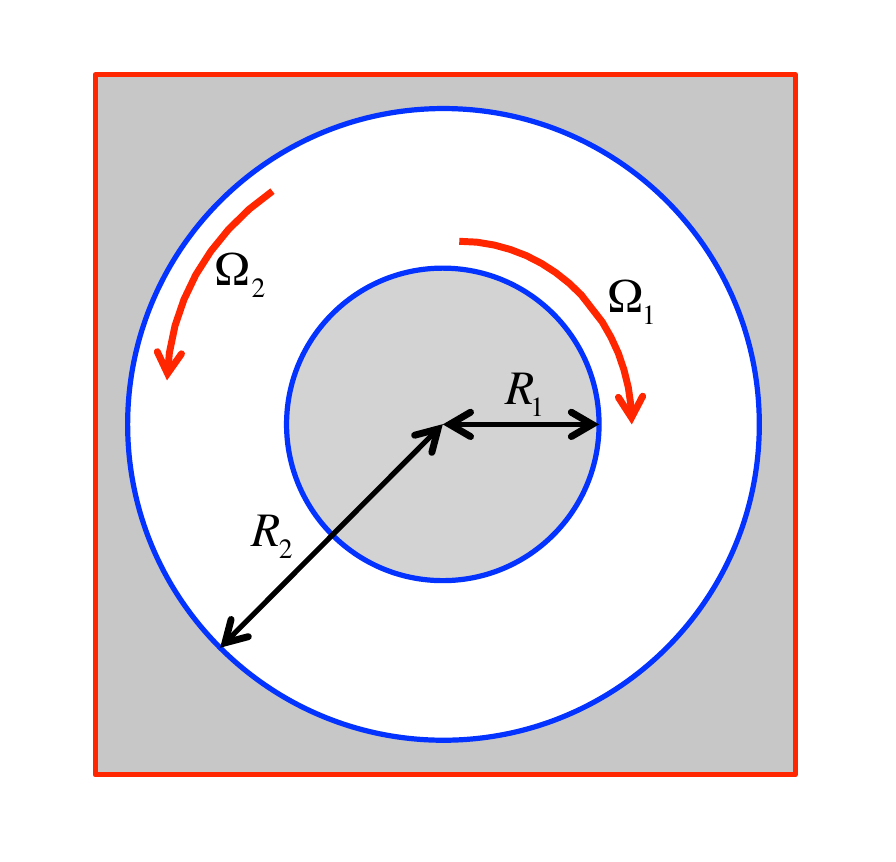}
\caption{A sketch of a Taylor-Couette flow}
\label{fig:TCflow}
\end{figure}

The purpose of the present study is to assess the performance of the boundary treatment schemes in general cases with curved and moving boundaries. For this purpose, the circular Couette flow, or Taylor-Couette flow between two concentric cylinders is employed. A sketch of this flow is shown in Fig.~\ref{fig:TCflow}.
The analytic solution of this flow is available in~\cite{he2015high}. We repeat it in Appendix A simply for readers' convenience.

In the simulations presented below, the inner cylinder is fixed while the outer cylinder rotates with an angular velocity that defines the flow Reynolds number $Re = \left(R_{2}-R_{1}\right)\Omega_{2}R_{2}/\nu = 45$. The ratio of the outer to inner
cylinder radius, $\gamma$, is set to 2. The simulations are conducted using the D2Q9 MRT collision model but run with a single relaxation time, {\it i.e.}, the equilibrium and the body force terms are defined in the moment space but all the relaxation times in matrix ${\bf S}$ are  identical. 

\begin{figure}
\centering
{\includegraphics[width=80mm]{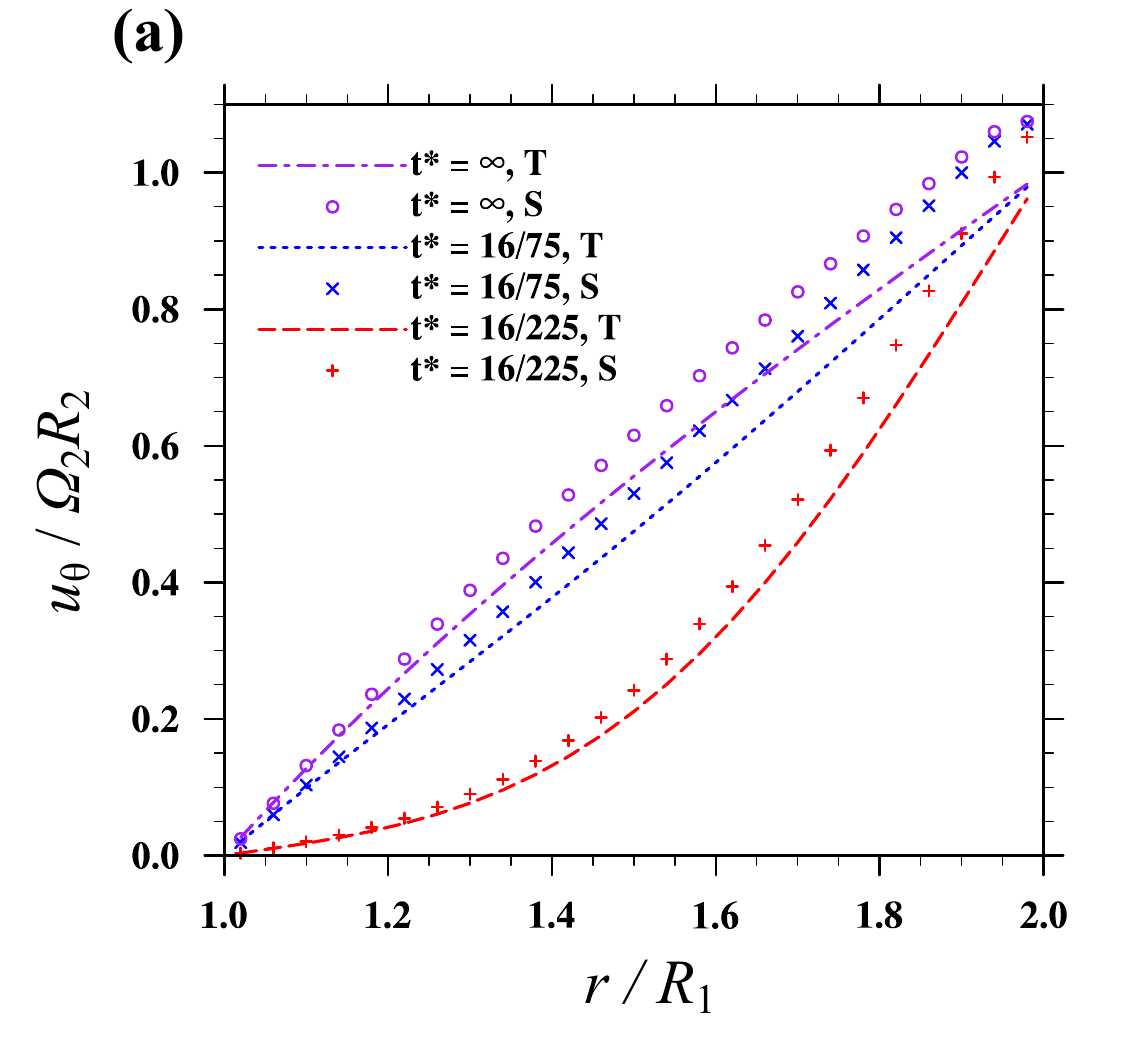}~~~~~\includegraphics[width=80mm]{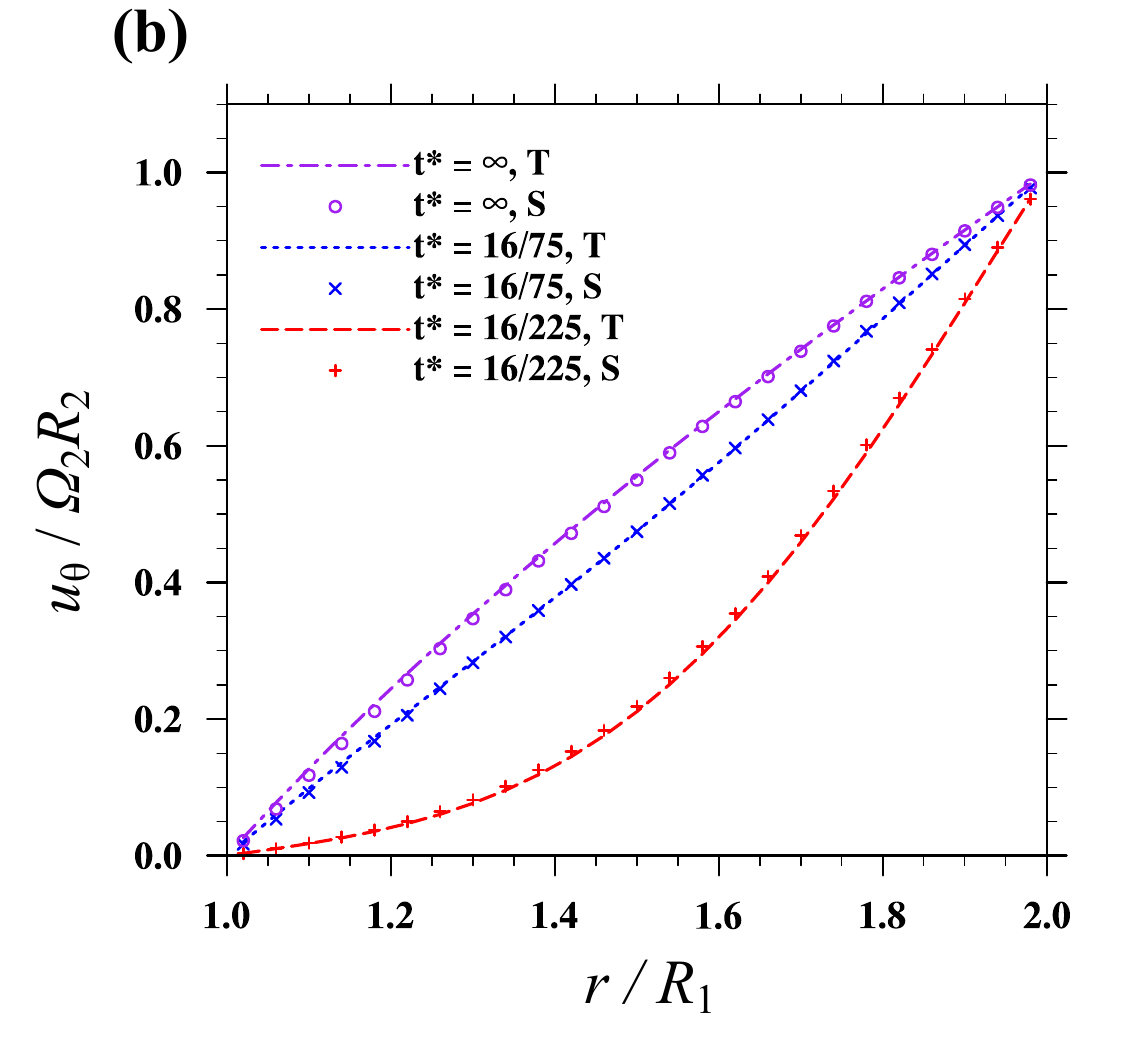}}
\vspace{0.1in}
\centering
{\includegraphics[width=80mm]{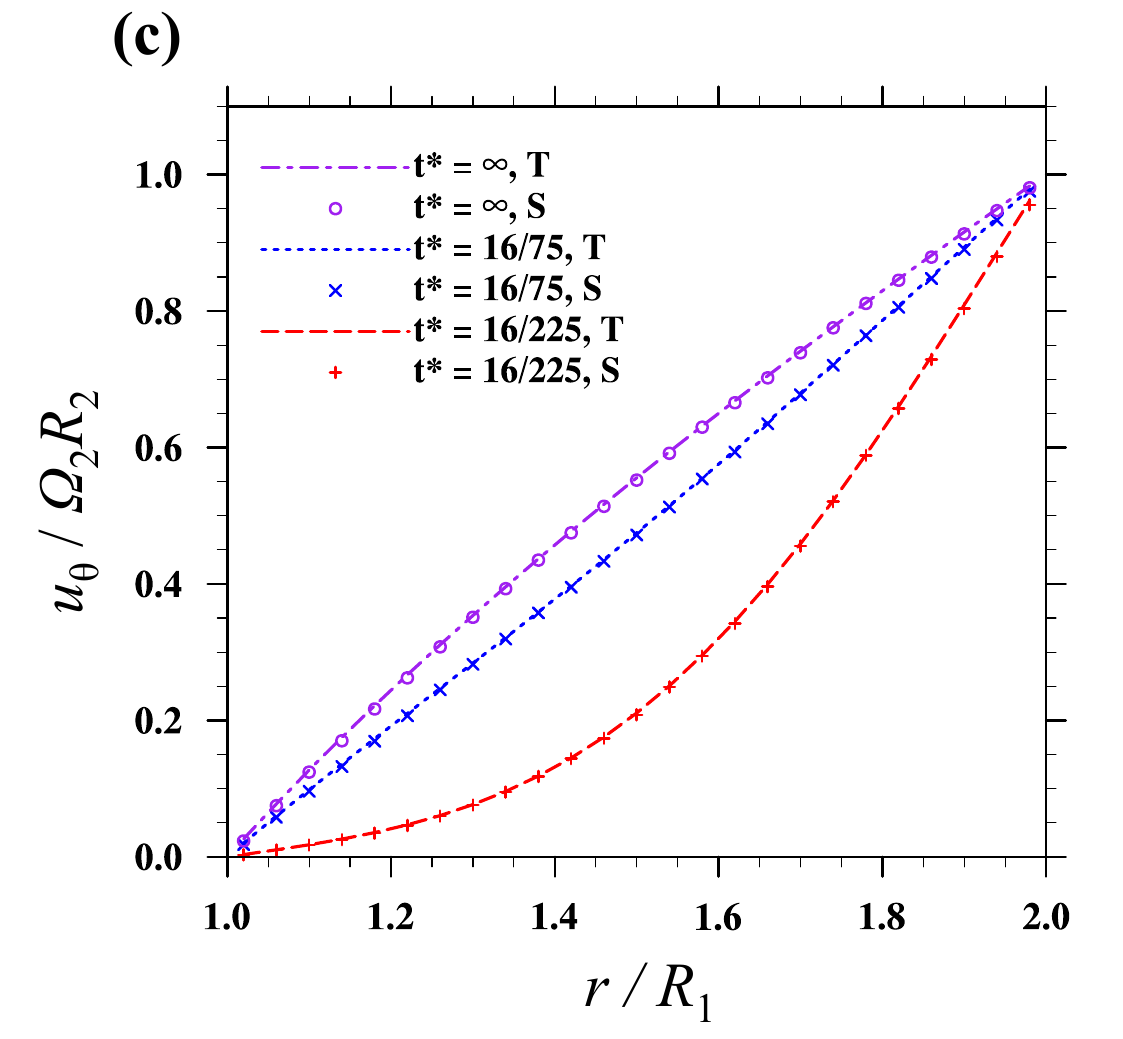}}
\caption{Velocity profiles of a transient Taylor-Couette flow: (a) LBM-IBM simulation with Breugem's IBM scheme with a retraction distance of $0.3\delta x$, the velocity on the edges of the computational domain is set as $u_{r} = 0, u_{\theta} = 0$; (b) same as (a), except that the velocity on the edges of the computational domain is set as $u_{r} = 0, u_{\theta} = \Omega_{2}r$; (c) LBM-IBB simulation with Bouzidi {\it et al}.'s quadratic interpolated bounce-back scheme.}
\label{fig:velprofiles}
\end{figure}

\begin{figure}
\centering
\includegraphics[width=90mm]{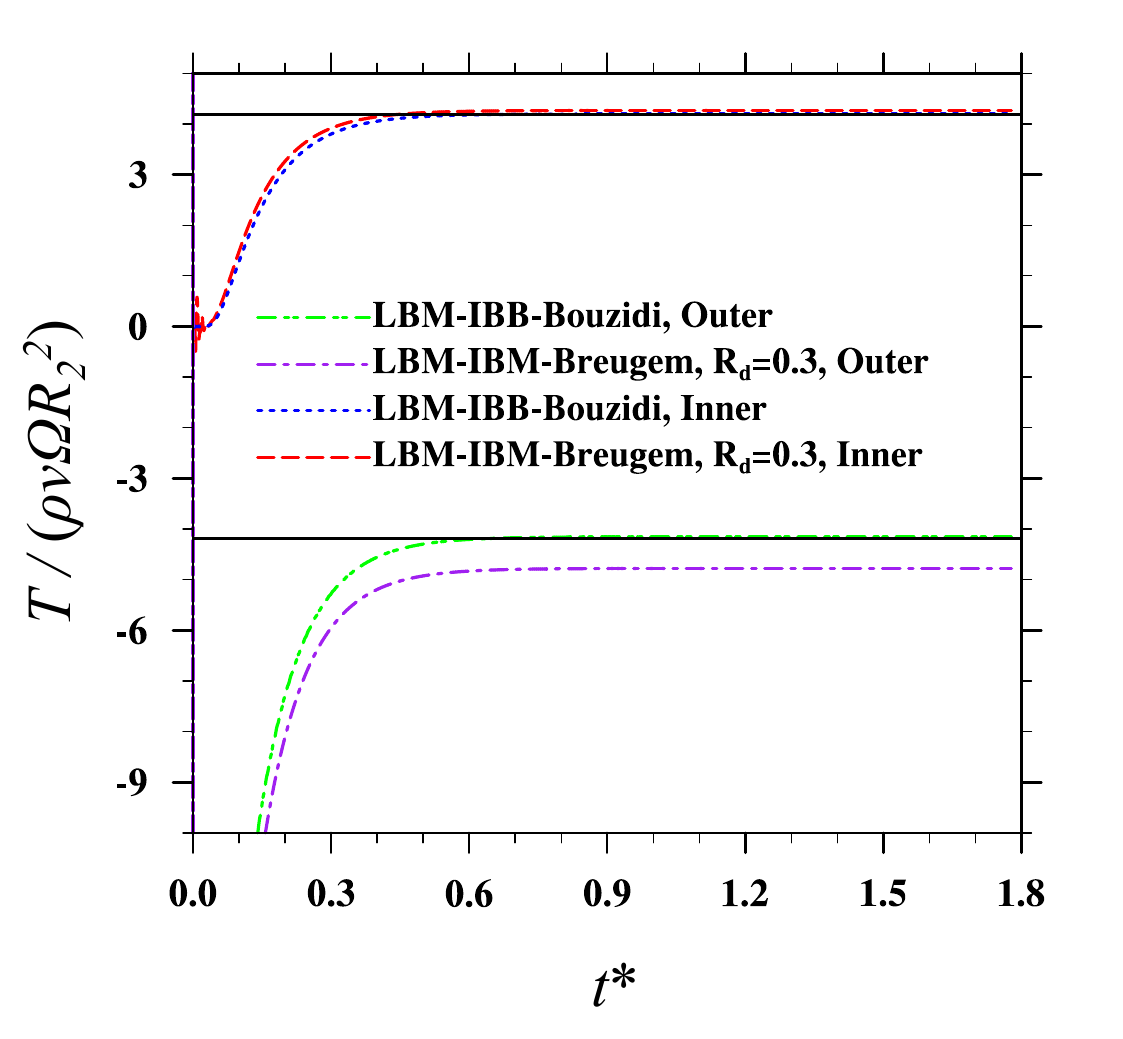}
\caption{The time-dependent torque acting on the inner and outer cylinders.}
\label{fig:timedependenttorque}
\end{figure}

Unlike in LBM-IBB simulations where the boundary treatment is purely determined by the information from the fluid region (white region in Fig.~\ref{fig:TCflow}), in the LBM-IBM simulations the whole domain, including the solid region (gray region in Fig.~\ref{fig:TCflow}), is filled with the same fluid and the flows outside and inside the solid region may be inter-connected through the N-S equations.  Therefore, how the flow in the solid region is treated may affect the flow within the fluid region. Specifically, in the Circular Couette flow, appropriate treatment of the boundary of the computational domain (red solid lines in Fig.~\ref{fig:TCflow}) plays an important role in ensuring the correctness of the results, especially when the outer cylinder is rotating. To demonstrate this point, we present the velocity profiles at different non-dimensional times ($t^{*} = t\nu/(R_2-R_1)^{2}$) from two LBM-IBM simulations, both use Breugem's IBM~\cite{breugem2012second} with a retraction distance $0.3\delta x$ to treat the two no-slip conditions on the cylinder surfaces, but with Dirichlet boundary (a) $u_{r} = 0, u_{\theta} = 0$, (b) $u_{r} = 0, u_{\theta} = \Omega_{2}r$ on the edges of the computational domain. The grid resolution used for the simulations is $R_{1} = 25\delta x$. The profiles are generated by averaging the velocity at the grid nodes sitting in 25 equal-width bins with the width of $dr = (R_{2}-R_{1})/25$. Obviously, with setting (a), the velocity profiles of the simulation deviate from the theoretical solution, while with setting (b), the velocity profiles match the theoretical solution quite well. This observation leads to the first remark that cautions must be given to the treatment of flow in the solid region when IBM is used. As we shall observe later in Fig~\ref{fig:timedependenttorque}, even setting (b) can result in a significant error in the hydrodynamic force evaluation. Unfortunately, the treatment on the edges of the computational domain is usually irrelevant to the physical description of the flow. The LBM-IBB simulation, on the other hand, does not suffer from the same problem. The construction on the unknown distribution functions at the boundary nodes purely depends on the information in the fluid region. The velocity profiles of the LBM-IBB simulation with Bouzidi et al.'s quadratic interpolation scheme are in good agreement with the theory, as shown in Fig.~\ref{fig:velprofiles}(c).

\begin{figure}
\centering
\includegraphics[width=80mm]{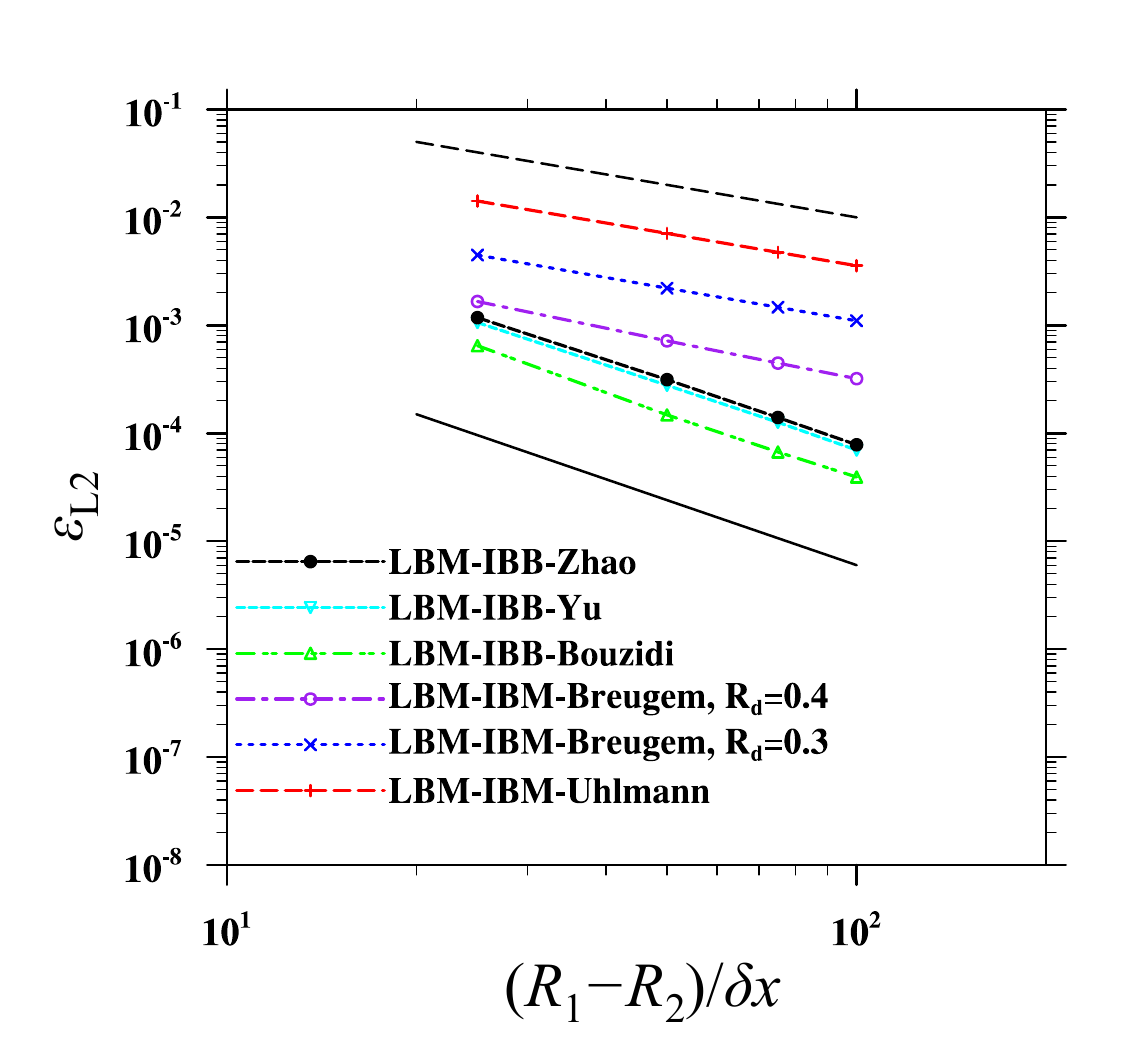}
\caption{Error convergence rates of the velocity field in the LBM-IBM and LBM-IBB simulations. The dash line and the solid line are references of slop -1 and -2, respectively. The same applied to all figures in the rest of the paper.}
\label{fig:velocitycovergence}
\end{figure}

To quantify the numerical error of the results in a simulation, the L1- and L2-norms, defined as 
\begin{equation}
    \varepsilon_{L1} = \frac{\sum_{{\bf x}}\left\| Q_{s}\left({\bf x}\right) - Q_{t}\left({\bf x}\right) \right\|}{\sum_{{\bf x}}\left\| Q_{t}\left({\bf x}\right) \right\|},~~~\varepsilon_{L2} = \frac{\sqrt{\sum_{{\bf x}}\left[ Q_{s}\left({\bf x}\right) - Q_{t}\left({\bf x}\right) \right]^2}}{\sqrt{\sum_{{\bf x}}\left[ Q_{t}\left({\bf x}\right)\right]^2}},
    \label{eq:L1L2norm}
\end{equation}
 are calculated, where $Q_{s}$ and $Q_{t}$ are the numerical result and theoretical result, respectively. The convergence rates of the L2-norm of the velocity fields at the steady state are presented in Fig.~\ref{fig:velocitycovergence}, for three LBM-IBM simulations, {\it i.e.}, with the IBM scheme of by Uhlmann (``LBM-IBM-Uhlmann''), and with the IBM scheme proposed by Breugem with two different retraction distances, $0.3\delta x$ and $0.4\delta x$ (``LBM-IBM-Breugem, $R_d = 0.3$'' and ``LBM-IBM-Breugem, $R_d = 0.4$''), as well as three LBM-IBB simulations, using the quadratic interpolation schemes by Bouzidi et al. (``LBM-IBB-Bouzidi'') and Yu et al. (``LBM-IBB-Yu''), and the single-node bounce-back scheme by Zhao \& Yong (``LBM-IBB-Zhao''). 
 The boundary force in the three LBM-IBM simulations are iterated for 5 times to ensure the representation of the no-slip boundary on the Lagrangian points is sufficiently accurate. 
 As clearly demonstrated in Fig.~\ref{fig:velocitycovergence}, the velocity fields in the three LBM-IBM simulations are always of first-order accuracy, while these from all the three LBM-IBB simulations are of second-order accuracy. 

The first-order accuracy of the LBM-IBM is a result of the fact that the delta-function used to interpolate information between the Eulerian and Lagrangian grids possesses second-order accuracy only for a smooth interface where the velocity gradient normal to the interface is continuous~\cite{lai2001remark,peskin2002immersed,uhlmann2005immersed}. While this remark is already quite well-known in IBM, we here provide a theoretical proof in the Appendix B. The idea of this proof is to assume the velocity prior to the boundary forcing process is exact, and examine what is the order of the error generated in the boundary forcing process.

Although the retraction of Lagrangian grid does not improve the order of accuracy of the velocity calculation in the LBM-IBM simulation, it does significantly reduce the magnitudes of the error at all resolutions (the results labeled Uhlmann in Fig.~\ref{fig:velocitycovergence} is equal to the case with zero retraction distance). Breugem (2012) examined the effect of the retraction distance in a few flow examples, such as a uniform flow passing a fixed sphere and the laminar pipe flow, and suggested that $R_d = 0.3\delta x$ was the general optimal retraction distance. Zhou \& Fan (2014) also confirmed such observation in LBM-IBM that an optimized retraction distance should be $0.3\delta x\le R_d \le 0.4\delta x$. The three-point delta-function of Roma {\it et al}.~\cite{roma1999adaptive} was adopted in both studies to draw this conclusion. Intuitively, since the physical fluid-solid interface is diffused at different levels by different delta-functions, the optimal retraction distance to offset such diffusion should be delta-function dependent. 

\begin{figure}
\centering
{\includegraphics[width=80mm]{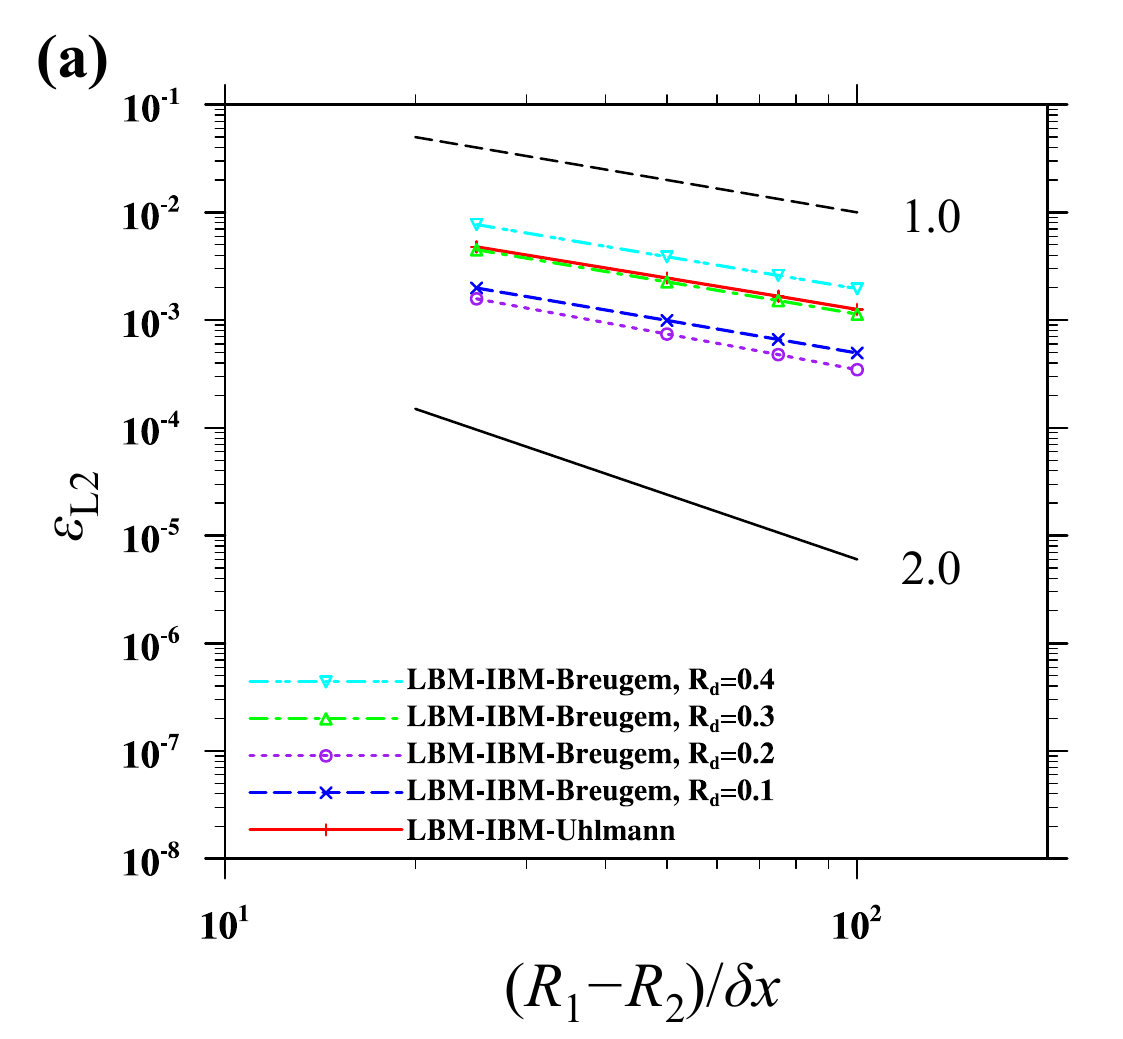}~~~~~\includegraphics[width=80mm]{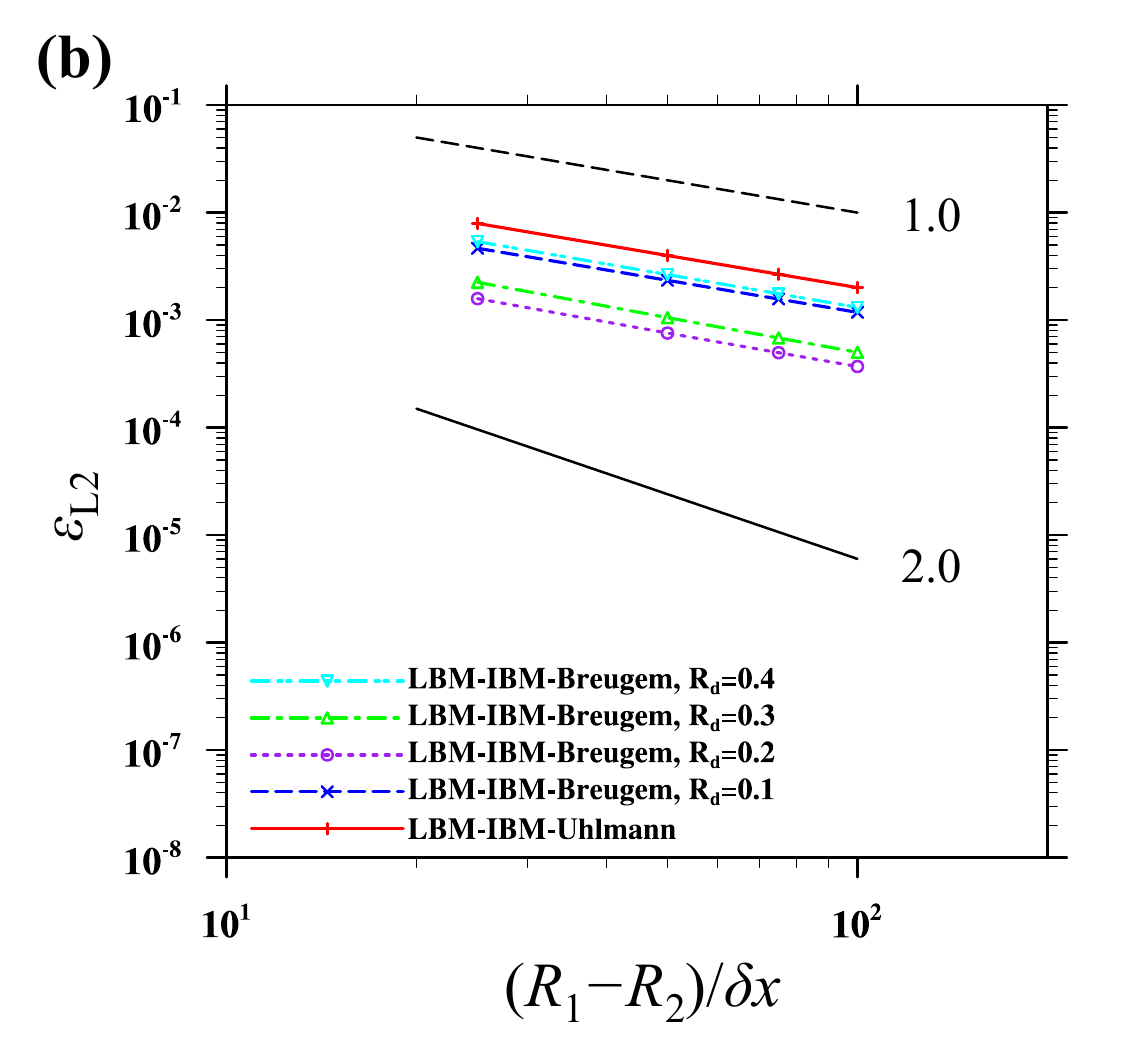}}
\vspace{0.1in}
\centering
{\includegraphics[width=80mm]{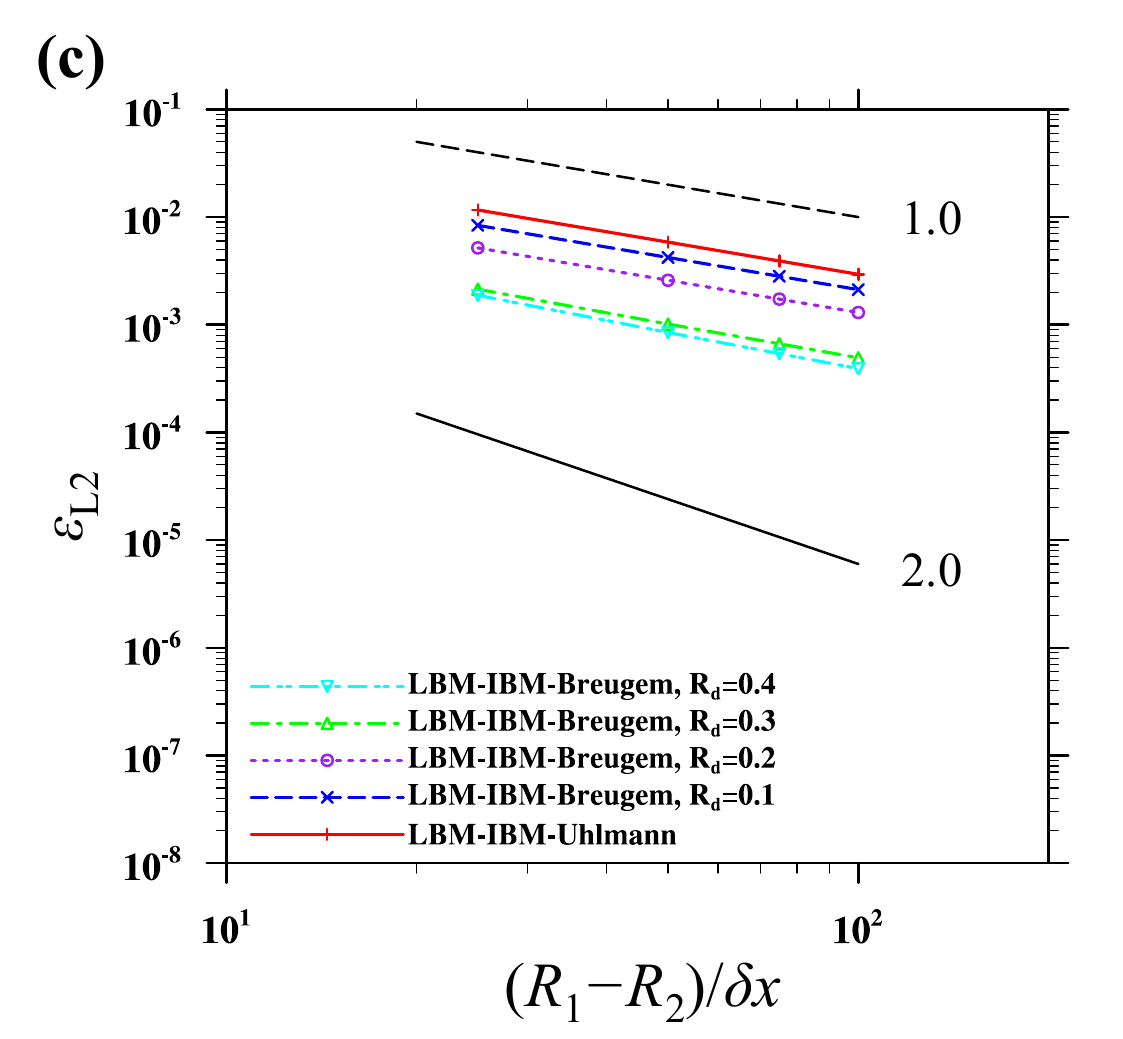}}
\caption{The convergence rates of velocity field with different delta-functions, (a) two-point linear delta-function, (b) three-point delta-function, (c) four-point delta-function.}
\label{fig:deltafunctions}
\end{figure}

To confirm this point, we simulate the same TC flow with different combinations of three delta-functions, {\it i.e.}, the four-point piecewise delta-function used above, the three-point piecewise delta-function by Roma {\it et al}., and the two-point linear delta-function, and five retraction distances $R_d = 0$, $R_d = 0.1\delta x$, $R_d = 0.2\delta x$, $R_d = 0.3\delta x$, and $R_d = 0.4\delta x$. It should be noted that the three-point piecewise delta-function and the two-point linear delta-function diffuse the physical fluid-solid interface less than their four-point counterpart, which may bring a negative impact on the numerical stability. In fact, with all the other simulation setup parameters being identical to what were
used earlier, switching to the three-point and two-point delta-functions made the code diverge. To ensure numerical stability with all combinations, a smaller flow Reynolds number $Re = (R_{2} - R_{1})\Omega_{2}R_{2}/\nu = 15$ is used instead. The convergence rates of the steady state velocity fields in different cases are shown in Fig.~\ref{fig:deltafunctions}. In each simulation, the boundary force is still iterated for 5 times to ensure better no-slip boundary representation. As shown in Fig.~\ref{fig:deltafunctions}, when the two-point linear delta-function is used, the retraction distances of $R_d = 0.1\delta x$ and $R_d = 0.2\delta x$ result in the most accurate velocity field. With more diffusive delta-functions, the optimized retraction distance becomes larger in magnitude. With the three-point delta-function, the optimized retraction distance is between $R_d = 0.2\delta x$ and $R_d = 0.3\delta x$; while with the four-point delta-function, the best result is observed
when $R_d = 0.3\delta x$ and $R_d = 0.4\delta x$. Another observation worth mentioning is that, for the current Reynolds number $Re = 15$, with four-point delta function (Fig.~\ref{fig:deltafunctions}(c)), the retraction distance of $R_d = 0.4\delta x$ only results in slightly more accurate velocity field than the retraction distance of $R_d = 0.3\delta x$. However, as shown in Fig.~\ref{fig:velocitycovergence}, when the Reynolds number increases to $Re = 45$, the  results improve much more significantly when  $R_d$ is increased from 0.3 to 0.4. A Reynolds number dependence of the optimized retraction distance may also  be expected.

Unlike LBM-IBM, the interpolated bounce-back schemes can preserve the second-order spatial accuracy when curved no-slip surfaces are present. This is because the interpolation schemes ensure the construction of the unknown distribution functions at the boundary grid points is of second- or higher-order spatial accuracy. Particularly, the single-node bounce-back scheme by Zhao \& Yong is able to achieve a second-order accuracy using only the information at the boundary node itself. This scheme is useful when  simulating flow in porous media, or flows with dense particle suspensions, where narrow gaps can form between two solid surfaces that disables multiple-point interpolations. With the contribution of Zhao \& Yong's bounce-back scheme, the no-slip boundary treatment via IBB should possess second-order accuracy in any situation. 

\begin{figure}
\centering
\includegraphics[width=90mm]{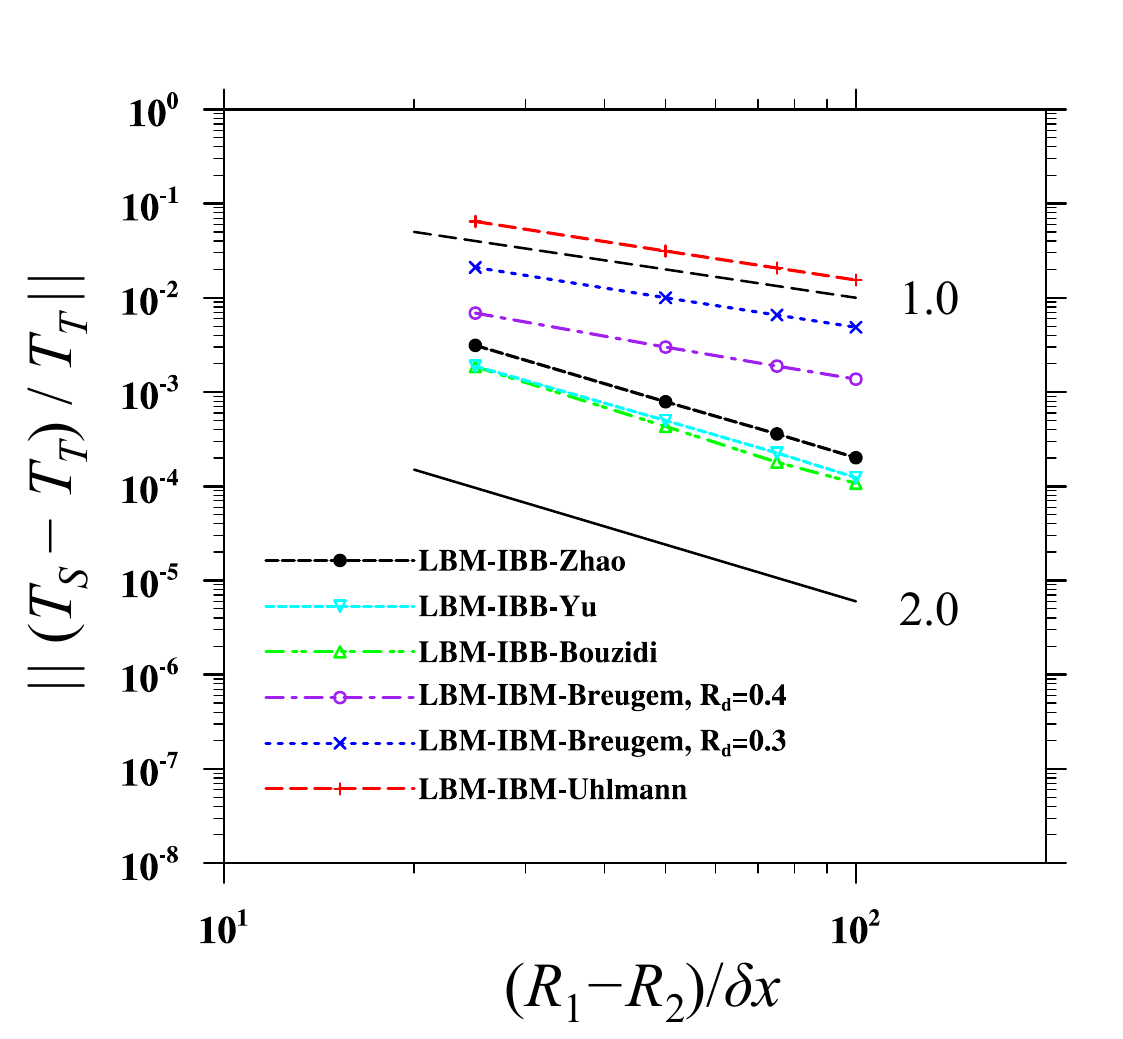}
\caption{The convergence rates of the torque evaluation error.}
\label{fig:convergerateforce}
\end{figure}

We next examine the accuracy of simulated hydrodynamic force in LBM-IBM and LBM-IBB. In LBM-IBM, the boundary force and torque have already been calculated at each Lagrangian grid, obtaining the total hydrodynamic force and torque acting on the solid objects simply amounts to summing up the contributions over all the Lagrangian grid points. When LBM-IBB is used, the hydrodynamic force and torque are calculated with Eq.~(\ref{eq:GIMEM}). A slight difference to note is that when LBM-IBB is used, the force calculated with the momentum exchange method contains a hydrostatic pressure contribution in the wall-normal direction since there is no fluid inside the solid domain. On the other hand, the force calculated in LBM-IBM contains only the viscous stress, as node points exist on both sides of the boundary. Fig.~\ref{fig:timedependenttorque} shows the torques acting on the inner and outer cylinders for the same cases shown in Fig.~\ref{fig:velprofiles}(b) and \ref{fig:velprofiles}(c).
The two solid black lines represent the analytic torque on the inner and outer cylinders at the steady state. While the torque results of LBM-IBB match well with the analytic results on both cylinders, the result of LBM-IBM has a significant derivation from the theory on the outer cylinder. This is again due to the poor treatment on edges of the computational domain. Rather than setting the Dirichlet boundary $u_{r} = 0, u_{\theta} = \Omega_{2}r$, a better boundary condition potentially reduces the error. However, this information is not available in the physical problem description. 
The accuracy of torque evaluation in different simulations is presented in Fig.~\ref{fig:convergerateforce}.
The results of the torque on the outer cylinder in LBM-IBM simulations are no longer included. Again, the torque evaluations in the three LBM-IBM simulations are still first-order accurate, with or without retracting the Lagrangian grid. This observation seems to conflict with the conclusion reported in the literature that the retraction of Lagrangian grid in IBM results in a second-order accurate total force/torque. As shown in Appendix B, the local velocity fields in IBM have only first-order accuracy, which constrains the accuracy of local force evaluation to be the first order. Whether the first-order error at each Lagrangian grid point can be canceled out to result in a second-order accurate total force/torque depends on the specific flow patterns. In a Taylor-Couette flow, the flow is azimuthally independent, which means the local error of hydrodynamic force calculation at each Lagrangian grid point should be the same. In this case, the first-order local errors cannot be cancelled out, as such the total hydrodynamic force remains to have only a first-order accuracy. On the other hand, in cases of a uniform flow passing a fixed cylinder or sphere, symmetric flow pattern may form around the cylinder/sphere. In such cases, the first-order local error contributed by each Lagrangian point may cancel out precisely to yield a second-order accuracy for the total force. The latter observation has been widely reported in the literature~\cite{peng2008comparative,breugem2012second,zhou2014second}, and also confirmed by our own simulation in Sec.~\ref{sec:uniform}. We emphasize that the hydrodynamic force/torque calculation in IBM cannot reach the second-order accuracy in general. 
 On the contrary, the torques calculated with momentum exchange method in the LBM-IBB simulations are always second-order accurate. This is because the bounced-back distribution function in Eq.~(\ref{eq:GIMEM}) are of second-order accuracy, same as the accuracy of the interpolated bounce-back schemes.

We last examine the calculation of the dissipation rate in different LBM-IBM and LBM-IBB simulations. The dissipation rate is an important quantity in turbulent flows that affects the energy budget in a flow. In turbulent flows, the dissipation rate is often defined as $\varepsilon= 2\nu s'_{ij}s'_{ij}$, where $s_{ij}$ is the velocity strain rate tensor, ``~$'$~" indicates its fluctuation part in the Reynolds decomposition~\cite{tennekes1972first}. Here in the laminar flow, the velocity is not decomposed and the dissipation rate is defined as $\varepsilon = 2\nu s_{ij}s_{ij}$ instead. In the framework of LBM, there are two different ways to calculate the strain rate tensor $s_{ij} = 0.5\left(\partial u_{i}/\partial x_{j} + \partial u_{j}/\partial x_{i}\right)$. The first way is to use a finite-difference approximation, as adopted in conventional CFD. To preserve the accuracy, a second- or higher-order finite-difference scheme is usually required. Alternatively, $s_{ij}$ in LBM can be calculated directly as a moment of the non-equilibrium distribution functions. According to the Chapman-Enskog expansion and taking the  D2Q9 MRT collision operator used in the simulation, the three components in $s_{ij}$ can be calculated as
\begin{subequations}
    \begin{align}
        &\frac{\partial u}{\partial x} = -\frac{s_e}{4\rho_{0}\delta t}\epsilon m_{e}^{(1)} - \frac{3s_{n}}{4\rho_{0}\delta t}\epsilon m_{n}^{(1)} - \frac{1}{4\rho_{0}}\left[\left(\frac{s_{e}}{2-s_{e}}\right)\psi_{e}+\left(\frac{3s_{n}}{2-s_{n}}\right)\psi_{n}\right],
        \label{eq:dissipationcalc:A}\\
        &\frac{\partial v}{\partial y} = -\frac{s_e}{4\rho_{0}\delta t}\epsilon m_{e}^{(1)} + \frac{3s_{n}}{4\rho_{0}\delta t}\epsilon m_{n}^{(1)} - \frac{1}{4\rho_{0}}\left[\left(\frac{s_{e}}{2-s_{e}}\right)\psi_{e}-\left(\frac{3s_{n}}{2-s_{n}}\right)\psi_{n}\right],
        \label{eq:dissipationcalc:B}\\
        &\frac{1}{2}\left(\frac{\partial u}{\partial y} + \frac{\partial v}{\partial x}\right) = -\frac{3s_{c}}{2\rho_{0}\delta t}\epsilon m_{c}^{(1)}-\frac{1}{2\rho_{0}}\left(\frac{3s_{c}}{2-s_{c}}\right)\psi_{c},
    \end{align}
    \label{eq:dissipationcalc}
\end{subequations}
where $s_{e}$, $s_{n}$ and $s_{c}$ are the relaxation parameters for the energy, normal stress and shear stress moments, respectively. $\epsilon m_{e}^{(1)} \approx m_{e} - m_{e}^{(eq)}$, $\epsilon m_{n}^{(1)} \approx m_{n} - m_{n}^{(eq)}$, $\epsilon m_{c}^{(1)} \approx m_{c} - m_{c}^{(eq)}$ are their corresponding leading-order non-equilibrium part. $\psi_{e}$, $\psi_{n}$, $\psi_{c}$ are the corresponding components in the mesoscopic forcing term ${\bf \Psi}$ in Eq.~(\ref{eq:MRTcollision}), whose definition can be found in~\cite{min2016inverse}. Compared to the finite-difference approximation, the mesoscopic method of calculating the strain rate tensor from the non-equilibrium moments (or distribution functions if LBGK collision operator is used) ensures a second-order accuracy even when the velocity field in the LBM simulation is of the same second-order accuracy~\cite{yong2012accuracy,peng2017lattice}, which makes it generally preferred.

\begin{figure}
\centering
\includegraphics[width=80mm]{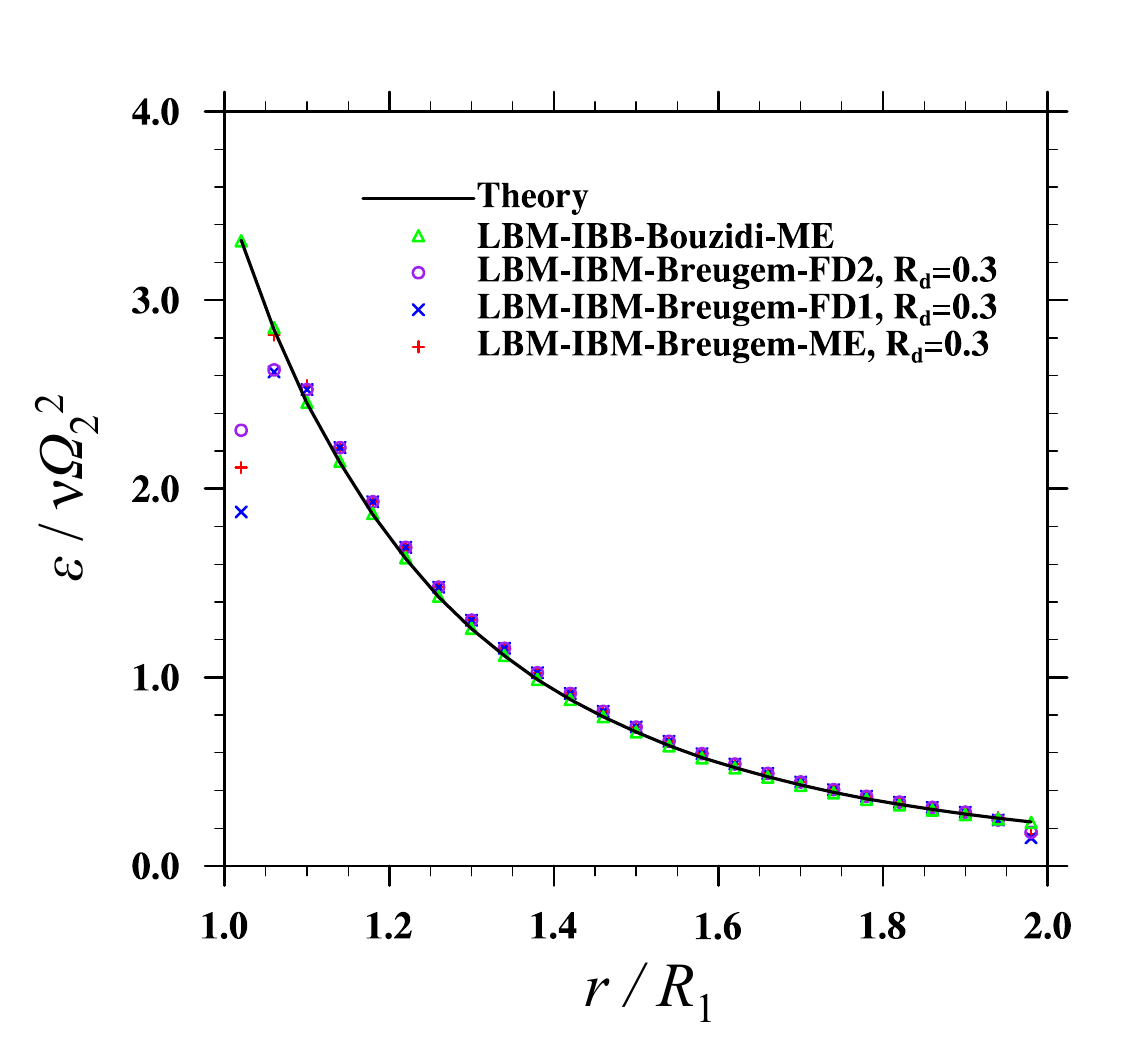}
\caption{The profiles of the dissipation rate.}
\label{fig:dissipationprofile}
\end{figure}

The profiles of dissipation rate in the two simulations shown in Fig.~\ref{fig:velprofiles} and ~\ref{fig:velprofiles}(c) are exhibited in Fig.~\ref{fig:dissipationprofile}(a). For the LBM-IBM simulation, the dissipation rate is calculated in three different ways, {\it i.e.}, 1) with the second-order central finite-difference scheme (FD1), 2) use the second-order central difference scheme in the bulk fluid region, but replace with a second-order upwind scheme near the two solid surfaces to exclude the grid points in the solid region from the calculation (FD2), and 3) from the non-equilibrium moments (ME). In the LBM-IBB simulation, for the sake of simplicity, only the mesoscopic method is employed. As shown in Fig.~\ref{fig:dissipationprofile}, no matter which method is employed to calculate the dissipation rate in the LBM-IBM simulation, the results are always significantly smaller than the theory. This is probably because IBM smooths out the sharp fluid-solid interfaces and reduces the local velocity gradient in the interface region. Excluding the grid points inside the solid volume improves the accuracy of dissipation rate calculation near the boundary but a large part of the error still remains. In the fluid region away from boundary ($1.2\le r/R_{1}\le1.8$), the dissipation rate results of the LBM-IBM simulations become acceptable, with only a slight over-prediction of the dissipation rate. 
The local dissipation rate result in LBM-IBB, on the other hand, is in excellent agreement with the theory. 
In the fluid region away from boundary ($1.2\le r/R_{1}\le1.8$), the calculated dissipation rate  from LBM-IBM is acceptable, but it is still worse than that in LBM-IBB. This indicates that the overall accuracy of IBB in terms of no-slip boundary treatment is much better than that in IBM.

\begin{figure}
\centering
\includegraphics[width=80mm]{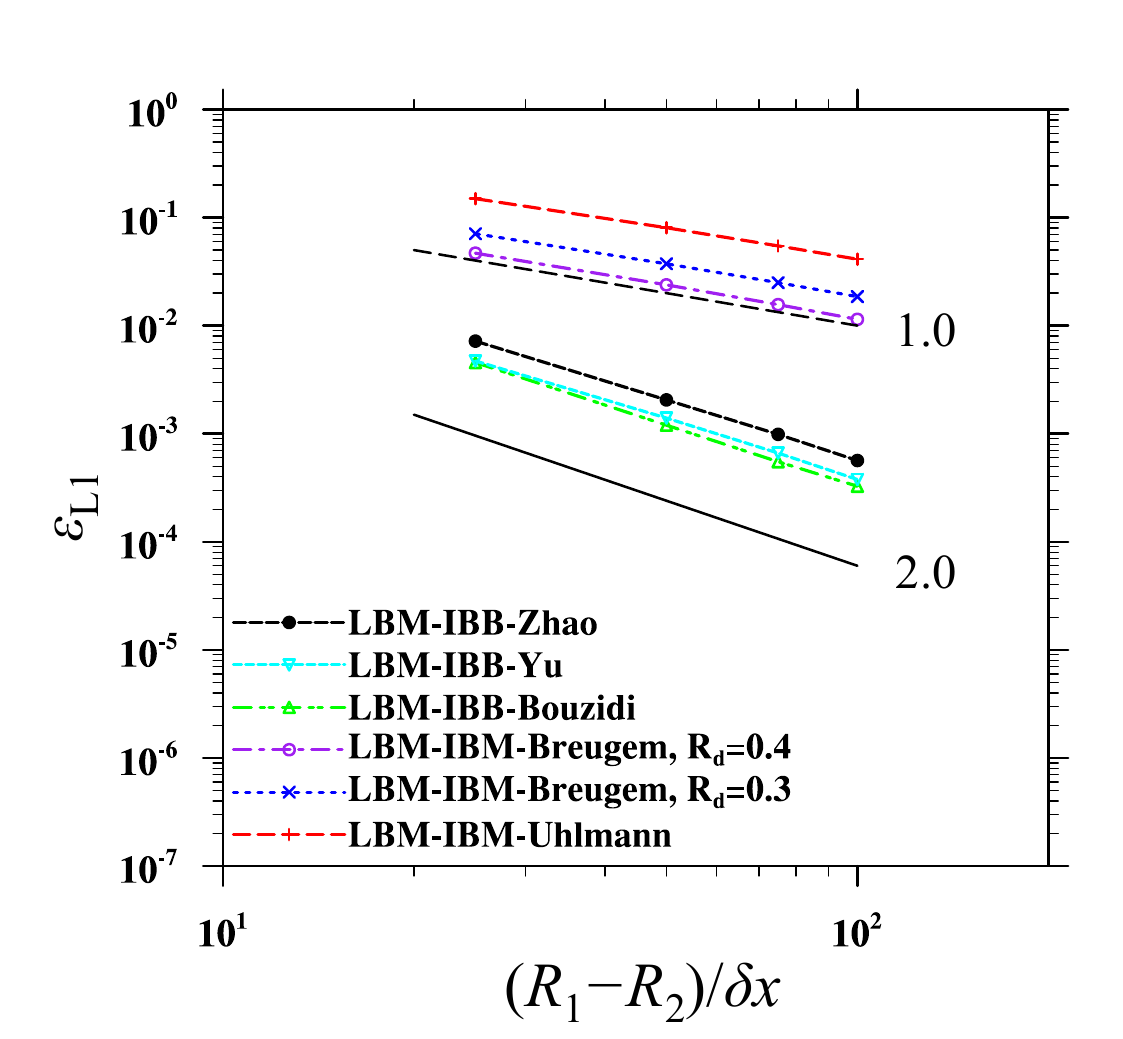}
\caption{The convergence rates of the dissipation rate calculation.
}
\label{fig:convergeratedissipation}
\end{figure}

\begin{figure}
\centering
\includegraphics[width=80mm]{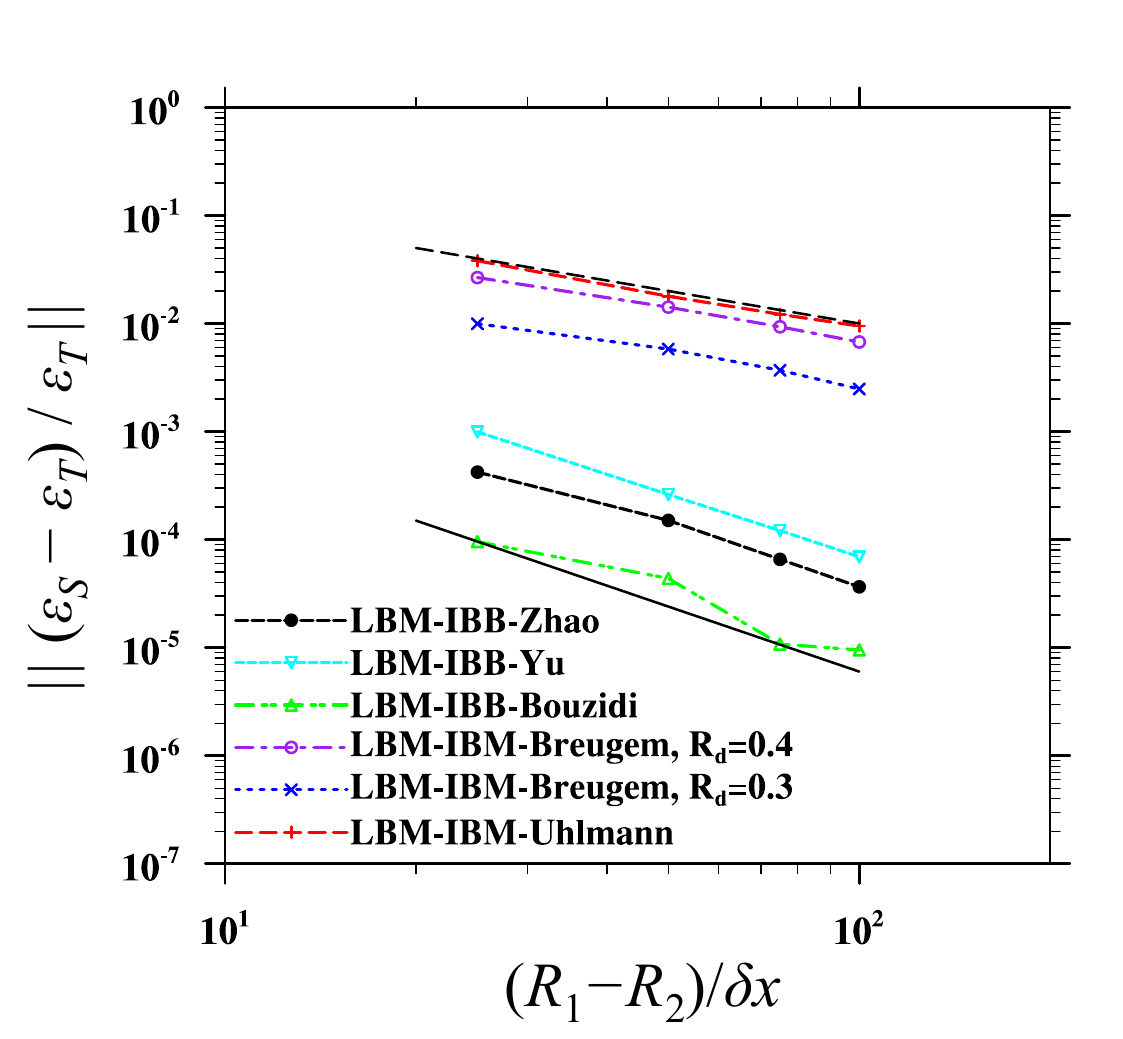}
\caption{The convergence rates of the total dissipation rate in the whole fluid domain.}
\label{fig:convergeratetotaldissipation}
\end{figure}

The convergence rates of the dissipation rate calculation in the LBM-IBM and LBM-IBB  are presented in Fig.~\ref{fig:convergeratedissipation}. Since the non-uniform distributions of the error in the LBM-IBM simulations (see Fig.~\ref{fig:dissipationprofile}) tend to amplify the L2 norm, only the L1 norm is presented. For conciseness, only the dissipation rate calculated by the mesoscopic way are presented. Clearly, the dissipation rate  in LBM-IBM is of only the first-order accuracy, while the dissipation rate in LBM-IBB is of the
second-order accuracy. The L1 error in the latter is about one to two orders of magnitude smaller.  While using IBM to treat the no-slip boundary can lead to significant numerical errors in dissipation rate results near the fluid-solid interfaces, the results of the total dissipation summing over the whole fluid domain tend to be more acceptable. The corresponding results are shown in Fig.~\ref{fig:convergeratetotaldissipation}. This is because the underestimated dissipation rates near the fluid-solid interfaces due to the diffused boundary are offset by their overestimated counterparts away from the interfaces, which can be seen in Fig.~\ref{fig:dissipationprofile}. Nevertheless, the above comparisons indicate that the regular definition of dissipation rate may need to be
improved in order to account for the diffused boundary effect in IBM. This aspect receives little attention in the past, thus further investigations are certainly required.

\subsection{Sedimentation of a cylinder in a vertical channel}\label{sec:sedimentation}
    Next we compare the performance of the interpolated bounce-back schemes and immersed boundary methods in calculating the force/torque on a moving solid object. For a solid object immersed in a viscous fluid, the governing equations for its translational motion and angular rotation read
    \begin{subequations}
        \begin{align}
        &\rho_{p}V_{p}\frac{d\vec{v}}{dt} = \oint_{\partial s}\left({\vec{\vec{\tau}}}\cdot{\vec{n}}\right)ds+\left(\rho_{p}-\rho_{f}\right)V_{p}{\vec g}+{\vec F}_{int}+\cdots,
        \label{eq:physicalparticlemotion:A}\\
        &I_{p}\frac{d\vec{\omega}}{dt} = \oint_{\partial s}{\vec r}\times\left({\vec{\vec{\tau}}}\cdot{\vec{n}}\right)ds+{\vec T}_{int}+\cdots,
         \label{eq:physicalparticlemotion:B}
        \end{align}
        \label{eq:physicalparticlemotion}
    \end{subequations}
    where $\rho_{p}$ and $\rho_{f}$ are the densities of the solid and fluid phases, respectively. $V_{p}$ is the volume of the solid object, ${\vec g}$ is the gravitational acceleration, ${\vec F}_{int}$ and ${\vec T}_{int}$ are the force and torque due to the interaction with other solid objects.  Other sources of force and torque may also be included. 
    In LBM-IBB simulations, the hydrodynamic force and torque are computed from the amount of momentum/angular momentum exchanges. 
    On the other hand, in LBM-IBM simulations, since the solid object is also filled with fluid, the fluid inertia inside the solid object appears in the momentum/angular momentum balances as
    \begin{subequations}
    \begin{align}
        &\oint_{\partial s}\left({\vec{\vec{\tau}}}\cdot{\vec{n}}\right)ds = \frac{d}{dt}\int_{\partial V}{\vec u}dV - \int_{\partial V}{\vec f}dV,
         \label{eq:rigidbodyassumption:A}\\
        & \oint_{\partial s}{\vec r}\times\left({\vec{\vec{\tau}}}\cdot{\vec{n}}\right)ds = \frac{d}{dt}\int_{\partial V}\left({\vec r}\times{\vec u}\right)dV - \int_{\partial V}\left({\vec r}\times{\vec f}\right)dV.
         \label{eq:rigidbodyassumption:B}
        \end{align}
        \label{eq:rigidbodyassumption}
    \end{subequations}
 where $\partial S$ and $\partial V$ are the surface and volume of a solid object.
   When calculating the hydrodynamic force/torque, the treatment of the fluid inertia inside the particle clearly plays an important role. A straightforward treatment is to assume the fluid inside the solid object is following rigid body motion, as did by Uhlmann~\cite{uhlmann2005immersed}. With such assumption, Eq.~(\ref{eq:physicalparticlemotion}) becomes
   \begin{subequations}
        \begin{align}
        &\left(\rho_{p}-\rho_{f}\right)V_{p}\frac{d\vec{v}}{dt} = - \int_{\partial V}{\vec f}dV+\left(\rho_{p}-\rho_{f}\right)V_{p}{\vec g}+{\vec F}_{int}+\cdots,
        \label{eq:uhlmannparticlemotion:A}\\
        &I_{p}\left(1-\frac{\rho_{f}}{\rho_{p}}\right)\frac{d\vec{\omega}}{dt} =  - \int_{\partial V}\left({\vec r}\times{\vec f}\right)dV+{\vec T}_{int}+\cdots.
         \label{eq:uhlmannparticlemotion:B}
        \end{align}
        \label{eq:uhlmannparticlemotion}
    \end{subequations}
   An obvious problem of Eq.~(\ref{eq:uhlmannparticlemotion}) is that the left-hand sides vanish when $\rho_{p}\approx\rho_{f}$. When the density ratio $\rho_{p}/\rho_{f}$ is below a limit, the simulations employing Eq.~(\ref{eq:uhlmannparticlemotion}) are not stable. To overcome such stability deficiency, Feng \& Michaelides~\cite{feng2009robust} proposed a specific time discretization of Eq.~(\ref{eq:uhlmannparticlemotion}) as
   \begin{subequations}
        \begin{align}
        &\rho_{p}V_{p}\frac{{\vec v}^{n+1}-{\vec v}^{n}}{\delta t} =  \rho_{f}V_{p}\frac{{\vec v}^{n}-{\vec v}^{n-1}}{\delta t}- \int_{\partial V}{\vec f}dV+\left(\rho_{p}-\rho_{f}\right)V_{p}{\vec g}+{\vec F}_{int}+\cdots,
        \label{eq:fengparticlemotion:A}\\
        &I_{p}\frac{{\vec \omega}^{n+1}-{\vec \omega}^{n}}{\delta t} =  I_{p}\frac{\rho_{f}}{\rho_{p}}\frac{{\vec \omega}^{n}-{\vec \omega}^{n-1}}{\delta t}- \int_{\partial V}\left({\vec r}\times{\vec f}\right)dV+{\vec T}_{int}+\cdots.
         \label{eq:fengparticlemotion:B}
        \end{align}
        \label{eq:fengparticlemotion}
    \end{subequations}
   Alternatively, one can directly compute the fluid inertia inside the solid object to avoid singularity when the density ratio is close to unity. Kempe {\it et al}.~\cite{kempe2012improved} used a level set functions to compute such terms as
   \begin{equation}
   \begin{split}
&\int_{\partial V}{\vec u}dV = \sum_{1}^{n_{x}}\sum_{1}^{n_{y}}\sum_{1}^{n_{z}}{\vec u}_{i,j,k}h^{3}\alpha_{i,j,k},\\
           &\int_{\partial V}{\vec r}\times{\vec u}dV = \sum_{1}^{n_{x}}\sum_{1}^{n_{y}}\sum_{1}^{n_{z}}{\vec r}_{i,j,k}\times{\vec u}_{i,j,k}h^{3}\alpha_{i,j,k},
      \end{split}
      \label{eq:kempeparticlemotion}
   \end{equation}
   where 
   \begin{equation}
   \alpha_{i,j,k} = \frac{\sum_{l=1}^{8}-\phi_{l}H\left(-\phi_{l}\right)}{\sum_{l=1}^{8}\|\phi_{l}\|},
       \label{eq:levelsetfunction}
   \end{equation}
   $\phi_{l}$ is a signed distance function,
   \begin{equation}
    \phi_{l} = \sqrt{\frac{\left({\vec{x}_{i,j,k}}-{\vec{x}_{c}}\right)^2}{a^2}+\frac{\left({\vec{y}_{i,j,k}}-{\vec{y}_{c}}\right)^2}{b^2}+\frac{\left({\vec{z}_{i,j,k}}-{\vec{z}_{c}}\right)^2}{c^2}} - 1
       \label{eq:signedDF}
   \end{equation}
  where $({\vec{x}_{c}},{\vec{y}_{c}},{\vec{z}_{c}})$ is the center location of a particle, $a$, $b$, $c$ are the lengths of the three axes of an ellipsoidal shaped particle. Apparently, one should obtain $\phi_{l}>0$ outside and $\phi_{l}<0$ inside the particle, $H(\phi_{l})$ is the Heaviside function. The summation is over the 8 corners of a three-dimensional grid cell, or 4 corners of a two-dimensional grid cell. In the following test, both Eq.~(\ref{eq:fengparticlemotion}) and Eq.~(\ref{eq:kempeparticlemotion}) will be examined in the LBM-IBM simulations of moving particles in viscous flows. 
   
   \begin{figure}
\centering
\includegraphics[width=60mm]{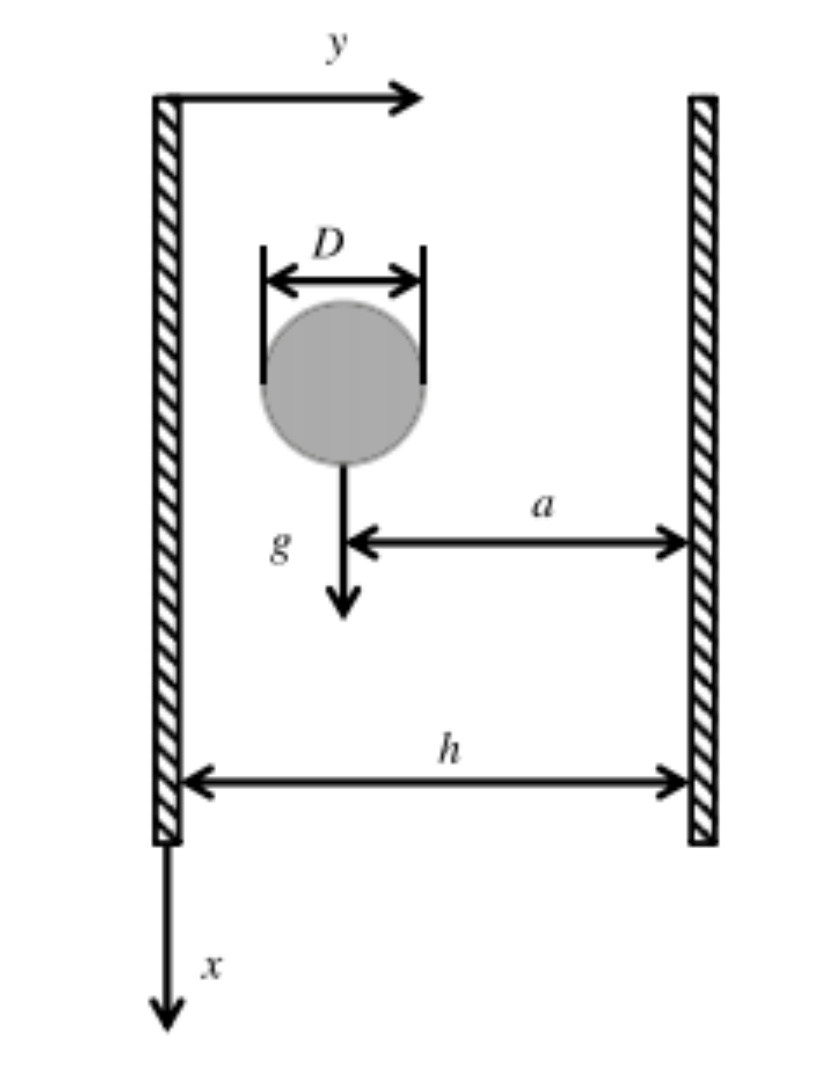}
\caption{A sketch of a cylinder settling in a quiescent flow.}
\label{fig:settlingcylinder}
\end{figure}
    
\begin{figure}
\centering
{\includegraphics[width=80mm]{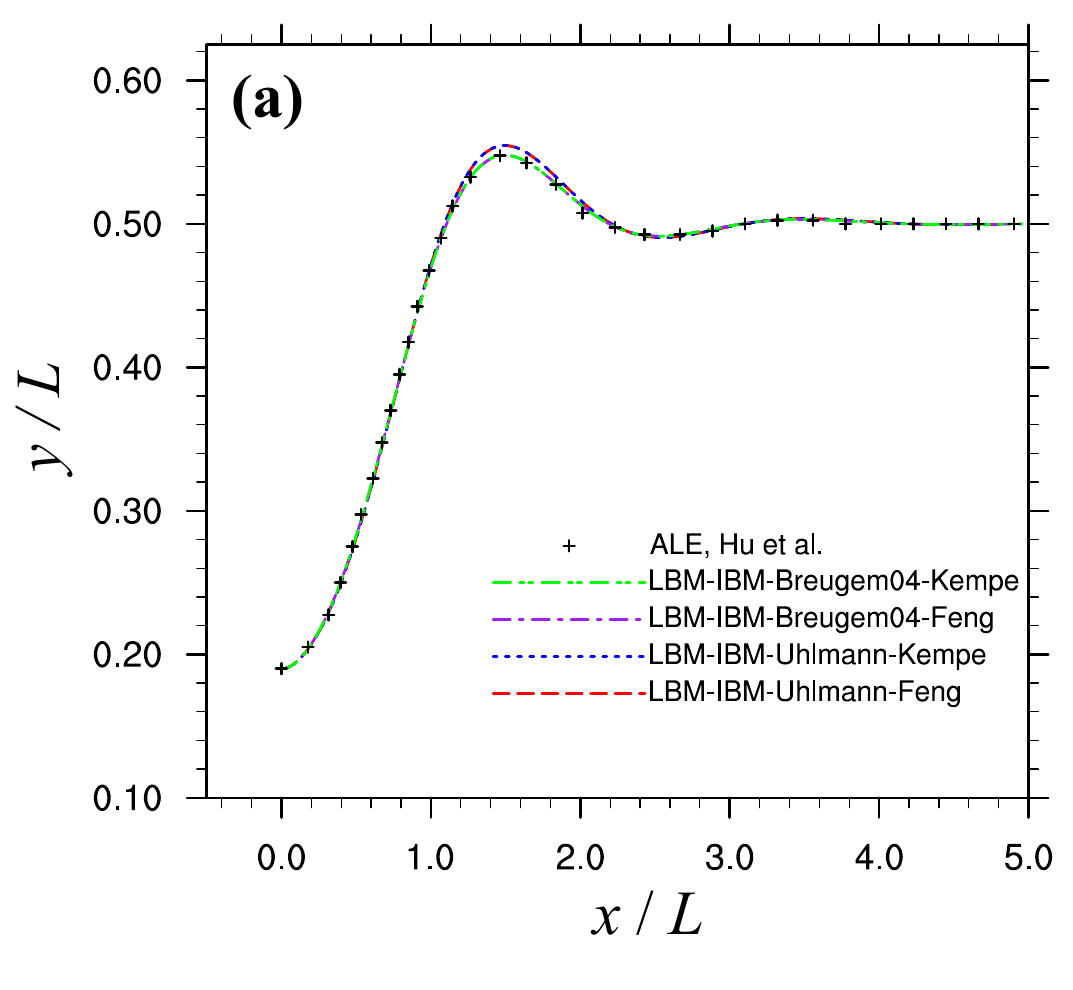}~~~~~\includegraphics[width=80mm]{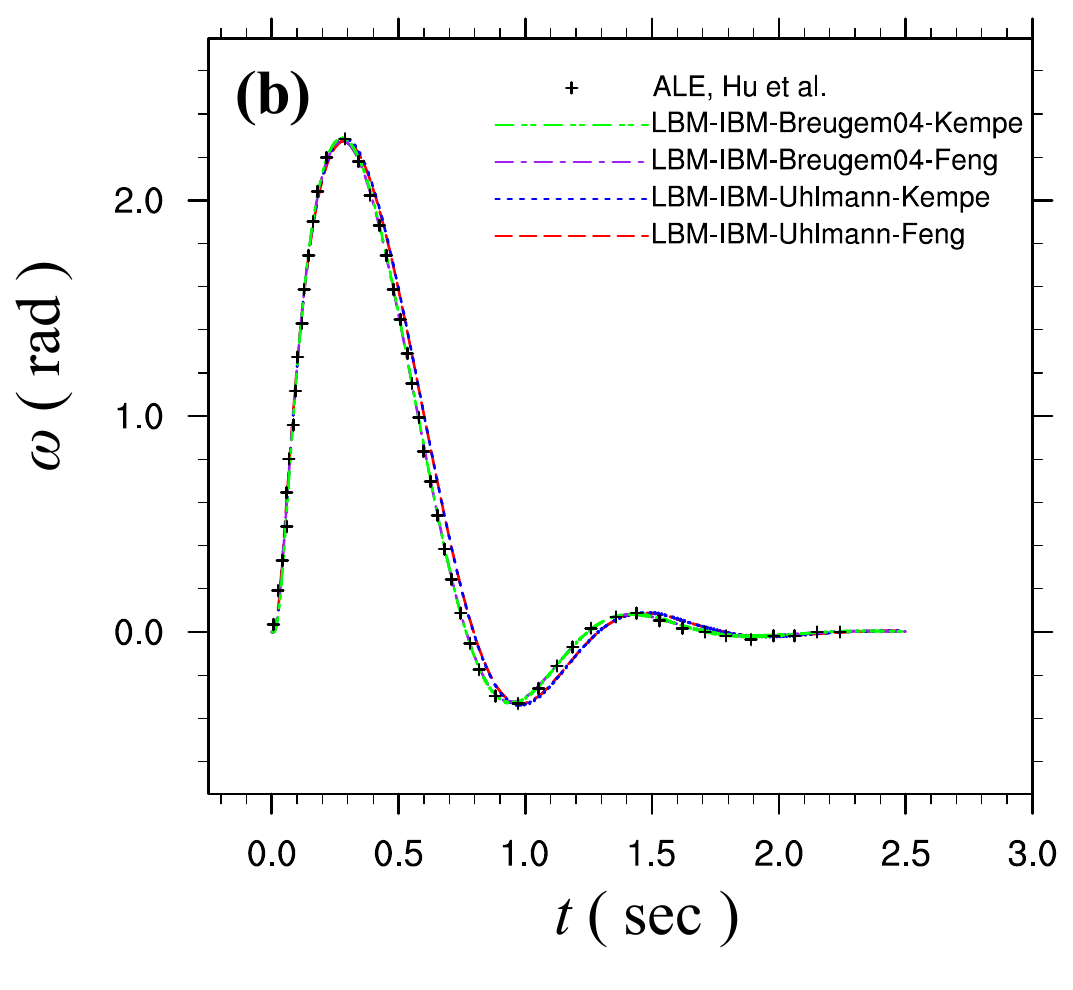}}
\vspace{0.1in}
\centering
{\includegraphics[width=80mm]{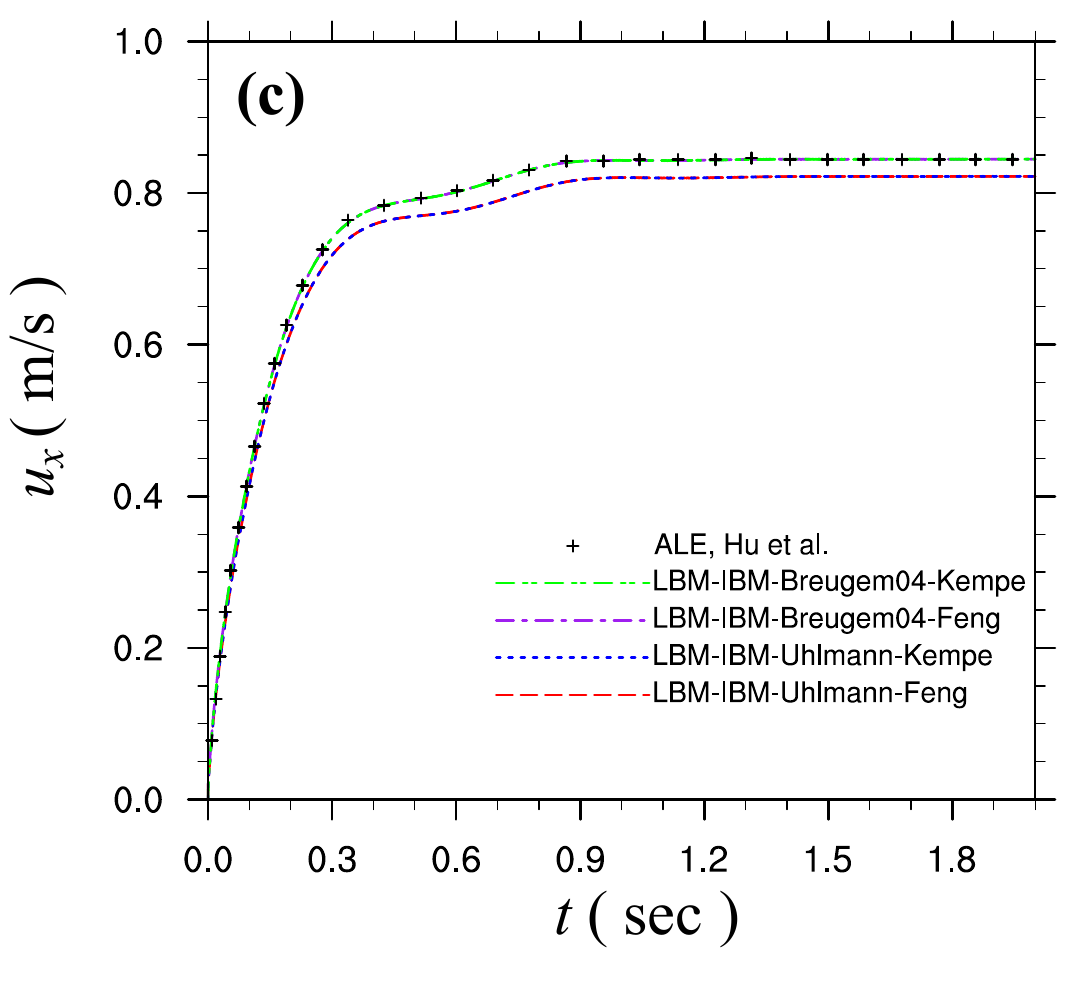}~~~~~\includegraphics[width=80mm]{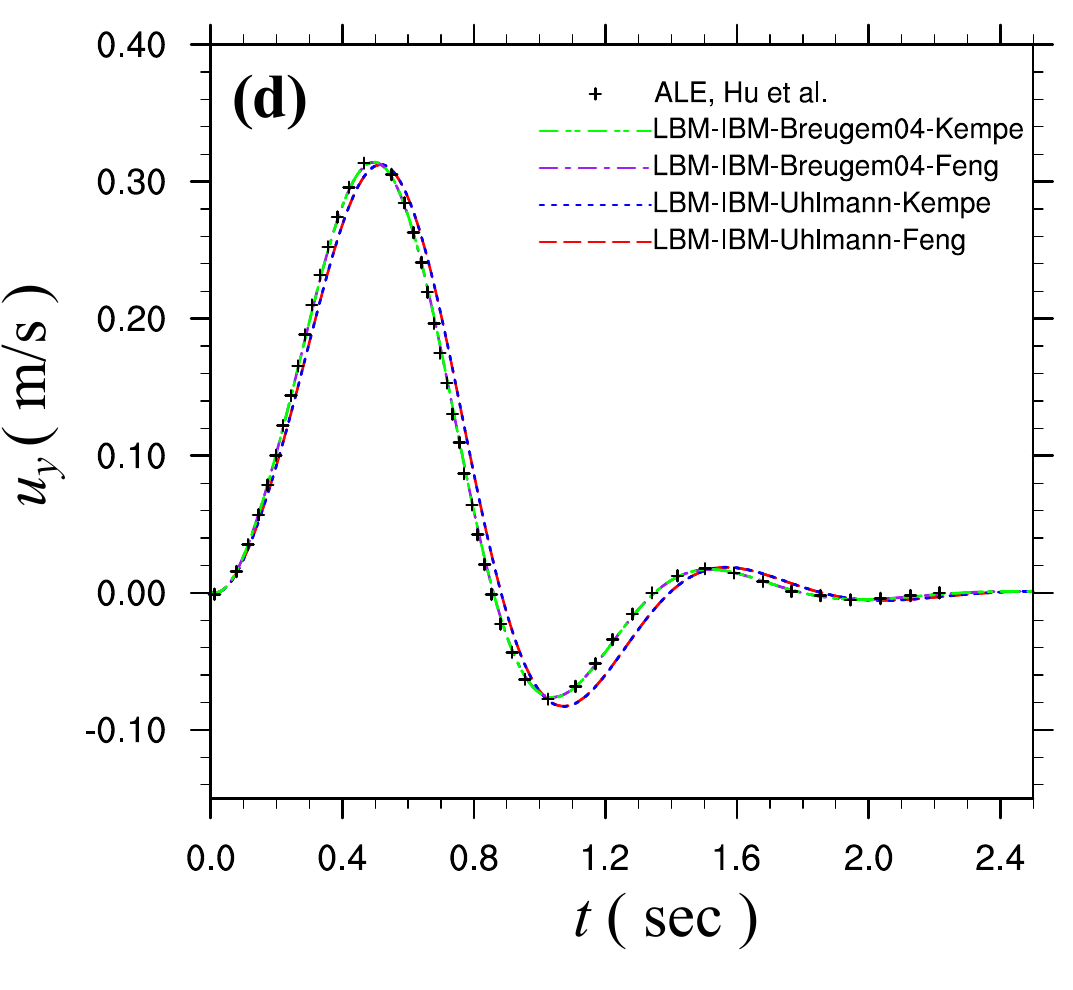}}
\caption{The effects of treatment of inertia of fluid inside cylinder on (a) particle trajectory, (b) particle angular velocity, (c) vertical translational velocity, (d) horizontal translational velocity.}
\label{fig:inertiaeffect}
\end{figure}

    The benchmark case chosen here is a cylinder settling in a vertical channel. A sketch of the flow is shown in Fig.~\ref{fig:settlingcylinder}. The parameters in physical units are chosen as $D = 0.1cm$, $L = 4cm$, 
$H = 0.4cm$, $a = 0.324cm$, $g = 980cm^2/s^2$, and the density ratio $\rho_{p}/\rho_{f} = 1.03$ to match the arbitrary Lagrangian Eulerian (ALE) simulation performed by Hu {\it et al.}~\cite{hu2001direct}. First, the two ways of considering the inertia of fluid inside the cylinder, {\it i.e.}, Eq.~(\ref{eq:fengparticlemotion}) and Eq.~(\ref{eq:kempeparticlemotion}) are compared in the LBM-IBM simulations. Eq.~(\ref{eq:uhlmannparticlemotion}) is not included as it results in instability with the current density ratio. The trajectory, angular velocity, vertical and horizontal translational velocities of the cylinder are presented in Fig.~\ref{fig:inertiaeffect}(a), \ref{fig:inertiaeffect}(b), \ref{fig:inertiaeffect}(c), and \ref{fig:inertiaeffect}(d), respectively. The results are obtained with a grid resolution of $D = 30\delta x$.  The results are not sensitive to how the inertia of the fluid in the cylinder is treated. Assuming the fluid inside the two-dimensional cylinder follows the rigid body motion appears to be safe.  Compared to Uhlmann's IBM with zero retraction distance, Breugem's IBM with a retraction distance of $r_{d} = 0.4\delta x$ clearly improves the accuracy of simulating the particle motion. Particularly, the terminal velocity of the cylinder with Uhlmann's IBM is obviously smaller than the benchmark result. This is because the diffused fluid-solid interface creates a larger effective hydraulic radius that over predicts the drag force. The retraction of Lagrangian grid points helps to offset such over prediction~\cite{breugem2012second}.  

\begin{figure}
\centering
{\includegraphics[width=80mm]{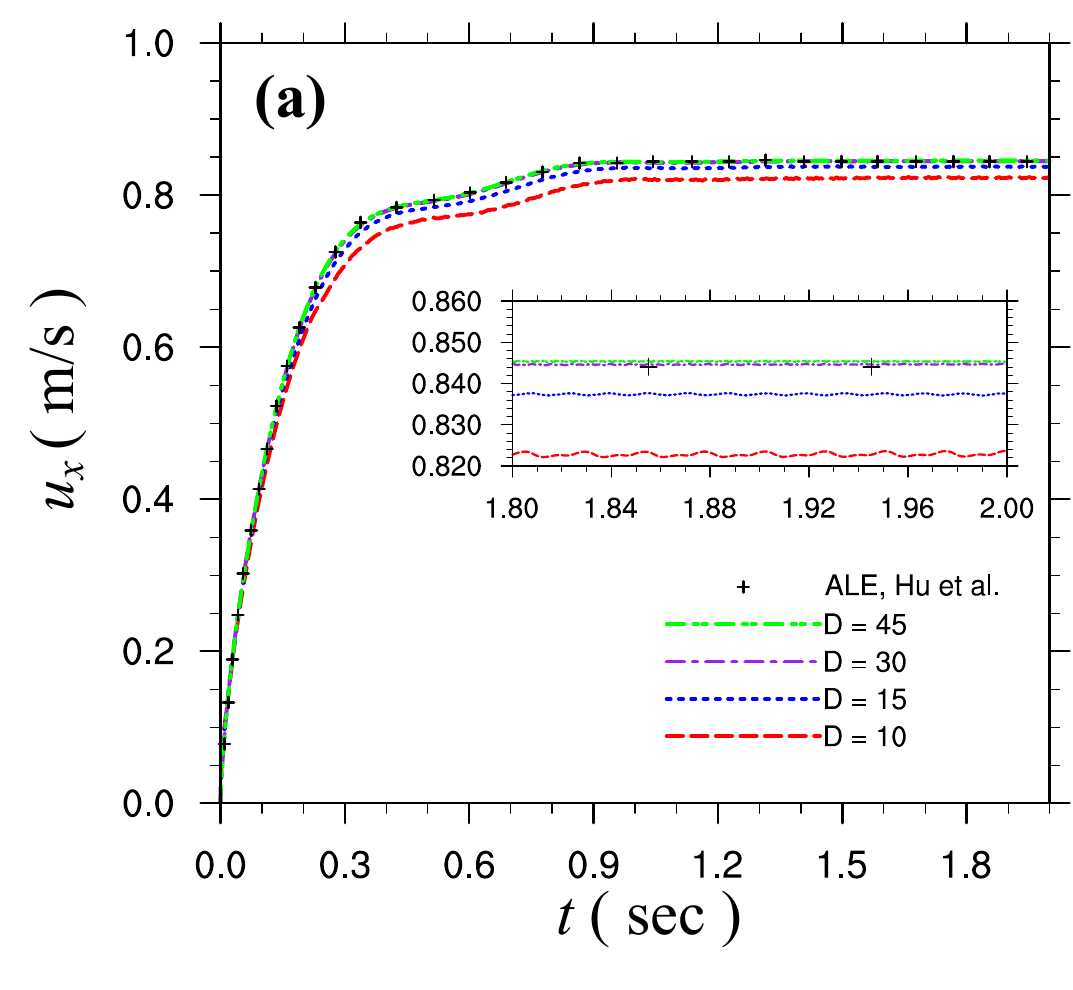}~~~~~\includegraphics[width=80mm]{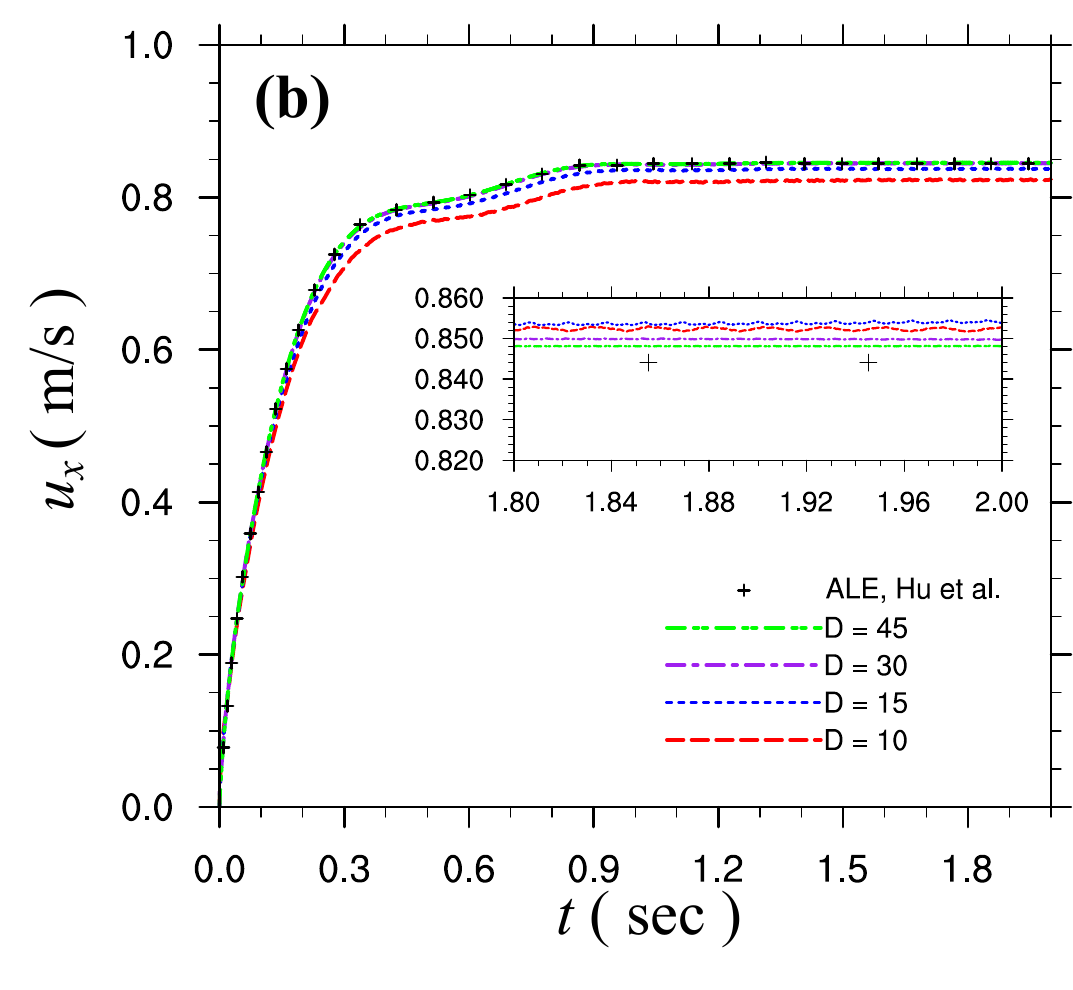}}
\vspace{0.1in}
\caption{The results of particle vertical translational velocity with different grid resolutions, (a) LBM-IBM with Breugem (2012)'s IBM, $r_{d} = 0.4\delta x$, (b) LBM-IBB with Bouzidi {\it et al.}'s quadratic interpolated bounce-back scheme.}
\label{fig:particleterminalvelocity}
\end{figure}

We next compare the performance of LBM-IBB and LBM-IBM in simulating the particle motion. The verticel velocity of the cylinder with Bouzidi {\it et al.}'s interpolated bounce-back scheme, and Breugem (2012)'s IBM with a retraction distance of $0.4\delta x$ are presented in Fig.~\ref{fig:particleterminalvelocity}(a) and ~\ref{fig:particleterminalvelocity}(b), respectively. Here we simulate the same flow with different grid resolutions from $D = 10\delta x$ to $D = 45\delta x$. The results of LBM-IBB simulation almost converge at the grid resolution of $D = 10\delta x$, while the results of LBM-IBM simulation reach the same accuracy from the grid resolution of $D = 15\delta x$. This is mainly due to the advantage of the second-order accuracy in IBB compared to the first-order accuracy in IBM. Assuming the ALE benchmark results are accurate, the grid-independent numerical error of the LBM-IBB simulation is slightly larger than that of the LBM-IBM. This benefit is likely brought by the adjustable retraction distance $r_{d}$ in the latter. 

\begin{figure}
\centering
{\includegraphics[width=80mm]{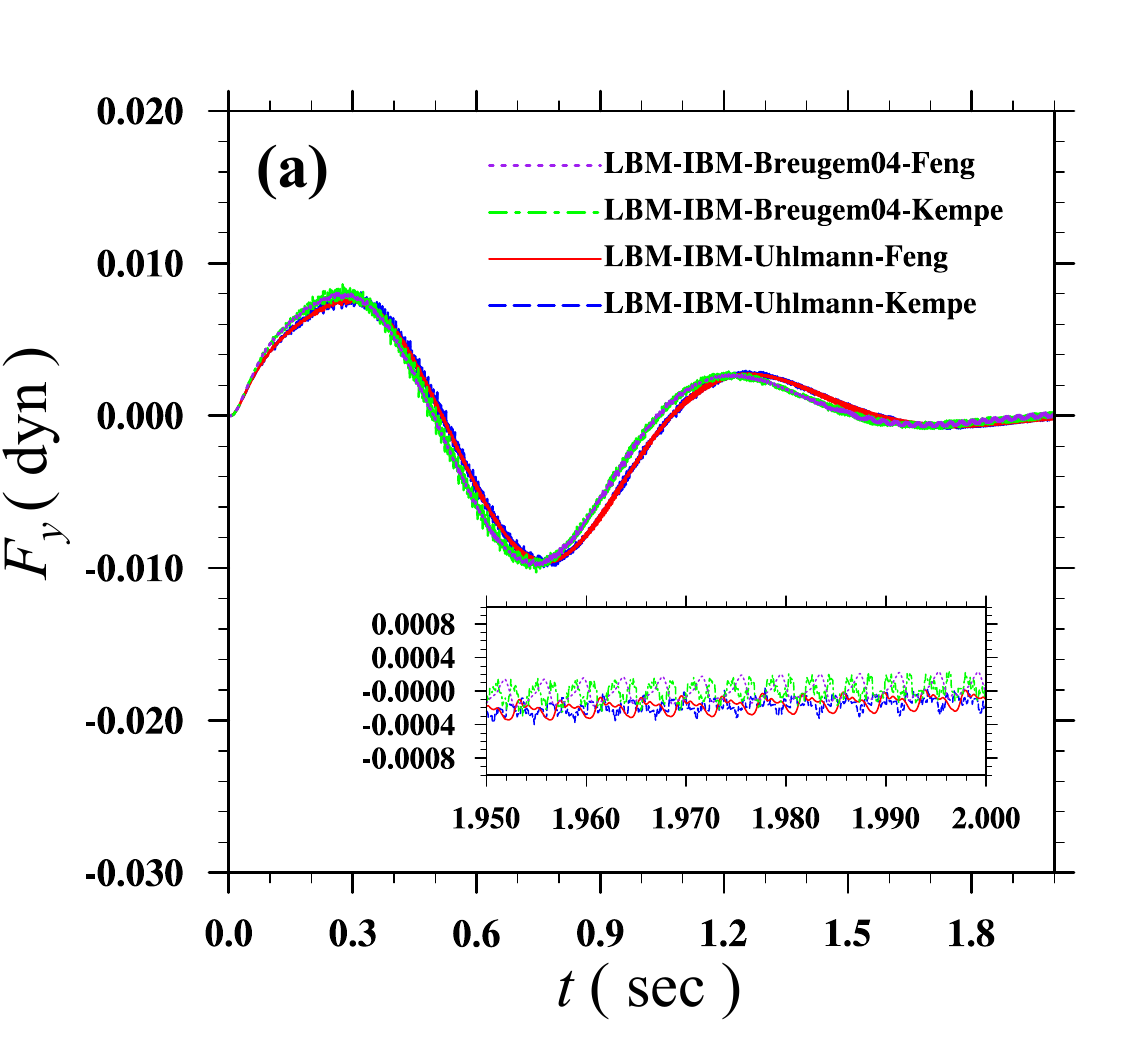}~~~~~\includegraphics[width=80mm]{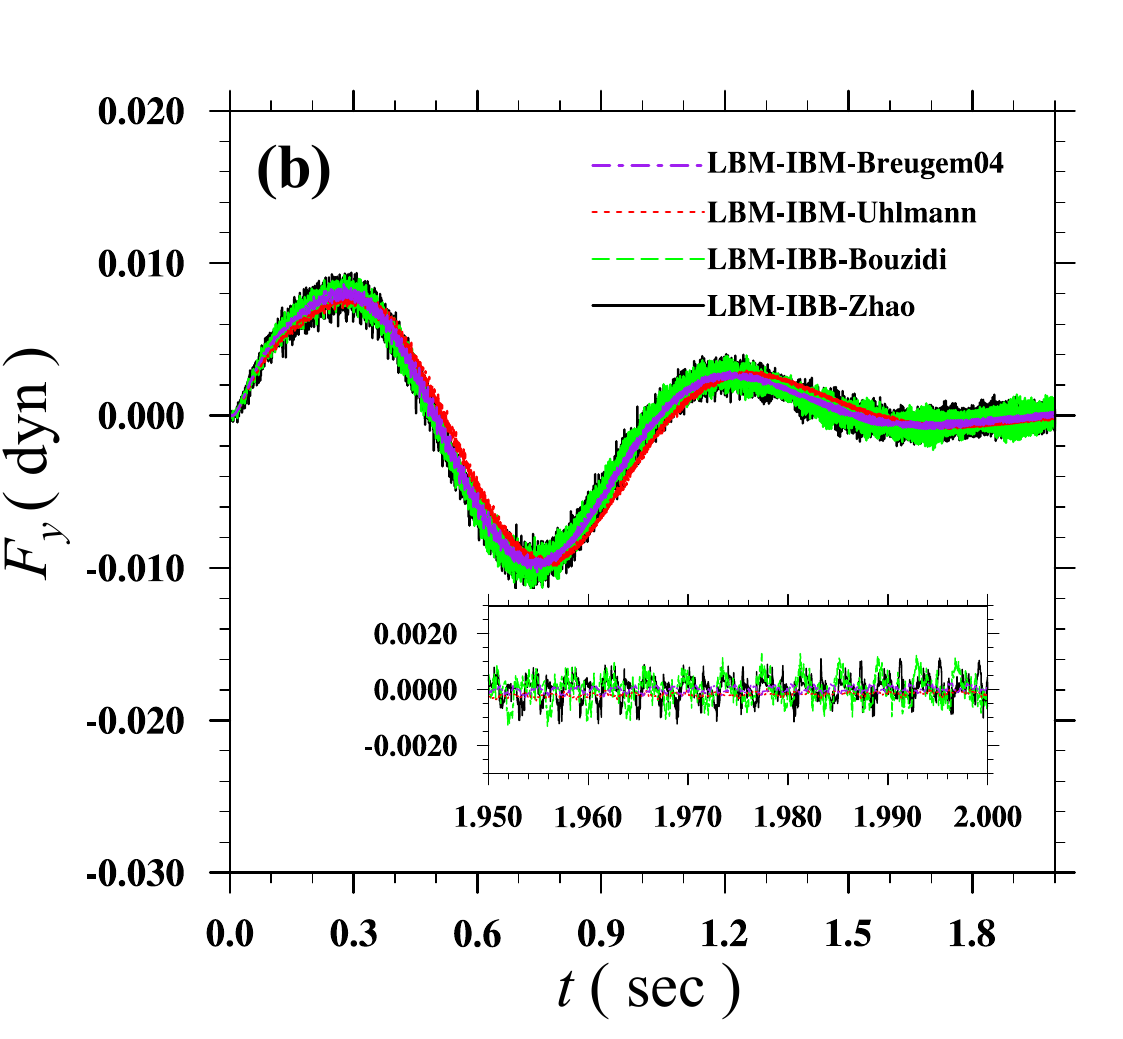}}
\vspace{0.1in}
\caption{The horizontal component of the hydrodynamic force acting on the particle, $D = 30$, (a) the effects of treatments on the inertia of fluid inside the solid region, (b) the comparison between LBM-IBB and LBM-IBM.}
\label{fig:forceevalution}
\end{figure}

At the end of this case, the level of ``grid locking" in the hydrodynamic force/torque evaluations is examined. The ``grid locking" means when a solid object crosses over the grid mesh, the calculated instantaneous hydrodynamic force/torque exerted on the solid object have a slight dependence on the configuration of the grid mesh and the solid object, and not being strictly Galilean invariant~\cite{breugem2012second}. The calculated instantaneous force and torque therefore present a high-frequency fluctuation which restores its initial value when the solid object displaces exactly one grid spacing. The term ``grid locking'' was dubbed by Breugem but the phenomenon was discovered much earlier in IBM, {\it e.g.}, in ~\cite{uhlmann2005immersed,yang2009smoothing}. The LBM-IBB simulations also suffer from the same problem, as discussed by Lallemand \& Luo~\cite{lallemand2003lattice}, Peng et al.~\cite{peng2016implementation} and Tao et al.~\cite{tao2016investigation}. Essentially, both the interpolation in IBB and the boundary diffusion in IBM have made the realization of no-slip condition on the sharp interface depending on the information of multiple grid points around, which helps to suppress the fluctuation in force and torque evaluation. In Fig.~\ref{fig:forceevalution}(a) the effects of the two schemes, {\it i.e.}, Feng \& Michaelides's scheme (Eq.~(\ref{eq:fengparticlemotion})) and Kempe {\it et al.}'s scheme (Eq.~(\ref{eq:kempeparticlemotion})), for treating the fluid inertia inside the solid volume in IBM are compared. At the initial stage, the scheme of Feng \& Michaelides presents a lower level of force fluctuation that the scheme of Kempe {\it et al.}, but the two schemes eventually lead to the same prediction of the force, as shown in the inserted zoom-in plot in Fig.~\ref{fig:forceevalution}(a). Fig.~\ref{fig:forceevalution}(b) shows the comparison of force evaluation among four simulations, two LBM-IBB simulations with the quadratic interpolation scheme of Bouzidi {\it et al.}(Eq.~(\ref{eq:Bouzidi})) and the single-node bounce-back scheme of Zhao \& Yong (Eq.~(\ref{eq:ZhaoYong})), and two LBM-IBM simulations with Uhlmann's IBM and Breugem's IBM with a retraction distance of $r_{d} = 0.4\delta x$. The two LBM-IBM simulations use Kempe {\it et al}.'s scheme to directly consider the inertia of fluid inside particle region. Compared to the IBB schemes, the IBM results clearly better suppress the force fluctuation. This benefit is a result that the delta-functions employed in IBM diffuse the sharp interface more in IBM than in the interpolation schemes used in IBB. In IBM, the force contributed from a single Lagrangian node depends the information from a maximum $4\times 4$ (in 2D) subdomain of the Eulerian mesh. On the contrary, in IBB, the force contributed by a single boundary link depends on the information from no more than three node points. The latter system therefore has a much less inertia to suppress the high-frequency fluctuations. The force fluctuation in an IBM simulation might be further suppressed by using more diffusive delta-functions with larger spans, as suggested in Ref.~\cite{yang2009smoothing}. However, those more diffusive delta-functions will introduce larger numerical viscosity and further reduce the accuracy of the IBM simulation in terms of averaged quantities. If the instantaneous force/torque computation is not of particular importance and the simulation has sufficient numerical stability, less diffusive delta-functions should be recommended. We notice that many finite-volume based IBM studies ({\it e.g.}, Ref.~\cite{uhlmann2005immersed,breugem2012second}) recommended the three-point delta-function by Roma {\it et al.}~\cite{roma1999adaptive}, perhaps due to the balance between its ability to suppress the force fluctuation and acceptable boundary diffusion. In our LBM based IBM simulations, however, we found the use of three-point delta-function leads to larger vulnerability for numerical instability (see Sec.~\ref{sec:CircularCouette}). We therefore recommend the four-point delta-function by Peskin~\cite{peskin2002immersed} instead.

Another source for the larger fluctuation in IBB is due to the requirement that  the distribution functions at a new grid point when it is uncovered by the cylinder need to be initialized (known as ``refilling"), since no distribution function is assigned to nodes inside the solid region when IBB is used. Proper refilling schemes may reduce fluctuation but its contribution cannot be removed~\cite{peng2016implementation}. The LBM-IBM, on the other hand, avoids the refilling process, as the whole computational domain is filled with fluid and assigned with distribution functions.

\subsection{Transient laminar pipe flow}\label{sec:laminarpipe}
We now move to discuss results for two three-dimensional problems, with the purpose to further support the remarks that have  been made using the two-dimensional flows discussed above. 
The first 3D flow is the transient laminar pipe flow. Strictly speaking, this flow is a two-dimensional flow, but is run on three-dimensional Cartesian grids. Under a constant driving force, the flow that is initially static in a circular pipe accelerates and reaches a steady state. The governing equation of this axi-symmetric flow reads as
\begin{equation}
\frac{\partial u_{z}}{\partial t} = \nu\frac{1}{r}\frac{\partial}{\partial r}\left(r\frac{\partial u_{z}}{\partial r}\right) + g,
\label{eq:laminarpipe}
\end{equation}
where $u_{z}$ is the streamwise velocity in a cylindrical coordinate system $\left(r,\theta,z\right)$, $g$ is the constant body force driving the flow. Applying periodic boundary condition in the streamwise direction and no-slip condition on the pipe wall, the above governing equation can be solved theoretically to obtain as~\cite{wang2008direct}
\begin{equation}
u_{z}\left(r,t\right) = u_{0}\left[\left(1-\frac{r^2}{R^2}\right)-\sum_{n=1}^{\infty}\frac{8J_{0}\left(\lambda_{n}r/R\right)}{\lambda_{n}^3J_{1}\left(\lambda_{n}\right)}exp\left(-\frac{\lambda_{n}^{2}\nu t}{R^2}\right)\right],
\label{eq:laminarpipesolution}
\end{equation}
where $u_{0}=gR^2/4\nu$ is the centerline velocity at the steady state, $J_{0}$ and $J_{1}$ are the Bessel function of the first kind $J_{\alpha}$ for integer orders $\alpha = 0$ and $\alpha = 1$, $\lambda_{n}$ is the $n$th root for $J_{0}$.

\begin{figure}
\centering
{\includegraphics[width=55mm]{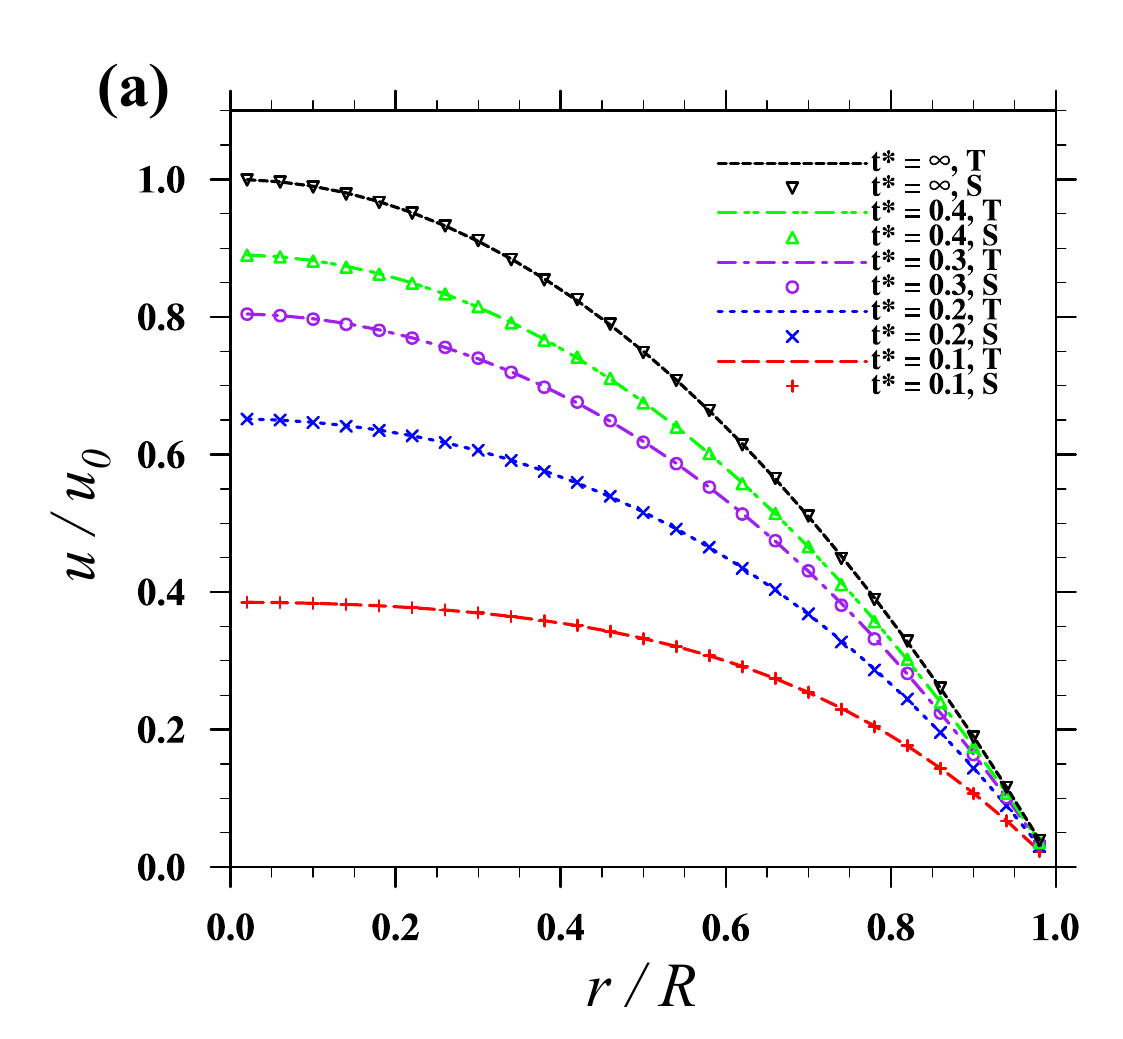}~~~~~\includegraphics[width=55mm]{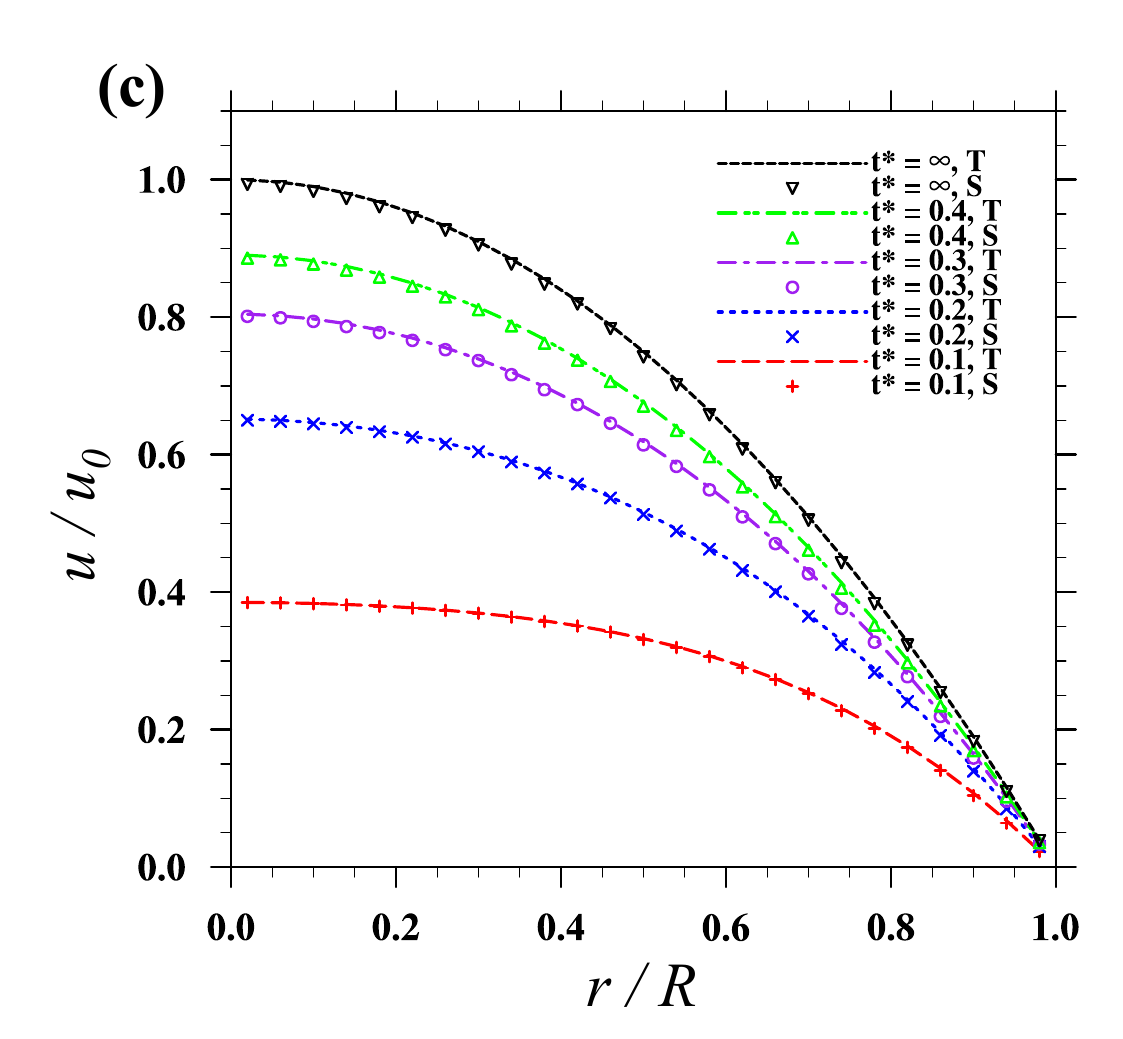}~~~~~\includegraphics[width=55mm]{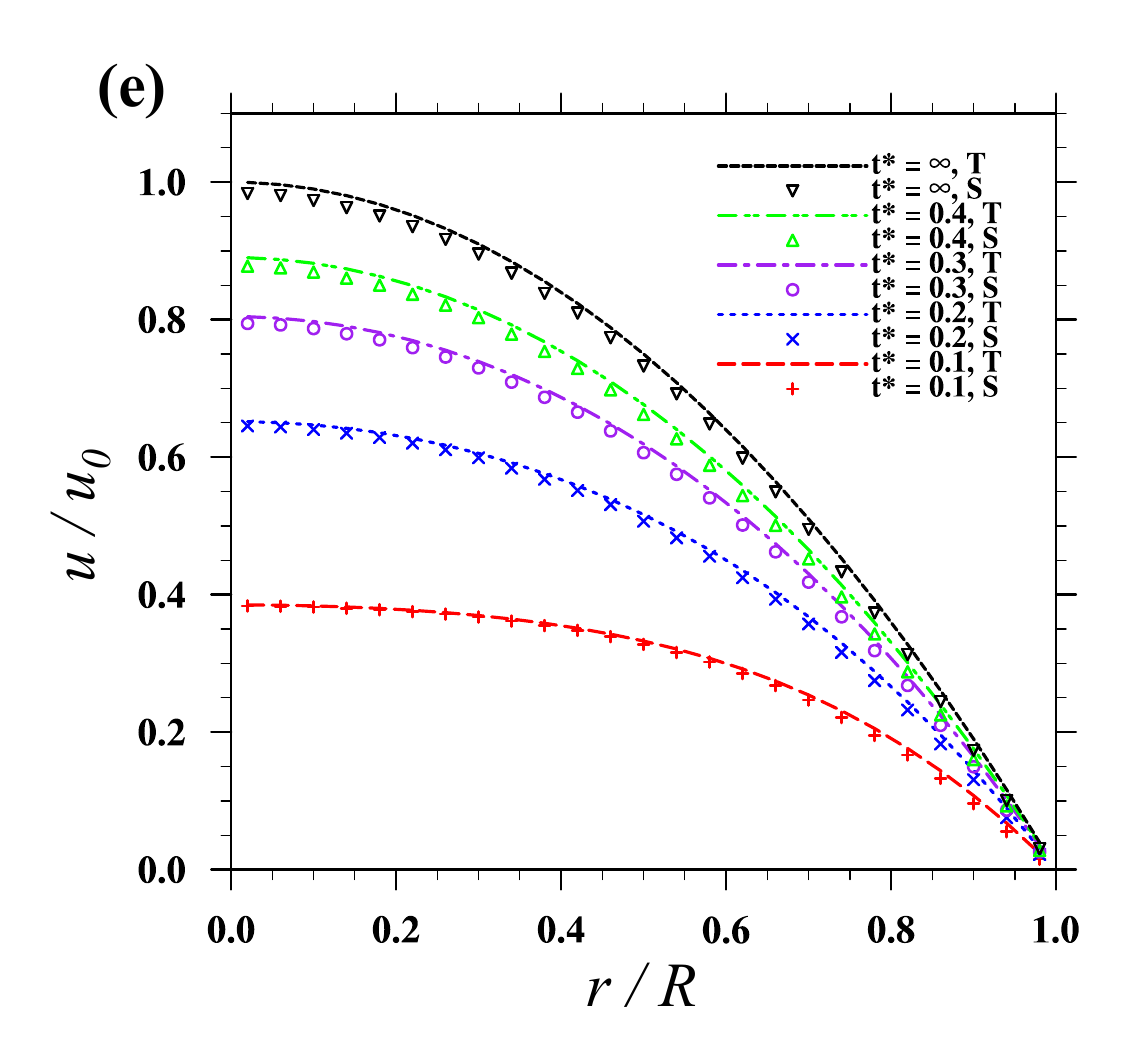}}
\vspace{0.05in}
\centering
{\includegraphics[width=55mm]{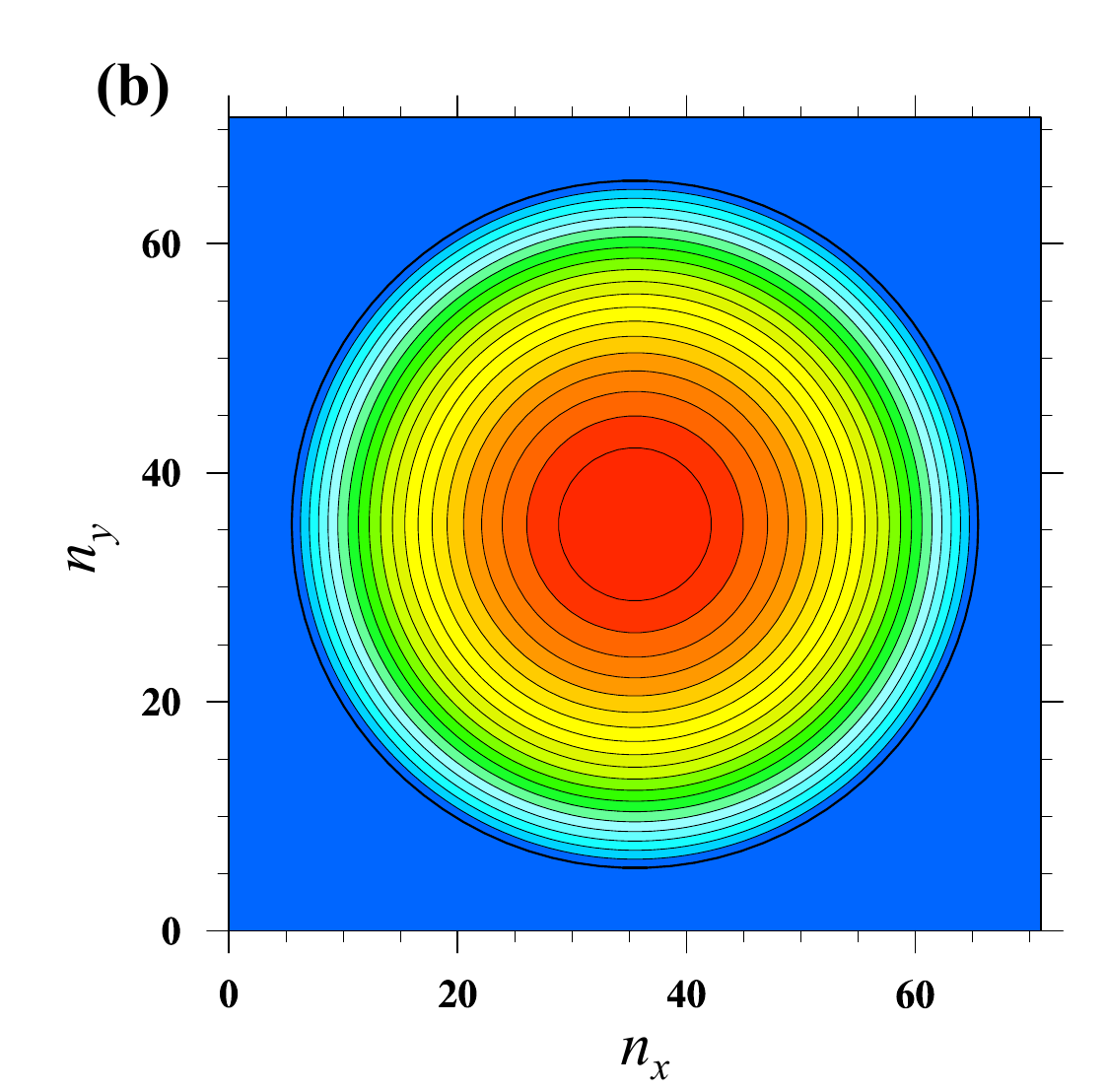}~~~~~\includegraphics[width=55mm]{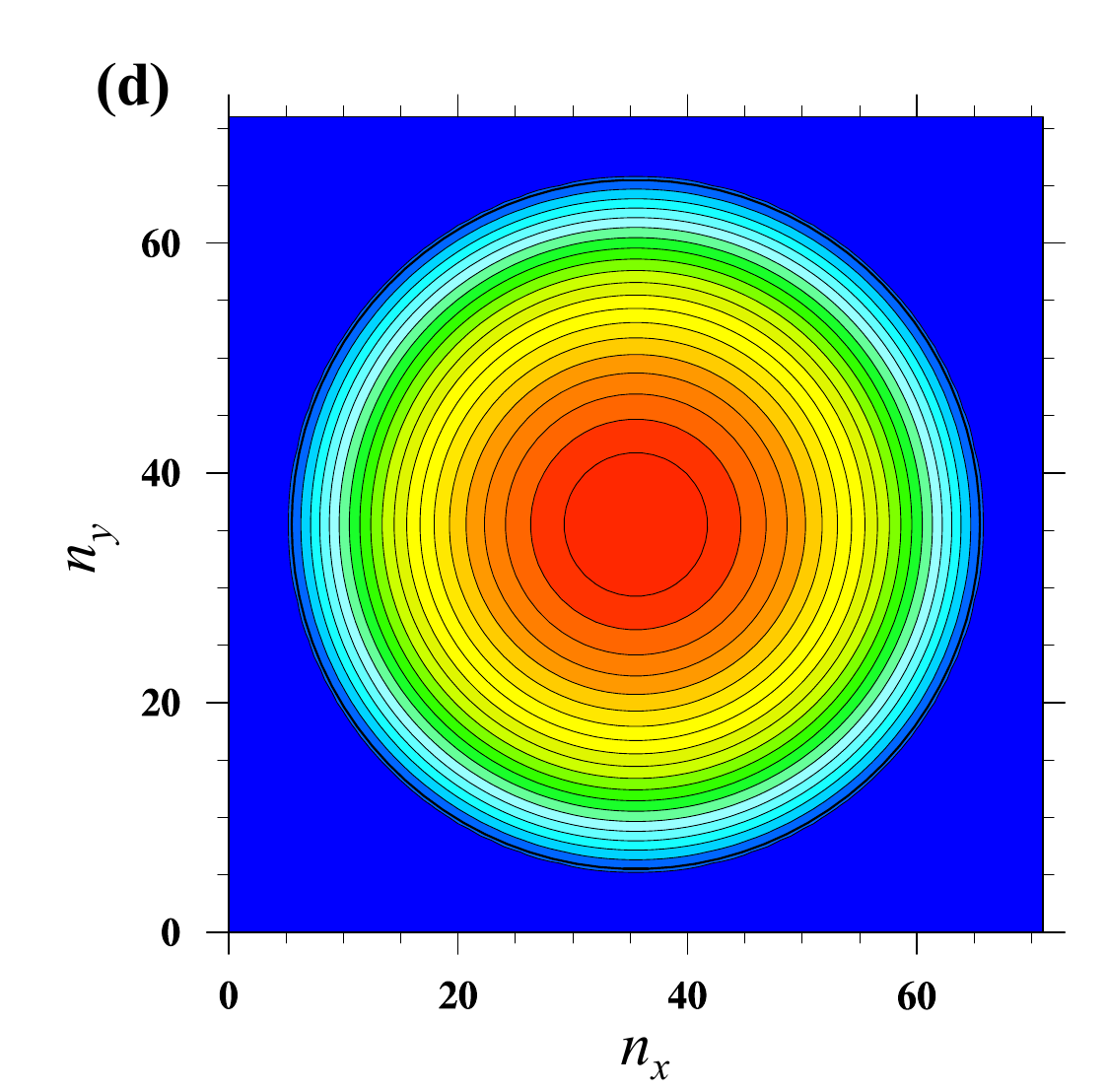}~~~~~\includegraphics[width=55mm]{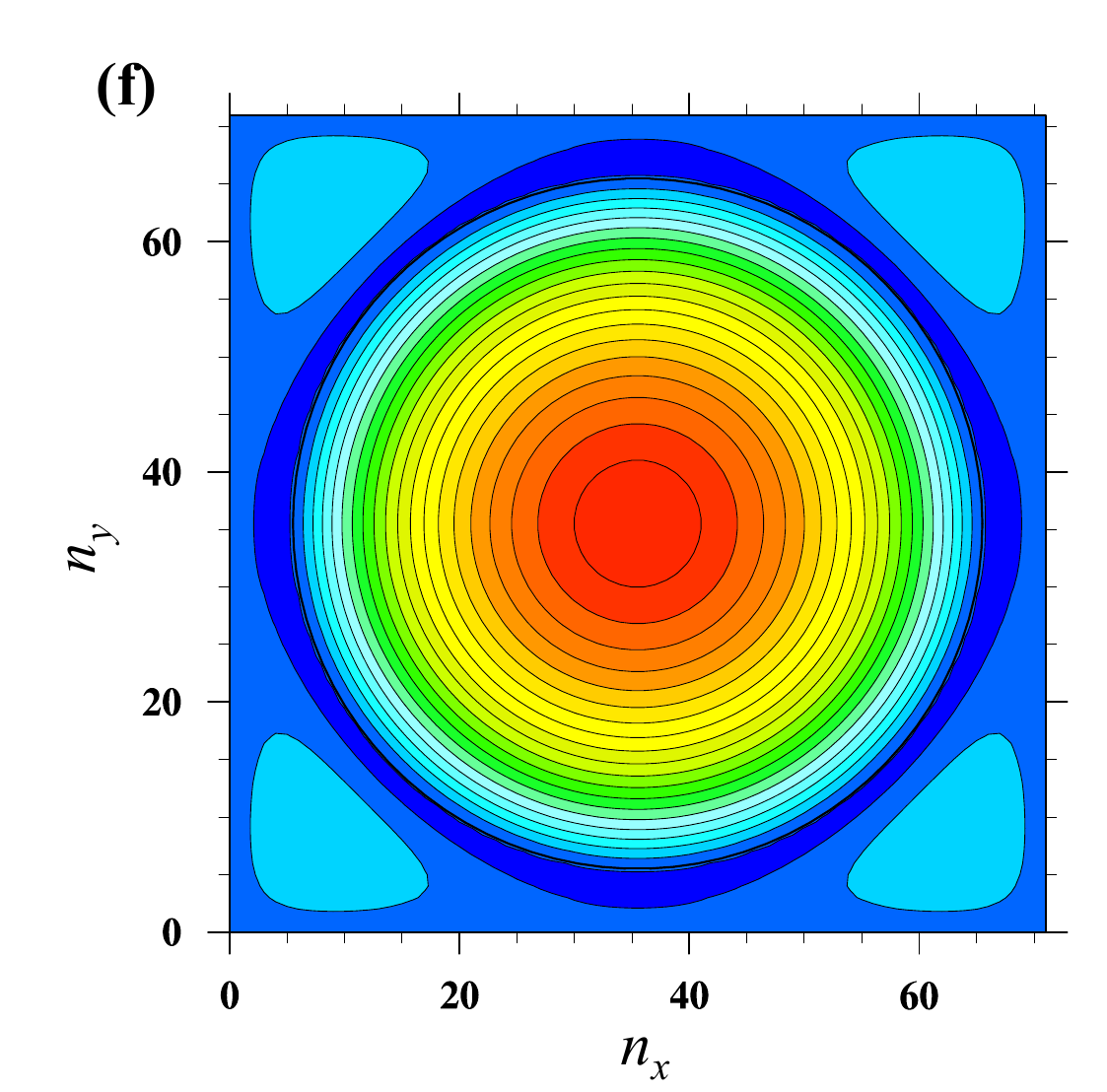}}
\vspace{0.05in}
\caption{The velocity contours and profiles of a laminar pipe flow at its steady state: (a) and (d), with Zhao \& Yong's bounce-back; (b) and (e), with Breugem's IBM with a retraction distance of $R_{d}=0.4$, the driving force only applied to the fluid region, (c) and (f), same as (b) and (e), but the driving force applied to the whole computational domain. }
\label{fig:laminarpipecontours}
\end{figure}

First, the velocity contours and profiles at the steady state with (a) Zhao \& Yong's bounce-back, (b) Breugem's IBM with $R_{d} = 0.4$, where the driving body force is applied only to the fluid domain, (c) same as (b) but the driving body force is applied to the whole computational domain are shown in Fig.~\ref{fig:laminarpipecontours}. The Reynolds number of the flow $Re = 2u_{0}R/\nu$ is 100, the radius of the pipe $R = 30\delta x$. The pipe is contained in a computational domain of $n_{x}\times n_{y}\times n_{z} = 72\times 72\times 16$. 
While the velocity profiles with Zhao \& Yong's bounce-back collapse well with the theoretical solutions at different times, those profiles with Breugem's IBM have slight visible deviations from the theory at later times. Comparing the velocity profiles in case (b) and case (c), we observe that applying the driving force in the fictitious fluid domain (physical domain occupied by solid phase) leads to larger derivation than restricting the driving force in the physical fluid domain. These again suggests that how to appropriately treat the flow in the fictitious domain affects the accuracy of flow in the physical fluid domain, as the two parts can directly exchange information via advection and diffusion through the N-S equations. 

\begin{figure}
\centering
\includegraphics[width=80mm]{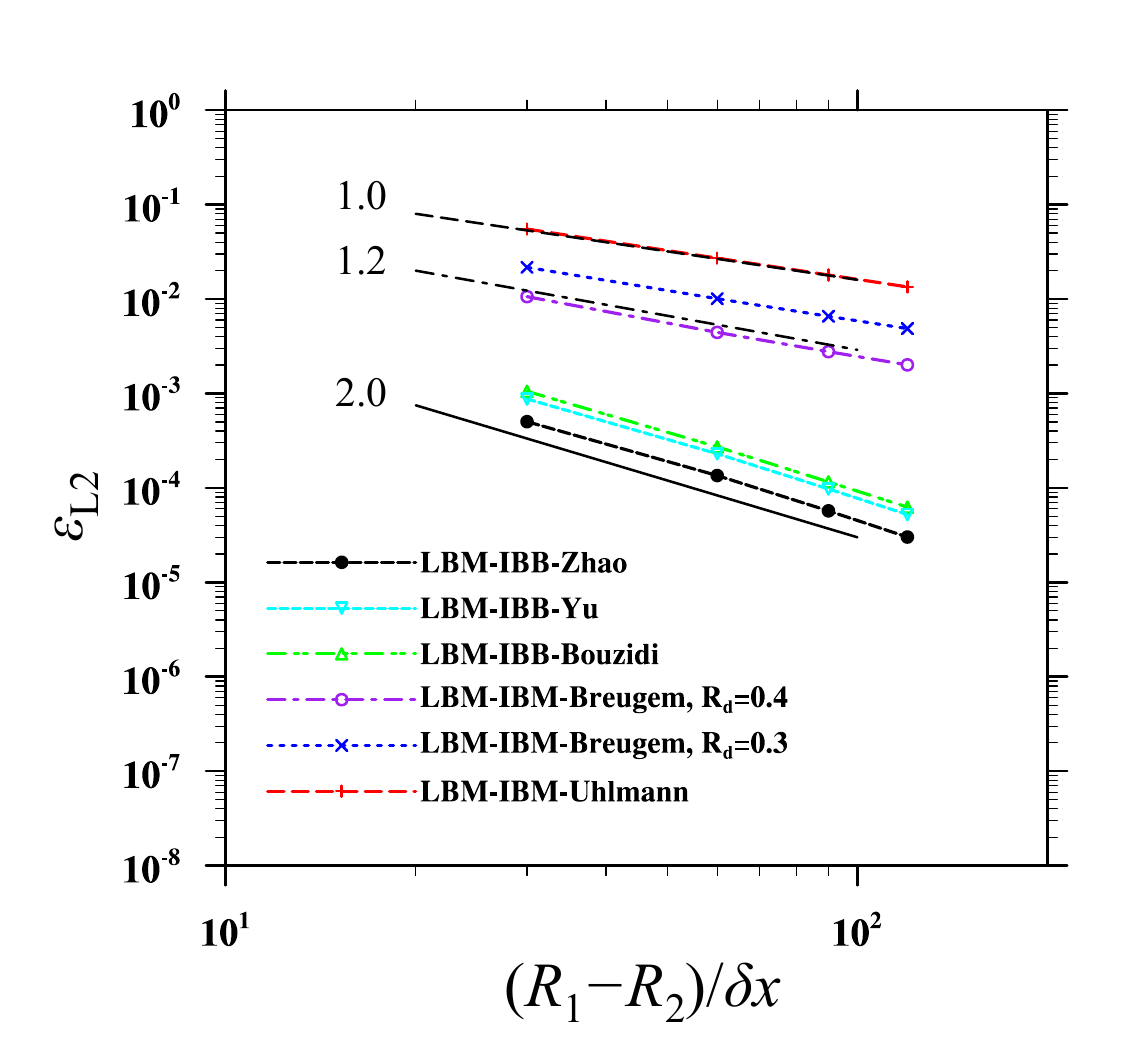}
\caption{The convergence rates of the velocity in a laminar pipe flow.}
\label{fig:velocityconvergencepipe}
\end{figure}

In order to better quantify the numerical errors in the laminar pipe flow simulations, the convergence rates of the L1 and L2 norms of the steady state velocity errors are calculated and presented in Fig.~\ref{fig:velocityconvergencepipe}. Here we examine six boundary treatment schemes, the linear interpolated bounce-back schemes of Bouzidi {\it et al.} and Yu {\it et al.}, Zhao \& Yong's bounce-back scheme, Uhlmann's IBM, Breugem's IBM with retraction distance of $R_{d} = 0.3\delta x$ and $0.4\delta x$. The boundary force in three simulations with IBMs are iterated for 5 times. Similar to the case of circular Couette flow, the numerical errors in the three cases with IBM generally have first-order convergence rate, in contrast to the second-order convergence rates in the three bounce-back cases. While retracting the Lagrangian grid to the solid side significantly reduces the magnitude of the numerical error, the convergence rate is only slightly improved, i.e., from 1.0 to 1.2 with the retraction distance of $R_{d} = 0.4\delta x$. 
According to our earlier derivation in Sec.~\ref{sec:CircularCouette}, as long as the actual boundary is diffused more than the retraction distance by the delta-function, the error induced by the interpolation should always involve the flow in both the fluid and solid regions. Therefore, the improved order of accuracy claimed in previous IBM studies, {\it e.g.}, in \cite{breugem2012second}, remains questionable or at least not generalizable, as we are unable to reproduce the second-order accuracy with the LBM-IBM here.

\subsection{Uniform flow passing a fixed sphere}\label{sec:uniform}

The last case we examine is a uniform stream passing a fixed sphere in an unbounded domain. A spherical particle with a diameter $D$ is fixed at $(x,y,z) = (6D,6D,6D)$ in a cuboid domain of a size $(Lx,Ly,Lz) = (24D,12D,12D)$. A uniform unidirectional upstream flow with ${\vec u} = (u_x,u_y,u_z) = (U_{i},0,0)$ enters the inlet ($x = 0$) of the domain and passes over the fixed particle. The four sides are set to be stress-free, {\it i.e.}, $u_z = 0, \partial u_{x}/\partial z = \partial u_{y}/\partial z = 0$ at $z = 0$ and $z = Lz$, and $u_y = 0, \partial u_{x}/\partial y = \partial u_{z}/\partial y = 0$ at $y = 0$ and $y = Ly$, to mimic the boundary condition in an infinitely large domain. The flow exits the domain with the following outflow boundary condition, $\partial (\rho_{0}{\vec u})/\partial t + U_{o}\partial (\rho_{0}{\vec u})/\partial x = 0$, where $U_{o}$ is the streamwise velocity at the outlet~\cite{lou2013evaluation}. 

\begin{figure}
\centering
\includegraphics[width=160mm]{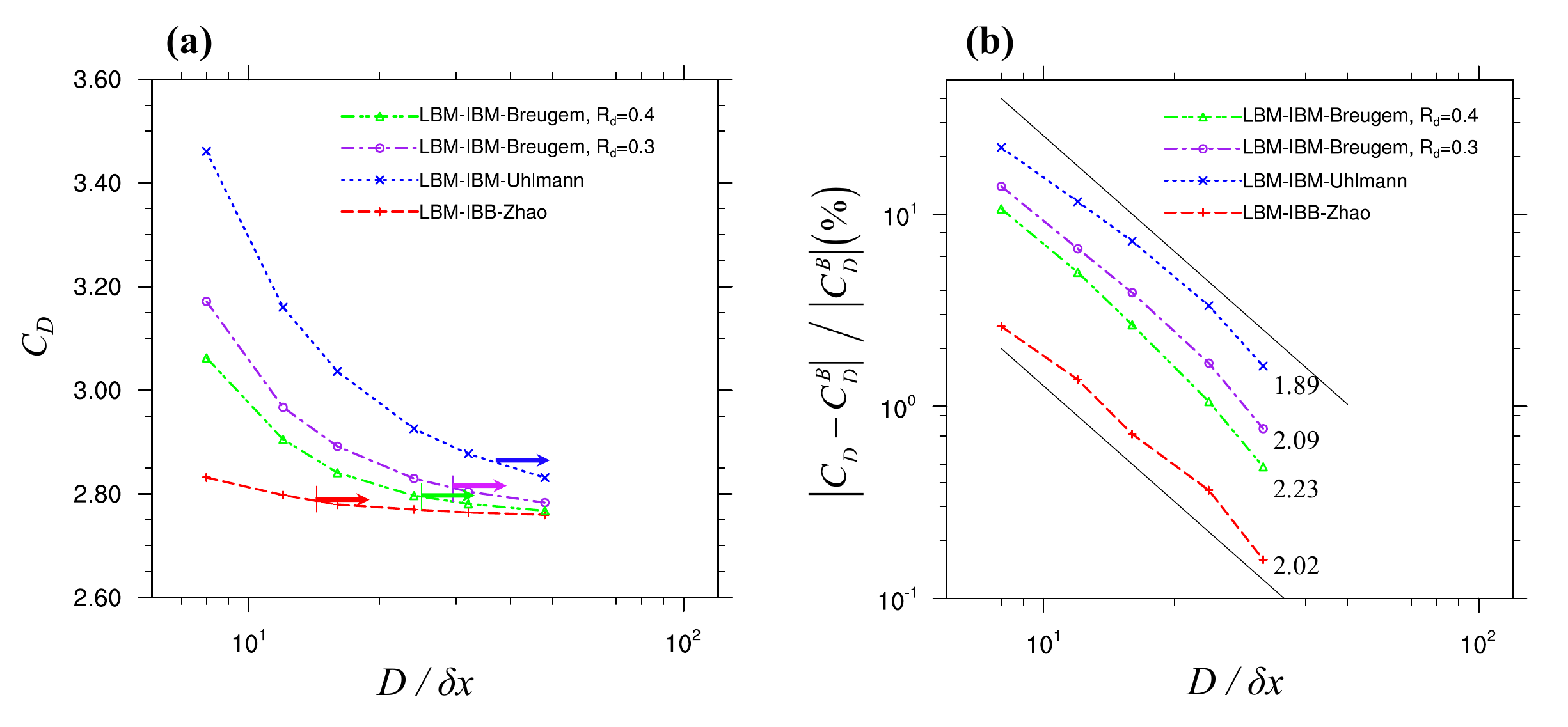}
\caption{The drag coefficients of a uniform flow passing a fixed sphere at $Re_{p} = 20$.}
\label{fig:dragcoefficientsRe20}
\end{figure}

\begin{figure}
\centering
\includegraphics[width=160mm]{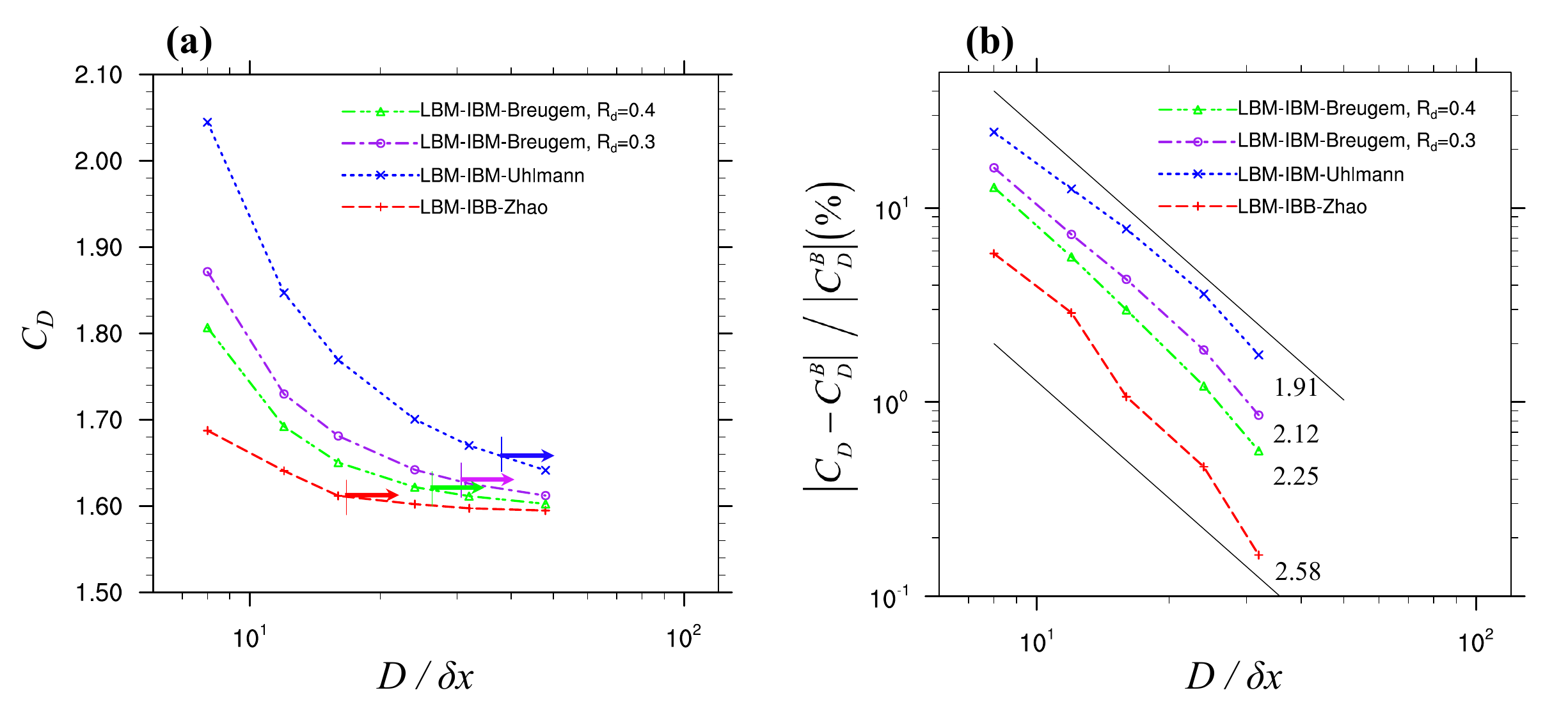}
\caption{The drag coefficients of a uniform flow passing a fixed sphere at $Re_{p} = 50$.}
\label{fig:dragcoefficientsRe50}
\end{figure}

\begin{figure}
\centering
\includegraphics[width=160mm]{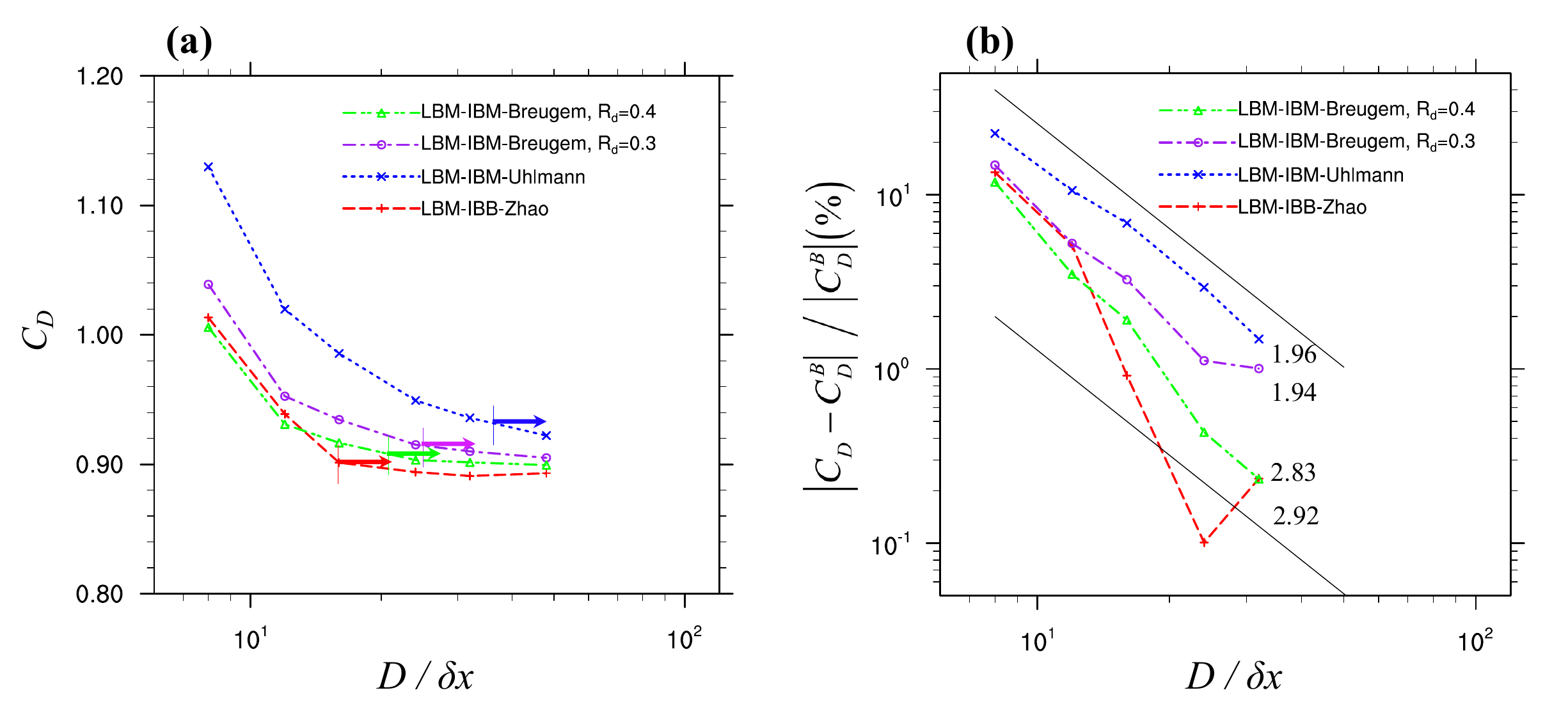}
\caption{The drag coefficients of a uniform flow passing a fixed sphere at $Re_{p} = 150$.}
\label{fig:dragcoefficientsRe150}
\end{figure}

The drag coefficients under three different particle Reynolds numbers, $Re_{p} = U_{i}D/\nu = 20$, 50, and 150 are examined. With these Reynolds numbers, the flow after the sphere is steady and axisymmetric with closed recirculating wake~\cite{jones2008simulation}. At each Reynolds number, we vary the grid resolution, {\it i.e.}, $D/\delta x$ from 8 to 48, and investigate the drag coefficient $C_{D} = 8F_{D}/(\rho_{f}Re_{p}^{2}\pi\nu^{2})$, with kinematic viscosity $\nu$ fixed when varying the grid resolution. The results of drag coefficient at $Re_{p} = 20$, $50$, and $150$ are presented in Fig.~\ref{fig:dragcoefficientsRe20}, Fig.~\ref{fig:dragcoefficientsRe50}, and Fig.~\ref{fig:dragcoefficientsRe150}, respectively. The vertical solid lines in each figure indicate the grid resolution gives an error of $1\%$ using the result of the current boundary treatment with the highest grid resolution as benchmark. At all three Reynolds numbers, Zhao \& Yong's bounce-back scheme always reaches the converged drag coefficient with the coarsest grid resolution among all four boundary treatments. This is perhaps due to the fact that Zhao \& Yong's bounce-back scheme has a second-order accuracy while the immersed boundary algorithms only have first-order accuracy. The retraction of Lagrangian grid points again results in much more accurate results compared to zero retraction distance. $R_{d} = 0.4\delta x$ always appears to be the optimal retraction distance when the four-point delta-function is employed. According to our results, We recommend that Breugem's Lagrangian grid retraction be used when IBM is used for no-slip boundary treatment.

Finally, if we define a ``sufficient" grid resolution as the grid resolution that gives $1\%$ relative error from the converged result, the sufficient grid resolutions for Zhao \& Yong's bounce-back at $Re_{p} = 20$, 50, and 150 are about $D\delta x = 14.3$, $16.7$ and $15.9$. The same quantities are $37.1$, $38.0$, $36.3$ with Uhlmann's IBM, $29.5$, $30.7$, $25.0$ with Breugem's IBM with $R_{d} = 0.3\delta x$, and $25.0$, $26.3$, $20.8$ with Breguem's IBM with $R_{d} = 0.4\delta x$. These results may provide a criterion to assess whether a grid resolution is fine enough to ensure trustworthy results when a certain scheme is adopted for no-slip boundary treatment in a three-dimensional particle-laden flow simulation, at the similar particle Reynolds number.

\section{Conclusions and remarks}\label{sec:conclusions}
In this work, we systematically assessed two categories of no-slip boundary treatment methods, which are the interpolated bounce-back schemes and the immersed boundary method, on an arbitrarily shaped surface in the context of the lattice Boltzmann method. Three representative interpolated bounce-back schemes, including a recently proposed single-node second-order bounce-back scheme~\cite{zhao2017single}, and two popular immersed boundary algorithms are selected. Their performances, especially the accuracy of resulting velocity, hydrodynamic force/torque, and the viscous dissipation rate are carefully benchmarked in four selected flows. In all the flows examined in the present study, the interpolated bounce-back schemes always lead to much more accurate results of velocity, force/torque, and dissipation rate than the immersed boundary algorithms. The immersed boundary algorithms, on the other hand, outperform in suppressing the fluctuations of the calculated hydrodynamic force/torque compared to the interpolated bounce-back schemes. The specific major observations of this present study are summarized as follows.

\begin{itemize}
    \item With immersed boundary algorithms, cautions should be taken to the treatment of the flow in the virtual fluid region, especially when the flow of interest is surrounded by solid objects, such as the Taylor-Couette flow and the circular pipe flow. Unfortunately, the information of how to specify the flow in the virtual fluid region may not be available in the description of physical problems.
    \item Our simulations confirm that the immersed boundary algorithms using the regularized delta-functions to interpolate information between the Eulerian and Lagrangian grids have only the first-order accuracy in flow velocity calculation. 
    This conclusion holds no matter whether the Lagrangian grid points are retracted towards the solid phase or not. 
    We prove with a theoretical analysis that information exchange between the Eulerian and Lagrangian grids via the regularized delta-function always induces an first-order error term as long as the velocity gradient is discontinuous across the solid-fluid interface. 
    On the other hand, the interpolated bounce-back schemes can ensure a second-order accuracy of
    the simulated velocity. The magnitudes of the velocity errors in the interpolated bounce-back schemes are also much smaller than their counterparts from immersed boundary algorithms.
    \item The {\it local}  hydrodynamic force and torque calculated with immersed boundary algorithms are only first-order accurate. These local first-order errors may cancel out to result in an apparent second-order accurate integral force/torque, as shown in Sec.~\ref{sec:uniform}. However, this cancellation may not be generalized. In the Taylor-Couette flow, the integral force is still first-order accurate. The forces calculated with interpolated bounce-back schemes and momentum exchange methods have a second-order accuracy in all the flows examined. 
    \item The most serious problem we find for the immersed boundary method is that the local dissipation rate can be significantly underestimated. This is because the sharp fluid-solid interface is diffused by the regularized delta-functions, which results in a  smaller velocity gradient near the interface. The same problem is not present with the interpolated bounce-back schemes, which can
    be viewed as a sharp-interface treatment
    for no-slip boundary.
    \item For moving particle problems, the high-frequency fluctuations in the force/torque are better suppressed in the immersed boundary methods than the interpolated bounce-back schemes. When the particle/fluid density ratio is close to unity, both Feng \& Michaelides's scheme and Kempe {\it et al.}'s scheme are suitable to update the particle motion.   
    \item We present convergence studies to find out the sufficient grid resolution associated with each boundary treatment method for 2D circular and 3D spherical particles. Since the interpolated bounce-back schemes have better accuracy than the immersed boundary method, its grid resolution requirement for a converged result is lower. For 2D circular particles, the sufficient grid resolutions for using interpolated bounce-back schemes and the immersed boundary method are $D/\delta x = 10$ and $D/\delta x = 15$, respectively. For 3D spherical particles, the sufficient grid resolutions become $D/\delta x = 15$ and $D/\delta x = 25$, respectively, for particle Reynolds number between 20 to 150. 
    Here the immersed boundary method refers to Breugem's IBM with an appropriate retraction distance. When Uhlmann's IBM with zero retraction distance is used, the sufficient grid resolution should be doubled. 
\end{itemize}

{\bf Acknowledgements}: This work has been supported by the National Natural Science Foundation of China (91852205 \& 91741101), and by the U.S. National Science Foundation (NSF) under 
grants CNS1513031 and CBET-1706130.
  Computing resources are 
provided by Center for Computational Science and Engineering of Southern University of Science and Technology and by National Center for Atmospheric Research through CISL-P35751014, and CISL-UDEL0001.

\section{Appendix A: Analytic solution of the Taylor-Couette flow}
The analytic solution of the transient Taylor-Couette flow was derived by He~\cite{he2015high}. Here we just repeat He's derivation for readers' convenience. The simplified N-S equations for the Taylor-Couette flow is written as
\begin{equation}
    \begin{split}
    &\frac{\partial u_{\theta}}{\partial t} = \nu\frac{\partial}{\partial r}\left[\frac{1}{r}\frac{\partial}{\partial r}\left(r u_{\theta}\right)\right],\\
    &u_{\theta}\left(t,R_{1}\right) = \Omega_{1}R_{1},~~~u_{\theta}\left(t,R_{2}\right) = \Omega_{2}R_{2},~~~u_{\theta}\left(t=0,r\right) = 0,
    \end{split}
    \label{eq:TCgoverning}
\end{equation}
where $u_{\theta}$ is the flow velocity in the angular direction, $\nu$ is the kinematic viscosity of the fluid, $t$ and $r$ are the time and radius coordinate, respectively. $R_{1}$ and $R_{2}$ are the radius of the inner and outer cylinders confining the flow, $\Omega_{1}$ and $\Omega_{2}$ are the corresponding angular velocities, respectively. The time-dependent solution of this flow can be expressed as~\cite{he2015high}:
\begin{equation}
    u_{\theta}\left(r,t\right) = u_{\theta}^{S} + \sum_{n=1}^{\infty}A_{n}e^{-\frac{\nu\lambda_{n}^2t}{\left(R_{2}-R_{1}\right)^2}}\left[J_{1}\left(\frac{\lambda_{n}r}{R_{1}}\right)-\frac{J_{1}\left(\lambda_{n}\right)}{Y_{1}\left(\lambda_{n}\right)}Y_{1}\left(\frac{\lambda_{n}r}{R_{1}}\right)\right]
    \label{eq:TCsolution}
\end{equation}
where $u_{\theta}^{S}$ is the steady state solution 
\begin{equation}
    u_{\theta}^{S} = \frac{1}{r}\frac{\Omega_{1}-\Omega_{2}}{R_{1}^{-2}-R_{2}^{-2}} + \frac{\Omega_{2}R_{2}^2-\Omega_{1}R_{1}^2}{R_{2}^2-R_{1}^2}r,
    \label{eq:steadystate}
\end{equation}
$J_{1}$ and $Y_{1}$ are the first-order Bessel function of the first and the second kind, $\lambda_{n}$ is the $n$th root satisfies
\begin{equation}
\begin{split}
    &J_{1}\left(\lambda_{n}\right)Y_{1}\left(\lambda_{n}\gamma\right) - J_{1}\left(\lambda_{n}\gamma\right)Y_{1}\left(\lambda_{n}\right) = 0,\\
    &0<\lambda_{1}<\lambda_{2}<\lambda_{3}<\cdots<\lambda_{n}<\cdots\rightarrow \infty.
    \end{split}
    \label{eq:definitionlambda}
\end{equation}
where $\gamma = R_{2}/R_{1}$ is the radii ratio between the outer and the inner cylinder. The $n$th coefficient of the series $A_{n}$ is
\begin{equation}
A_{n} = \frac{\int_{1}^{\gamma}\left(-u_{\theta}^{S}\right)\frac{r}{R_{1}}\left[J_{1}\left(\frac{\lambda_{n}r}{R_{1}}\right)-\frac{J_{1}\left(\lambda_{n}\right)}{Y_{1}\left(\lambda_{n}\right)}Y_{1}\left(\frac{\lambda_{n}r}{R_{1}}\right)\right]dr}{\int_{1}^{\gamma}\frac{r}{R_{1}}\left[J_{1}\left(\frac{\lambda_{n}r}{R_{1}}\right)-\frac{J_{1}\left(\lambda_{n}\right)}{Y_{1}\left(\lambda_{n}\right)}Y_{1}\left(\frac{\lambda_{n}r}{R_{1}}\right)\right]^{2}dr} = \frac{\rm{top}}{\rm{bottom}}.
    \label{eq:series}
\end{equation}

The integrals in Eq.~(\ref{eq:series}) can be calculated as
\begin{subequations}
\begin{align}
    \begin{split}
        {\rm top} = & -\frac{1}{\lambda_{n}^{2}}\left[
        c_{2}J_{0}\left(\lambda_{n}\right)\lambda_{n}
        -2c_{1}J_{1}\left(\lambda_{n}\right)
        +c_{1}J_{0}\left(\lambda_{n}\right)\lambda_{n}
        +2\alpha c_{1}Y_{1}\left(\lambda_{n}\right)\right .\\
        &-\alpha c_{1}Y_{0}\left(\lambda_{n}\right)\lambda_{n}
        -c_{2}\alpha Y_{0}\left(\lambda_{n}\right)\lambda_{n} 
        -c_{2}J_{0}\left(\lambda_{n}\gamma\right)\lambda_{n}
        +2c_{1}\gamma J_{1}\left(\lambda_{n}\gamma\right) \\
        & \left .- c_{1}J_{0}\left(\lambda_{n}\gamma\right)\lambda_{n}\gamma^{2}
        -2\alpha c_{1}\gamma Y_{1}\left(\lambda_{n}\gamma\right)
        +\alpha c_{1}Y_{0}\left(\lambda_{n}\gamma\right)\lambda_{n}\gamma^{2}
        +c_{2}\alpha Y_{0}\left(\lambda_{n}\gamma\right)\lambda_{n}\right]
    \end{split}
    \label{eq:topbottom:A}\\
    \begin{split}
        {\rm bottom} = & -\frac{1}{2\lambda_{n}}\left\{ 
        \lambda_{n} \left[J_{1}\left(\lambda_{n}\right)\right]^{2} 
        -2J_{0}\left(\lambda_{n}\right)J_{1}\left(\lambda_{n}\right) 
        +\left[J_{0}\left(\lambda_{n}\right)\right]^2\lambda_{n}\right.\\
        &-2\alpha J_{1}\left(\lambda_{n}\right)Y_{1}\left(\lambda_{n}\right)\lambda_{n}
        +2\alpha Y_{0}\left(\lambda_{n}\right)J_{1}\left(\lambda_{n}\right)
        -2\alpha Y_{0}\left(\lambda_{n}\right)J_{0}\left(\lambda_{n}\right)\lambda_{n}\\
        &+2\alpha J_{0}\left(\lambda_{n}\right)Y_{1}\left(\lambda_{n}\right)
        +\alpha^{2}\left[Y_{1}\left(\lambda_{n}\right)\right]^{2}\lambda_{n}
        -2\alpha^{2}Y_{0}\left(\lambda_{n}\right)Y_{1}\left(\lambda_{n}\right)\\
        &+\alpha^{2}\left[Y_{0}\left(\lambda_{n}\right)\right]^{2}\lambda_{n}
        -\gamma^{2}\left[J_{1}\left(\lambda_{n}\gamma\right)\right]^{2}\lambda_{n}
        +2\gamma J_{0}\left(\lambda_{n}\gamma\right)J_{1}\left(\lambda_{n}\gamma\right)\\
        &-\gamma^{2}\left[J_{0}\left(\lambda_{n}\gamma\right)\right]^{2}\lambda_{n}
        +2\alpha\gamma^{2}J_{1}\left(\lambda_{n}\gamma\right)Y_{1}\left(\lambda_{n}\gamma\right)\lambda_{n}
        -2\gamma\alpha Y_{0}\left(\lambda_{n}\gamma\right)J_{1}\left(\lambda_{n}\gamma\right)\\
        &+2\alpha\gamma^{2}Y_{0}\left(\lambda_{n}\gamma\right)J_{0}\left(\lambda_{n}\gamma\right)\lambda_{n}
        -2\gamma\alpha J_{0}\left(\lambda_{n}\gamma\right)Y_{1}\left(\lambda_{n}\gamma\right)
        -\alpha^{2}\gamma^{2}\left[Y_{1}\left(\lambda_{n}\gamma\right)\right]^{2}\lambda_{n}\\
        &\left .2\gamma\alpha^{2}Y_{0}\left(\lambda_{n}\gamma\right)Y_{1}\left(\lambda_{n}\gamma\right)
        -\alpha^{2}\gamma^{2}\left[Y_{0}\left(\lambda_{n}\gamma\right)\right]^{2}\lambda_{n}\right\}
    \end{split}
        \label{eq:topbottom:B}
\end{align}
\label{eq:topbottom}
\end{subequations}
where
\begin{equation}
    c_{1} = \frac{\Omega_{2}R_{2}\gamma-\Omega_{1}R_{1}}{\gamma^{2}-1},~~~c_{2} = \frac{\Omega_{1}R_{1}\gamma^2-\Omega_{2}R_{2}\gamma}{\gamma^2 - 1},~~~\alpha = \frac{J_{1}\left(\lambda_{n}\right)}{Y_{1}\left(\lambda_{n}\right)}.
    \label{eq:coeff}
\end{equation}
\section{Appendix B: a theoretical examination on the order of accuracy of immersed boundary method}

\begin{figure}
\centering
\includegraphics[width=90mm]{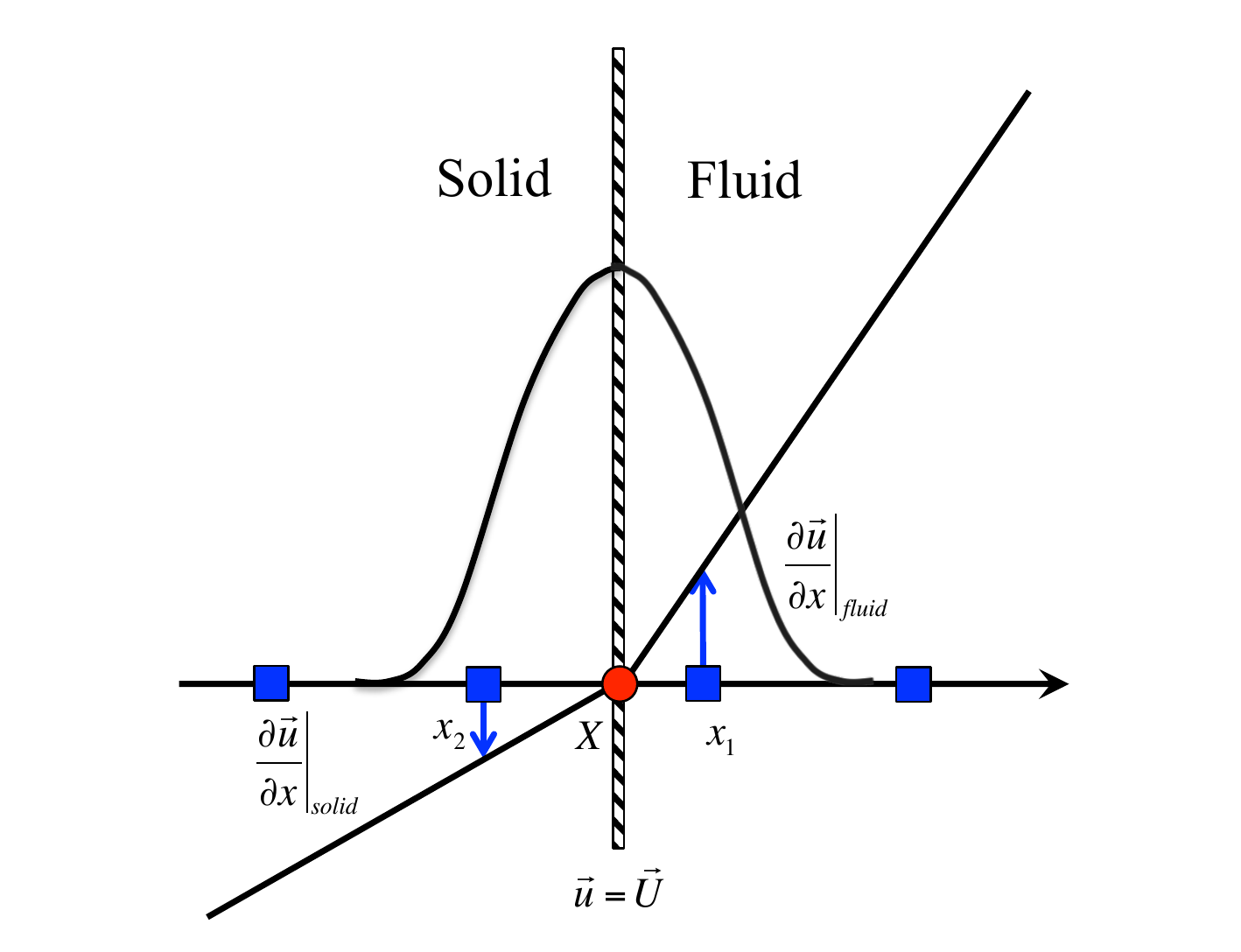}
\caption{The grid arrangement at a fluid-solid interface.}
\label{fig:boundaryconfig}
\end{figure}

A configuration of Lagrangian-Eulerian grid system for a one-dimensional fluid-solid interface is sketched in Fig.~\ref{fig:boundaryconfig}, where $x_{1}$ and $x_{2}$ are two Eulerian grid points on the fluid side and the solid side of the Lagrangian grid point $X$. The interpolation of unforced velocity from the Eulerian grid to the Lagrangian grid and the redistribution of the boundary force the other way around take place at $x_{1}$, $x_{2}$, and $X$. Let us assume the velocity field prior to applying the boundary forcing is accurate, {\it i.e.},
\begin{equation}
    \vec{u}\left(x_{1}\right) = \vec{u}_{exact}\left(x_{1}\right),~~~~~\vec{u}\left(x_{2}\right) = \vec{u}_{exact}\left(x_{2}\right).
    \label{eq:Eulerianvelocity}
\end{equation}
The boundary force on the Lagrangian point $\vec{\cal F}(X)$ would be
\begin{equation}
    \vec{\cal F}\left(X\right)\delta t = \rho \left[ \vec{U}\left(X\right) - \vec{u}\left(X\right) \right] = \rho \left[ \vec{U}\left(X\right) - \phi_{1}\vec{u}_{exact}\left(x_{1}\right) - \phi_{2}\vec{u}_{exact}\left(x_{2}\right) \right],  
    \label{eq:velinterpolation}
\end{equation}
where $\phi_{1} = \phi\left(x_{1}-X\right)$ and $\phi_{2}=\phi\left(x_{2}-X\right)$ are the weighting factors obtained from the delta-function, $\phi_{1} + \phi_{2} = 1$. The boundary forces distributed back to $x_{1}$ and $x_{2}$ are
\begin{subequations}
\begin{align}
    \vec{\cal F}\left(x_{1}\right)\delta t = \phi_{1}\vec{\cal F}\left(X\right)\delta t = \phi_{1}\rho \left[ \vec{U}\left(X\right) - \phi_{1}\vec{u}_{exact}\left(x_{1}\right) - \phi_{2}\vec{u}_{exact}\left(x_{2}\right) \right],
    \label{eq:Eulerianforce:A}\\
    \vec{\cal F}\left(x_{2}\right)\delta t = \phi_{2}\vec{\cal F}\left(X\right)\delta t = \phi_{2}\rho \left[ \vec{U}\left(X\right) - \phi_{1}\vec{u}_{exact}\left(x_{1}\right) - \phi_{2}\vec{u}_{exact}\left(x_{2}\right) \right],
    \label{eq:Eulerianforce:B}
\end{align}
\label{eq:Eulerianforce}
\end{subequations}
After applying the boundary forcing, the velocity at $x_{1}$ and $x_{2}$ are
\begin{equation}
\tilde{\vec{u}}\left(x_{1}\right) = \vec{u}_{exact}\left(x_{1}\right) + \frac{1}{\rho}\vec{\cal F}\left(x_{1}\right)\delta t,~~~~~\tilde{\vec{u}}\left(x_{2}\right) = \vec{u}_{exact}\left(x_{2}\right) + \frac{1}{\rho}\vec{\cal F}\left(x_{2}\right)\delta t.
\label{eq:postforcingvel}
\end{equation}
Since the velocity field before applying the boundary forcing is already exact, the errors introduced by the boundary force at $x_{1}$ and $x_{2}$ are simply
\begin{subequations}
\begin{align}
    \Delta\vec{u}_{1} = \tilde{\vec{u}}\left(x_{1}\right)-\vec{u}_{exact}\left(x_{1}\right) = \phi_{1}\rho \left[ \vec{U}\left(X\right) - \phi_{1}\vec{u}_{exact}\left(x_{1}\right) - \phi_{2}\vec{u}_{exact}\left(x_{2}\right) \right],
    \label{eq:velcityerror:A}\\
    \Delta\vec{u}_{2} = \tilde{\vec{u}}\left(x_{2}\right)-\vec{u}_{exact}\left(x_{2}\right) = \phi_{2}\rho \left[ \vec{U}\left(X\right) - \phi_{1}\vec{u}_{exact}\left(x_{1}\right) - \phi_{2}\vec{u}_{exact}\left(x_{2}\right) \right].
\end{align}
\label{eq:velocityerror}
\end{subequations}
Performing a Taylor expansion for $\vec{u}_{exact}\left(x_{1}\right)$ and  $\vec{u}_{exact}\left(x_{2}\right)$ with respect to $X$, {\it i.e.},
\begin{equation}
\begin{split}
     &\vec{u}_{exact}\left(x_{1}\right) = \vec{U}\left(X\right) + \frac{d \vec{u}}{d x}\mid_{fluid}\left(x_{1}-X\right) + O\left(\Delta x^2\right),\\
     &\vec{u}_{exact}\left(x_{2}\right) = \vec{U}\left(X\right) + \frac{d \vec{u}}{d x}\mid_{solid}\left(x_{2}-X\right) + O\left(\Delta x^2\right),
     \end{split}
    \label{eq:Taylorexpansion}
\end{equation}
Substitute Eq.~(\ref{eq:Taylorexpansion}) to Eq.~(\ref{eq:velocityerror}), we shall obtain
\begin{equation}
    \begin{split}
       & \Delta\vec{u}_{1} = \phi_{1}\left\{-\phi_{1}\left[\frac{d \vec{u}}{d x}\mid_{fluid}\left(x_{1}-X\right)+O\left(\Delta x^2\right)\right]-\phi_{2}\left[\frac{d \vec{u}}{d x}\mid_{solid}\left(x_{2}-X\right)+O\left(\Delta x^2\right)\right]\right\},\\
        & \Delta\vec{u}_{2} = \phi_{2}\left\{-\phi_{1}\left[\frac{d \vec{u}}{d x}\mid_{fluid}\left(x_{1}-X\right)+O\left(\Delta x^2\right)\right]-\phi_{2}\left[\frac{d \vec{u}}{d x}\mid_{solid}\left(x_{2}-X\right)+O\left(\Delta x^2\right)\right]\right\}.
    \end{split}
    \label{eq:finalerror}
\end{equation}
Note that the delta-function should have the property $\sum\left(x-X\right)\phi\left(x-X\right) = 0$~(The 4-point cosine delta-function employed frequently does not possess this property strictly, but it does not affect the argument, {\it i.e.}, $\phi_{1}\left(x_{1}-X\right)+\phi_{2}\left(x_{2}-X\right) = 0$. To simplify the notation, denote $\phi_{1}\left(x_{1}-X\right) = c$, $\phi_{2}\left(x_{2}-X\right) = -c$, $c\sim O(\Delta x)$, Eq.~(\ref{eq:finalerror}) becomes
\begin{equation}
  \Delta\vec{u}_{1} = -c\phi_{1}\left(\frac{d \vec{u}}{d x}\mid_{fluid} - \frac{d \vec{u}}{d x}\mid_{solid}\right) + O\left(\Delta x^2\right),~~~~\Delta\vec{u}_{2} = -c\phi_{2}\left(\frac{d \vec{u}}{d x}\mid_{fluid} - \frac{d \vec{u}}{d x}\mid_{solid}\right) + O\left(\Delta x^2\right).
    \label{eq:velocityerrororder}
\end{equation}
Therefore, only when the velocity gradient is continuous, according to the Taylor expansion
\begin{equation}
     \frac{d \vec{u}}{d x}\mid_{solid} = \frac{d \vec{u}}{d x}\mid_{fluid} + O\left(\Delta x\right),
    \label{eq:continousvelgradient}
\end{equation}
$\Delta\vec{u}_{1}$ and $\Delta\vec{u}_{2}$ in Eq.~(\ref{eq:velocityerrororder}) can then have a second-order accuracy. When the velocity gradient is discontinuous, the boundary forcing process shown above always induces a first-order error to the velocity field.
Generally speaking, the velocity gradient is, unfortunately, not continuous across a fluid-solid interface, thus IBM using the delta-function degrades the accuracy to the first-order. 

In IBM, The hydrodynamic force $\vec{F}$ on a solid object is calculated as
\begin{equation}
    \vec{F}\delta t = \sum_{l}\vec{\cal F}\left(X_{l}\right)\delta t\Delta V_{l},
    \label{eq:torque}
\end{equation}
where $X_{l}$ is the location of the $l$th Lagrangian grid point,  $\Delta V_{l}$ is the control volume of $X_{l}$. Similarly, we can define the exact hydrodynamic force as
\begin{equation}
    \vec{F}_{exact}\delta t = \sum_{l}\vec{\cal F}_{exact}\left(X_{l}\right)\delta t\Delta V_{l},
    \label{eq:torqueexact}
\end{equation}
The error of hydrodynamic force in IBM is simply the difference between the two, {\it i.e.},
\begin{equation}
\begin{split}
   & \Delta \vec{F} \delta t= \vec{F}\delta t - \vec{F}_{exact}\delta t = \sum_{l}\left[\vec{\cal F}\left(X_{l}\right)-\vec{\cal F}_{exact}\left(X_{l}\right)\right]\delta t\Delta V_{l}\\
   & = \sum_{l}\left\{\left[\vec{U}\left(X_{l}\right)-\phi_{1}\vec{u}\left(x_{1}\right)-\phi_{2}\vec{u}\left(x_{2}\right)\right]-\left[\vec{U}\left(X_{l}\right)-\phi_{1}\vec{u}_{exact}\left(x_{1}\right)-\phi_{2}\vec{u}_{exact}\left(x_{2}\right)\right]\right\}\Delta V_{l}\\
   &=-\sum_{l}\left[\phi_{1}\Delta\vec{u}\left(x_{1}\right)+\phi_{2}\Delta\vec{u}\left(x_{2}\right)\right]\Delta V_{l}.
    \end{split}
    \label{eq:torqueerror}
\end{equation}

Substituting $\Delta\vec{u}\left(x_{1}\right)$ and $\Delta\vec{u}\left(x_{2}\right)$ obtained in Eq.~(\ref{eq:velocityerrororder}) results in
\begin{equation}
\begin{split}
   \Delta \vec{F} \delta t =-\sum_{l}\left[\left(c\phi_{1}^{2}+c\phi_{2}^{2}\right)\left(\frac{d \vec{u}}{d x}\mid_{fluid} - \frac{d \vec{u}}{d x}\mid_{solid}\right) + O\left(\Delta x^2\right)\right]\Delta V_{l},
    \end{split}
    \label{eq:torqueerrorfinal}
\end{equation}
where $c\phi_{1}^{2}+c\phi_{2}^{2}$ is always positive (or negative) and on the order of $\Delta x$. Evidently, for the local first-order error in Eq.~(\ref{eq:torqueerrorfinal}) to cancel out in the summation, the difference of the velocity derivatives across the fluid-solid interface must be follow certain patterns, or at least being positive on some Lagrangian nodes and being negative on the others. While we do observe this situation in the case of a uniform flow passing a fixed sphere, which has also been reported in the literature~\cite{peng2008comparative,breugem2012second,zhou2014second}, this observation may not be generalized. In the case of Taylor-Couette flow, the hydrodynamic force calculation with IBM is only first-order accurate.



\bibliographystyle{elsarticle-num}

\bibliography{elsarticle-template}

\end{document}